\documentclass[12pt]{article}
\pdfoutput=1
\usepackage{jheppub}

\usepackage{amsmath,bbm,array,amsfonts,graphicx,wrapfig,lscape,float,mathtools,multirow,longtable,rotating,subfigure,ulem,cancel}

\usepackage[section]{placeins}

\newcommand{\be}{\begin{equation}}
\newcommand{\ee}{\end{equation}}
\newcommand{\beq}{\begin{equation}}
\newcommand{\beql}[1]{\begin{equation}\label{#1}}
\newcommand{\eeq}{\end{equation}}
\newcommand{\ba}{\begin{array}}
\newcommand{\ea}{\end{array}}
\newcommand{\bea}{\begin{eqnarray}}
\newcommand{\beal}[1]{\begin{eqnarray}\label{#1}}
\newcommand{\eea}{\end{eqnarray}}
\newcommand{\ben}{\begin{enumerate}}
\newcommand{\een}{\end{enumerate}}
\newcommand{\bean}{\begin{eqnarray*}}
\newcommand{\eean}{\end{eqnarray*}}

\newcommand{\btab}[1]{\begin{tabular}{#1}}
\newcommand{\etab}{\end{tabular}}

\newcommand{\comment}[1]{}

\let\n=\nu

\newcommand{\qed}{\nobreak \ifvmode \relax \else
      \ifdim\lastskip<1.5em \hskip-\lastskip
      \hskip1.5em plus0em minus0.5em \fi \nobreak
      \vrule height0.75em width0.5em depth0.25em\fi}

\numberwithin{equation}{section}

\title{Highest Weight Generating Functions for Hilbert Series}

\author{Amihay Hanany,}
\author{Rudolph Kalveks}

\affiliation{
Theoretical Physics Group, The Blackett Laboratory,
Imperial College London, \\
Prince Consort Road, London SW7 2AZ, United Kingdom
}

\emailAdd{a.hanany@imperial.ac.uk, rudolph.kalveks09@imperial.ac.uk}

\preprint{Imperial/TP/14/AH/07}

\abstract{We develop a new method for representing Hilbert series based on the highest weight Dynkin labels of their underlying symmetry groups. The method draws on plethystic functions and character generating functions along with Weyl integration. We give explicit examples showing how the use of such \textit{highest weight generating functions} (``HWGs") permits an efficient encoding and analysis of the Hilbert series of the vacuum moduli spaces of classical and exceptional SQCD theories and also of the moduli spaces of instantons. We identify how the HWGs of gauge invariant operators of a selection of classical and exceptional SQCD theories result from the interaction under symmetrisation between a product group and the invariant tensors of its gauge group. In order to calculate HWGs, we derive and tabulate character generating functions for low rank groups by a variety of methods, including a general character generating function, based on the Weyl Character Formula, for any classical or exceptional group.\\

~\today}

\begin{document}

\maketitle
\section{Introduction}
Hilbert series (``HS") provide an important tool in the analysis of the moduli spaces associated with various field theories. One key area of interest is the study of the moduli spaces of the gauge invariant operators (``GIOs") constructed from fields transforming in some combination of flavour, colour and/or other symmetry group representations. Such product group structures arise in supersymmetric (``SUSY") quiver gauge theories.

Considerable progress has been made in recent years under the auspices of the Plethystics Program \cite{Benvenuti:2006qr, Feng:2007ur} in the systematic calculation and analysis of the Hilbert series of the moduli spaces of SUSY quiver gauge theories. The Plethystics Program deploys plethystic functions and Weyl integration to construct generating functions for the Hilbert series of objects transforming under various representations of classical or exceptional Lie groups or product groups. 

The Hilbert series for these SUSY quiver gauge theories enumerate holomorphic operators, such as the chiral ring of BPS operators characterising the moduli space of a supersymmetric vacuum. Theories that have been well studied include SQCD for classical gauge groups \cite{Hanany:2008kn, Gray:2008yu, Hanany:2008sb}, the moduli spaces of one and two instantons \cite{Benvenuti:2010pq, Hanany:2012dm} and the master space of ${\cal N} =1$ SUSY gauge theories based on particular Calabi-Yau manifolds \cite{Forcella:2008bb, Forcella:2008eh}. In SQCD, the theories are without superpotentials $(W \equiv 0)$ and the gauge group structures are based on those specified by the chiral scalar field definitions; in the other cases, such as SUSY quiver theories for instanton moduli spaces, the theories have non-trivial superpotentials $W$ and the vacuum conditions can place F-flatness constraints on the field representations and/or give rise to hidden symmetry groups. In all cases, however, the Hilbert series associated with the moduli spaces are amenable to analysis by similar methods.

Key analytical tools within the Plethystics Program include the Plethystic Exponential (ÒPEÓ), the Fermionic Plethystic Exponential (ÒPEFÓ), Plethystic Logarithm (ÒPLÓ) and Weyl integration. The PE symmetrises polynomials and can be used to generate symmetric tensor products of group representations described in terms of their characters. Weyl integration, combined with the orthogonality property of the inner products of characters, makes it possible to project out irreducible representations (``irreps") from such generating functions. The PL provides an inverse function. Such procedures can be used to identify GIOs, which are necessarily singlets of the gauge group, and the results can be arranged into Hilbert series and their generating functions that encode information about the GIOs at a given level of field counting. 

Hilbert series can be expressed in refined form, with fields described in terms of class functions built from the characters of the irreps of the symmetry group. Alternatively, Hilbert series can be expressed in unrefined form, where they simply count representations according to their dimensions. Unrefined Hilbert series are the most straightforward to calculate, but their limitation to dimensional information entails that they do not fully encode moduli space data, for example, about the specific representations in which fields transform. On the other hand, while refined Hilbert series do fully encode moduli spaces, their generating functions can be cumbersome to deploy, with complicated plethystic procedures being necessary to extract character expansions.

Our aim in this paper is to outline a methodology for constructing highest weight generating functions (``HWGs") for the character expansions of Hilbert series, from which the latter can readily be extracted, either in refined or unrefined form. We wish to obtain HWGs that faithfully and efficiently encode the combinatorics of objects transforming in representations of the chosen symmetry group. We shall show that, just like Hilbert series, HWGs can be constructed as polynomials and rational functions of polynomials. While the prime objective here is to facilitate the analysis of the moduli spaces of SUSY gauge theories, we shall also show how HWG techniques can in principle be applied to other aspects of the combinatorics of group and product group representations, such as the calculation of Littlewood-Richardson and similar coefficients.

Section 2 of this paper outlines the theory underlying the construction of HWGs. This draws upon the standard plethystic functions and their use in symmetrising and antisymmetrising characters, as summarised in Appendix 1. We show how Weyl integration, as summarised in Appendix 2, can be used together with the plethystic functions to construct generating functions for the characters of irreps of classical and exceptional groups. We also show how the well-known Weyl Character Formula can be recast as a generating function for the characters of irreps of any classical or exceptional group.  These character generating functions are instrumental in deriving HWGs and can also be useful for other group theoretic analyses. In Section 2.4, we show how the character generating functions can be used to construct HWGs that encode the Littlewood-Richardson and similar coefficients governing the decomposition of the tensor products of irreps. In Section 2.5, we calculate the HWGs for symmetrisations of objects transforming in the basic irreps of lower rank $SU(N)$ simple groups. In Section 2.6 we examine the relationship between the invariant tensors of a group and the plethystics of product groups, since this will be helpful in understanding the structure of HWGs for the GIOs of product groups. 

In Sections 3 and 4 we apply the HWG methodology to explicate and extend some established results regarding SQCD and the moduli spaces of selected instanton theories, respectively. Section 5 summarises the key new results and perspectives and concludes with a comparison of some geometric properties of HWGs and Hilbert series.

\paragraph{Notation and Terminology}

Some preliminary comments on notation are in order. We present the characters of groups either in the generic form ${\cal X}_{Group}$ or, more specifically, using Dynkin labels such as ${[ {{n_1}, \ldots , {n_r}}]_{Group}}$, where $r$ is the rank of the group (dropping group subscripts if no ambiguities arise). We may refer somewhat interchangeably to \textit{series}, such as $1 + f + {f^2} +  \ldots $, by their \textit{generating functions} $1/\left( {1 - f} \right)$. We rely on the use of coordinates/variables to distinguish the different types of generating functions, as indicated in Table \ref{table1}.
\begin{table}[htdp]
\caption{Types of Generating Function}
\begin{center}
\begin{tabular}{c}
$\begin{array}{*{20}{c}}
{Generating~Function}&\vline& {{g^{Group}}\left( {coordinates} \right)}\\
\hline
{HWG}&\vline& {{g^{Group}}\left( {{t_j},{m_i}} \right)}\\
{Character}&\vline& {{g^{Group}}\left( {{m_i},{\cal X}} \right)}\\
{Refined~HS~(character~notation)}&\vline& {{g^{Group}}\left( {{t_j},{\cal X}} \right)}\\
{Refined~HS~(coordinates)}&\vline& {{g^{Group}}\left( {{t_j},{x_i}} \right)}\\
{Unrefined~HS~(distinct~counting)}&\vline& {{g^{Group}}\left( {{t_j}} \right)}\\
{Unrefined~HS}&\vline& {{g^{Group}}\left( t \right)}
\end{array}$
\end{tabular}
\end{center}
\label{table1}
\end{table}

These different types of generating function are related and can be considered as a hierarchy in which the HWG, character and refined HS generating functions fully encode the group theoretic information. We label Cartan subalgebra (``CSA") coordinates for weights within characters by $x$ or $y$, using subscripts when necessary; these coordinates are unimodular. We generally label object counting variables with $t$, and Dynkin label counting variables with $m$, although we may also use other letters, where this is helpful. We choose such counting variables to have a complex modulus of less than unity, which is essential for the residue calculations involved in Weyl integration to be valid. We follow established practice in referring to both counting variables and coordinates as ``fugacities".

\section{Highest Weight Generators}
\subsection{Representation of Dynkin Labels using Monomials}

The canonical classification of the irreps of a Lie group is carried out using Dynkin labels. These label the states or weights within an irrep; each irrep has a unique highest weight and can therefore be identified by the Dynkin labels of its highest weight state. We can map the Dynkin labels $\left[ {{n_1}, \ldots ,{n_r}} \right]$ of the irreps of a group of rank $r$ onto the complex manifold ${\mathbb C}^{r}$ by introducing the Dynkin label fugacities $\left\{ {{m_1}, \ldots, {m_r}} \right\}$ and establishing a correspondence:
\begin{equation} \label{eq:HWG1}
 [{{n_1}, \ldots, {n_r}}] \Leftrightarrow \prod\limits_{i = 1}^r {m_i^{{n_i}}}. 
\end{equation}

If we choose the fugacities $\{ {{m_1}, \ldots , {m_r}}\}$  to have absolute values less than unity, then each point on the infinite lattice of Dynkin labels corresponds to a unique point lying inside the unit complex disk on ${\mathbb C}^{r}$. This lattice is freely generated since the Dynkin labels can be chosen independently of each other. Now consider a second lattice on ${\mathbb C}^{N}$, with its points representing the possible combinations of $N$ objects that transform in some representations of the group. Each such combination of objects will transform in some (generally reducible) representation of the group and there will  be a non-trivial relationship between these lattices. Such relationships can be described most simply in terms of the dimensions of the representations, and the resulting polynomials of fugacities on ${\mathbb C}^{N}$ are termed unrefined Hilbert series:

\begin{equation} \label{eq:hwhs2a}
\begin{aligned}
HS{\left( t_i \right)} &= \sum \limits_{k_i}{{a_{k_1, \ldots, k_N}}}~{{t_1}^{k_1} \ldots {t_N}^{k_N}}.\\
\end{aligned}
\end{equation}

We can, however, also choose to represent such series in other ways. If instead of dimensions, we use characters composed of monomials of CSA coordinates, then we obtain a \textit{refined} Hilbert series:

\begin{equation} \label{eq:hwhs2b}
\begin{aligned}
HS{\left( {{x_i},t_j} \right)}  &= \sum \limits_{k_i} {{c_{k_1, \ldots, k_N}}\left( {{x_1}, \ldots, {x_r}} \right)}~{{t_1}^{k_1} \ldots {t_N}^{k_N}},\\
& {\text {where}}\\
 a_{k_1, \ldots, k_N} &= {c_{k_1, \ldots, k_N}}\left(1, \ldots, 1 \right).\\
\end{aligned}
\end{equation}

Alternatively, if we use the Dynkin labels of irreps of the group, then we obtain series expansions in terms of monomials in the Dynkin label fugacities: 
\begin{equation} \label{eq:hwhs2c}
\begin{aligned}
HWG{\left( {m_i,t_j} \right)}  &= \sum \limits_{{n_i},k_j}{{b_{n_1,\ldots,n_r ,k_1, \ldots, k_N}}~{{m_1}^{n_1} \ldots {m_r}^{n_r}}~ {{t_1}^{k_1} \ldots {t_N}^{k_N}}}.\\
\end{aligned}
\end{equation}

While such series expansions necessarily represent lattices on ${\mathbb C}^{N+r}$, they can often be obtained from highest weight generating functions (``HWGs") based on a small number of monomials that encode very concisely the relationships between the combinations of objects and the irreps in which they transform.

\subsection{Character Generating Functions}

As a further preliminary to developing the theory of HWGs, it is useful to recall the Peter Weyl Theorem \cite{Fuchs:1997bb}. This entails that, in addition to being orthonormal under Weyl integration, the characters of a compact group form a complete basis for the class functions of the group. Since the PEs and PEFs of characters, being functions of characters, belong to the class functions of the group, it is always possible, in principle, to decompose such a PE or PEF into a sum of characters, each with some polynomial coefficient.

As summarised in Appendix 1, we can use the PE and PEF functions to symmetrise or antisymmetrise the characters of representations. It is well known that any irrep of a group can be obtained from a small number of basic irreps by some combination of symmetrisations and antisymmetrisations. It is interesting to explore how this relationship can be encoded most succinctly.

We can illustrate this for classical A series Lie groups using the PEF. Thus, taking the PEF of the fundamental irrep gives, for example,
\begin{equation} \label{eq:hwhs2}
\begin{aligned}
  SU(2): PEF[ {[1]t}] &= 1 + [1]t + {t^2},\\
  SU(3): PEF[ {[1,0]t}] &= 1 + [1,0]t + [0,1]{t^2} + {t^3},\\ 
\end{aligned}
\end{equation}
where we have represented the characters of the singlets [0] and [0,0] by unity. The pattern generalises to higher rank groups within the A series, for which we can obtain all the basic irreps [0,\ldots ,1,\ldots ,0] either from the fundamental (or antifundamental) by antisymmetrisation. In the case of the C series, this role is played by the fundamental representation, while in the case of the B and D series, this role is played by the vector and spinor representations together. In all cases, including Exceptional groups, we can use the PEF to generate all the basic irreps carrying a single unit Dynkin label starting from a small subset of the irreps of lowest dimension.

Given such a set of basic irreps of a group, with Dynkin labels [1,0,\ldots ,0] through [0,\ldots ,0,1], we can construct irreps with higher Dynkin labels. These higher irreps can all be obtained by symmetrisation of the basic irreps, followed by taking tensor products between these series. So, we can usefully define character generating functions ${g^G}( {{m_i},{\cal{X}}})$, which encode the relationship between the full set of irreps, as identified by their Dynkin labels, and the action of the PE on the basic irreps:

\begin{equation}\label{eq:hwhs2.3}
{g^G}\left( {{m_i},{{\cal X}}} \right) \equiv \sum\limits_{{n_i} = 0}^\infty  {{{\left[ {{n_1}, \ldots , {n_r}} \right]}_G}} \prod\limits_{j= 1}^r {m_j^{{n_j}}}  \equiv {P^G}\left( {{m_i},{{\cal X}}} \right)PE\left[ {\sum\limits_{i = 1}^r {\underbrace {\left[ {0, \ldots , {1_i}, \ldots , 0} \right]}_{{\rm{1~in~ith~slot}}}{m_i}} } \right].
\end{equation}

The numerators ${P^G}(m_i,\cal{X})$, which have been defined \textit {implicitly}, are necessarily class functions of characters, since both the sum on the LHS of \ref{eq:hwhs2.3} and the PE on the RHS are class functions. For low rank groups, $P^G( m_i,\cal X)$ takes a simple form, although some considerable analytical work can be required to find it for higher rank groups. A general procedure for finding $P^G( m_i,\cal X)$ can be outlined as follows. 

Firstly, we rearrange \ref{eq:hwhs2.3} to obtain:
\begin{equation}\label{eq:hwhs2.9}
{P^G}({m_i},{\cal X}) =  \sum\limits_{{n_i} = 0}^\infty  \left([n_1, \ldots , n_r] \prod\limits_{j = 1}^r {m_j^{{n_j}}} \right) PE\left[ { - \sum\limits_{k = 1}^r {\left[ {0, \ldots , {1_k}, \ldots , 0} \right]{m_k}} } \right].
\end{equation}
The class function $P^G( m_i,\cal X)$ is thus a product of two series, of which the first is infinite and the second is finite.

For $SU(2)$, we can find $P^{SU(2)}( m_i,\cal X)$ directly: we take $x$ as the CSA coordinate and use $\left[ 1 \right] \equiv \left( {x + 1/x} \right)$   as the character of the fundamental. It then follows that:
\begin{equation}\label{eq:hwhs2.10}
\begin{aligned}
{P^{SU\left( 2 \right)}} &  \equiv PE\left[ { - \left[ 1 \right]m} \right]\sum\limits_{n = 0}^\infty  {\left[ n \right]{m^n}} \\
 &  = \left( {1 - m/x} \right)\left( {1 - xm} \right)\sum\limits_{n = 0}^\infty  {\left[ n \right]{m^n}} \\
 &  = \left[ 0 \right] + \sum\limits_{n = 0}^\infty  {\underbrace {\left( {\left[ {n + 2} \right] - \left[ {n + 1} \right]\left[ 1 \right] + \left[ n \right]} \right)}_{{{ = 0}}}{m^{n + 2}}} \\
 &  = 1.
\end{aligned}
\end{equation}
The sum in the RHS term vanishes for all n as a consequence of the multiplication law for $SU(2)$. Thus, we obtain the simple generating function for the series of characters of $SU(2)$:
\begin{equation}\label{eq:hwhs2.11}
{g^{SU\left( 2 \right)}}\left( {m,{{\cal X}}} \right) \equiv \sum\limits_{n = 0}^\infty  {\left[ n \right]} {m^n} = PE\left[ {\left[ 1 \right]m} \right] = 1 + \left[ 1 \right]m + \left[ 2 \right]{m^2} +  \ldots. 
\end{equation}

For $SU(3)$, we can follow a similar route:
\begin{equation}\label{eq:hwhs2.12}
\begin{aligned}
{P^{SU\left( 3 \right)}} &  \equiv PE\left[ { - \left[ {1,0} \right]{m_1} - \left[ {0,1} \right]{m_2}} \right]\sum\limits_{{n_i} = 0}^\infty  {\left[ {{n_1},{n_2}} \right]{m_1}^{{n_1}}{m_2}^{{n_2}}} \\
 &  = \left( {1 - {m_1}x} \right)\left( {1 - {m_1}y} \right)\left( {1 - {m_1}/xy} \right)\left( {1 - {m_2}/x} \right)\left( {1 - {m_2}/y} \right)\left( {1 - {m_2}xy} \right)\sum\limits_{{n_i} = 0}^\infty  {\left[ {{n_1},{n_2}} \right]{m_1}^{{n_1}}{m_2}^{{n_2}}} \\
 &  = \left( \begin{array}{r}
\left( {m_2^3m_1^3 - m_1^3 + m_2^2m_1^2 + {m_2}{m_1} - m_2^3 + 1} \right)\left[ {0,0} \right]\\
 + \left( {m_2^2 - m_2^2m_1^3 - {m_2}m_1^2 + m_2^3{m_1} - {m_1}} \right)\left[ {1,0} \right]\\
 + \left( {{m_2}m_1^3 - m_2^3m_1^2 + m_1^2 - m_2^2{m_1} - {m_2}} \right)\left[ {0,1} \right]\\
 + {m_1}{m_2}\left( {{m_1}{m_2} + 1} \right)\left[ {1,1} \right]\\
 - m_1^2{m_2}\left[ {0,2} \right]\\
 - {m_1}m_2^2\left[ {2,0} \right]
\end{array} \right)\sum\limits_{{n_i} = 0}^\infty  {\left[ {{n_1},{n_2}} \right]{m_1}^{{n_1}}{m_2}^{{n_2}}} \\
 &  = \left( {1 - {m_1}{m_2}} \right)\left[ {{\rm{0,0}}} \right] + \underbrace {\sum\limits_{{n_1} > 0||{n_2} > 0}^\infty  {\left( { \ldots ,  \ldots , } \right)\left[ { \ldots ,  \ldots , } \right]{m_1}^{{n_1}}{m_2}^{{n_2}}} }_{RHS}\\
 &  = 1 - {m_1}{m_2}.
\end{aligned}
\end{equation}
Once again, it can be shown that the RHS term vanishes and so we obtain the generating function for the characters of $SU(3)$:
\begin{equation}\label{eq:hwhs2.13}
{g^{SU\left( 3 \right)}}\left( {{m_1},{m_2},{{\cal X}}} \right) \equiv \sum\limits_{{n_i} = 0}^\infty  {\left[ {{n_1},{n_2}} \right]{m_1}^{{n_1}}{m_2}^{{n_2}}}  = \left( {1 - {m_1}{m_2}} \right)PE\left[ {\left[ {1,0} \right]{m_1}} \right]PE\left[ {\left[ {0,1} \right]{m_2}} \right].
\end{equation}

More generally, we use the completeness principle to expand $P^G( m_i,\cal X)$ as a superposition of characters, each with polynomial coefficients $P_{\left[ A \right]}^G\left( {{m_i}} \right)$  , where $[A]$ denotes the collective Dynkin labels corresponding to an irrep:
\begin{equation}\label{eq:hwhs2.14}
{P^G}\left( {{m_i},{{\cal X}}} \right) \equiv \sum\limits_{\left[ A \right]}^{} {P_{\left[ A \right]}^G\left( {{m_i}} \right)\left[ A \right]}. 
\end{equation}
We can then use Weyl integration, as described in Appendix 2, to project out the polynomial coefficients $P_{\left[ A \right]}^G\left( {{m_i}} \right)$, one at a time, from:
\begin{equation}\label{eq:hwhs2.15}
P_{\left[ A \right]}^G\left( {{m_i}} \right) = \oint\limits_G {d\mu \left[ A \right]^*} \sum\limits_{{n_i} = 0}^\infty  {\left( {\left[ {{n_1}, \ldots , {n_r}} \right]\prod\limits_{j = 1}^r {m_j^{{n_j}}} } \right)} PE\left[ { - \sum\limits_{k = 1}^r {\left[ {0, \ldots , {1_k}, \ldots , 0} \right]{m_k}} } \right].
\end{equation}
Without digressing into the mechanics of efficient algorithms for computing these coefficients, using \textit{Mathematica}, for example, we compile the results of such calculations in Table \ref{table2}.

\begin{table}[htdp]
\caption{$P^G$ for Low Rank Classical and Exceptional Groups}
\begin{center}
\begin{tabular}{|c|c|}

\hline
$Group$ &$ {{P^G}\left( {{m_i},{{\cal X}}} \right)}$ \\
\hline
${{A_1} \cong {B_1} \cong {C_1}}$&$1$ \\
\hline
${{A_2}}$&${\left( {1 - {m_1}{m_2}} \right)}$\\
\hline
${B_2}$&${\left( {1 - m_1^2 + {m_1}m_2^2 - m_1^3m_2^2} \right) + {m_1}{m_2}\left( {{m_1} - 1} \right)\left[ {0,1} \right]}$\\
\hline
${{C_2}}$&   ${As ~B_2 ~with~ {m_1} \Leftrightarrow {m_2}~and~ [1,0] \Leftrightarrow [0,1]}$\\
\hline
${{D_2} \cong {A_1} \otimes {A_1}}$&1\\
\hline
${{G_2}}$&{See Appendix 3}\\
\hline
${{A_3}}$ & 

$\begin{array}{c}
{\left( {1 - m_2^2} \right)\left( {1 - {m_1}{m_3} + {m_2}{m_3}^2 + {m_1}^2{m_2} - {m_1}{m_2}^2{m_3} + {m_1}^2{m_2}^2{m_3}^2} \right)\left[0,0,0 \right]}\\
{ + {m_2}\left( { - {m_3} + {m_1}{m_2} - {m_1}^2{m_2}{m_3} + {m_1}{m_2}^2{m_3}^2} \right)\left[ {1,0,0} \right]}\\
{ + {m_2}\left( { - {m_1} + {m_2}{m_3} - {m_1}{m_2}{m_3}^2 + {m_1}^2{m_2}^2{m_3}} \right)\left[ {0\,,0\,,1} \right]}\\
{ + {m_1}{m_2}{m_3}\left( {1 - m_2^2} \right)\left[ {0,1,0} \right]}
\end{array}$

\\
\hline
${{D_3}}$&   ${As ~A_3 ~with~ {m_1} \Leftrightarrow {m_2}~and~ [1,0,0] \Leftrightarrow [0,1,0]}$\\
\hline
${{A_4}}$&{See Appendix 4}\\
\hline

\end{tabular}
\end{center}
\label{table2}
\end{table}
The results up to $B_2$, $C_2$, $D_2$ and $A_3$ correspond to those obtained by other methods \cite{Hanany:2008qc}. The numerators tabulated in Appendix 3 and Appendix 4 should be understood as sums of characters with the polynomial coefficients given, similar to the numerator for $A_3$ in Table \ref{table2}. Importantly, for all the Classical and Exceptional groups, the class function $P^G$ is finite and has palindromic properties. The finite nature of $P^G$ is a consequence of relations between the characters generated from each basic irrep and, indeed, follows as a corollary of the construction of character generating functions based on the Weyl Character Formula, as described later in this section.

A polynomial in $\left\{ {{t_1}, \ldots , {t_r}} \right\}$  is defined as palindromic of degree $\left( {{d_1}, \ldots , {d_r}} \right)$ if each monomial term $t_1^{{n_1}} \ldots  t_r^{{n_r}}$ in the polynomial is paired with a term $t_1^{{d_1} - {n_1}} \ldots  t_r^{{d_r} - {n_r}}$ with the same coefficients \cite{Hanany:2008sb}. The polynomial is anti-palindromic if the coefficients in each pair are of opposite sign. Under this definition, the class function ${P^{B_2}}$ in Table \ref{table2}, for example, is anti-palindromic of degree (3,2).

The generating functions for characters of irreps can be simplified to give generating functions for the dimensions of irreps by replacing characters with dimensions. The general form of these relationships follows from \ref{eq:hwhs2.3} and \ref{hwhs8.2} as:
\begin{equation}
\label{eq:hwhs2.16}
{g^G}\left( {{m_i}} \right) = \sum\limits_{{n_i} = 0}^\infty  {Dim \left[ {{n_1}, \ldots , {n_r}} \right]} \prod\limits_{i = 1}^r {m_i^{{n_i}}}\\
 = {P^G}\left( {{m_i},Dim {{\cal X}}} \right)/\prod\limits_{i = 1}^r {{{\left( {1 - {m_i}} \right)}^{Dim[0,\ldots,1_i,\ldots,0]}}}. 
\end{equation}
We then obtain the generating functions for dimensions summarised in Table \ref{table3}, with $P^{A4}$ as set out in Appendix 5.

\begin{table}[htdp]
\small
\caption{ Generating Functions for Dimensions of Low Rank Classical and Exceptional Groups}
\begin{center}
\begin{tabular}{|c|c|c|}
\hline
${Group}$&${{g^G}\left( {{m_i}} \right)}$& ${ \begin{array}{c}Degree ~of \\Palindromic\\Numerator\end{array}}$\\
\hline
${{A_1} \cong {B_1} \cong {C_1}}$&   $\frac{1}{{{\left( {1 - m} \right)}^2}}$   &(0)\\
\hline
${{A_2}}$& $\frac{\left( {1 - {m_1}{m_2}} \right)}{{{\left( {1 - {m_1}} \right)}^3}{{\left( {1 - {m_2}} \right)}^3}}$       &(1,1)\\
\hline
${{B_2}}$&
$\frac
{\left( {1 +m_1 - 4{m_1}{m_2}  + {m_1}m_2^2 + {m_1^2} m_2^2} \right)}{{{\left( {1 - {m_1}} \right)}^4}{{\left( {1 - {m_2}} \right)}^4}}$
&
(2,2)\\
\hline
${{C_2}}$&   ${As ~B_2 ~with~ {m_1} \Leftrightarrow {m_2}}$   &(2,2)\\
\hline
${{D_2}}$&  $\frac{1}{{{\left( {1 - {m_1}} \right)}^2}{{\left( {1 - {m_2}} \right)}^2}}$   &(0)\\
\hline
${{G_2}}$&

$\frac{{\left( {\begin{array}{*{20}{c}}
{1 + {m_1} + 8{m_2} + 8m_2^2 + m_2^3}\\
 - 26{m_1}{m_2} - 41{m_1}m_2^2 - 6{m_1}m_2^3 + 15m_1^2{m_2}+ 78m_1^2m_2^2\\
{ + 15m_1^2m_2^3 - 6m_1^3{m_2} - 41m_1^3m_2^2 - 26m_1^3m_2^3}\\
{ + m_1^4{m_2} + 8m_1^4m_2^2 + 8m_1^4m_2^3 + m_1^3m_2^4 + m_2^4m_1^4}
\end{array}} \right)}}{{{{\left( {1 - {m_1}} \right)}^6}{{\left( {1 - {m_2}} \right)}^6}}}$
&(4,4)\\
\hline
${{A_3}}$&

$\frac{{\left( {\begin{array}{c}
1 + {m_2} - {m_1}{m_3} - 4{m_1}{m_2} - 4{m_2}{m_3}\\
 + 5{m_1}{m_2}{m_3} + m_1^2{m_2} + {m_2}m_3^2\\
 + m_2^2m_3^2 + m_1^2m_2^2 + 5{m_1}{m_3}m_2^2\\
 - 4m_1^2m_2^2{m_3} - 4{m_1}m_2^2m_3^2 - {m_1}m_2^3{m_3} + m_1^2m_2^2m_3^2 + m_1^2m_2^3m_3^2 \end{array}} \right)}}
{{{{\left( {1 - {m_1}} \right)}^4}{{\left( {1 - {m_2}} \right)}^5}{{\left( {1 - {m_3}} \right)}^4}}}$

&(2,3,2)\\
\hline
${{D_3}}$&   ${As ~A_3 ~with~ {m_1} \Leftrightarrow {m_2}}$     &(3,2,2)\\
\hline
${{A_4}}$
&$\frac
{{P^{A4}}\left( {{m_1},{m_2},{m_3},{m_4}} \right)}{{{\left( {1 - {m_1}} \right)}^5}{{\left( {1 - {m_2}} \right)}^7}{{\left( {1 - {m_3}} \right)}^7}{{\left( {1 - {m_4}} \right)}^5}}$
&(3,5,5,3)\\
\hline

\end{tabular}
\end{center}
\label{table3}
\end{table}%

\begin{sidewaystable}
\small
\caption{ Generating Functions for Characters of Low Rank Unitary Groups}
\begin{center}
\begin{tabular}{|c|c|c|c|}
\hline
${Group} $&$ {{P^G}\left( {{m_i},{{\cal X}}} \right)}$&${}$&${PE ~of~ basic~ irreps} $\\
\hline
 ${U\left( 1 \right)}$&$ 1$&$ \times $&$ {PE\left[ m \right]}$ \\
\hline
$ {U\left( 2 \right)}$&$ 1$&$ \times $&$ {PE\left[ {\left[ {1,0} \right]{m_1} + \left[ {0,1} \right]{m_2}} \right]}$ \\
\hline
$ {U\left( 3 \right)}$&$ {\left[ {0,0,0} \right] - {m_1}{m_2}\left[ {0,0,1} \right]}$&$ \times $&$ {PE\left[ {\left[ {1,0,0} \right]{m_1} + \left[ {0,1,0} \right]{m_2} + \left[ {0,0,1} \right]{m_3}} \right]} $\\
\hline
$ {U\left( 4 \right)}$
&
$ {\begin{array}{*{20}{c}}
1&{\left[ {0,0,0,0} \right]}\\
{ - {m_1}{m_2}}&{\left[ {0,0,1,0} \right]}\\
{ + \left( {{m_1}^2{m_2} - {m_1}{m_3} - {m_2}^2} \right)}&{\left[ {0,0,0,1} \right]}\\
{ + {m_2}({m_1}{m_2} - {m_3})}&{\left[ {1,0,0,1} \right]}\\
{ + {m_1}{m_2}{m_3}}&{\left[ {0,1,0,1} \right]}\\
{ + {m_2}^2{m_3}}&{\left[ {0,0,1,1} \right]}\\
{ + {m_2}\left( {{m_3}^2 - {m_1}^2{m_2}^2} \right)}&{\left[ {0,0,0,2} \right]}\\
{ - {m_1}^2{m_2}^2{m_3}}&{\left[ {1,0,0,2} \right]}\\
{ - {m_1}{m_2}^3{m_3}}&{\left[ {0,1,0,2} \right]}\\
{ + {m_1}{m_2}^2{m_3}({m_1}{m_2} - {m_3})}&{\left[ {0,0,1,2} \right]}\\
{ + {m_2}^2{m_3}\left( {{m_1}^2{m_3} + {m_1}{m_2}^2 - {m_2}{m_3}} \right)}&{\left[ {0,0,0,3} \right]}\\
{ + {m_1}{m_2}^3{m_3}^2}&{\left[ {1,0,0,3} \right]}\\
{ - {m_1}^2{m_2}^4{m_3}^2}&{\left[ {0,0,0,4} \right]}
\end{array}}$
&
$ \times $
&
$ {PE\left[ {\left[ {1,0,0,0} \right]{m_1} + \left[ {0,1,0,0} \right]{m_2} + \left[ {0,0,1,0} \right]{m_3}} + \left[ {0,0,0,1} \right]{m_4}\right]} $\\
\hline

\end{tabular}
\end{center}
\label{table3a}
\end{sidewaystable}

\FloatBarrier

The palindromic properties of ${P^G}\left({{m_i},Dim {{\cal X}}} \right)$  can clearly be seen. Interestingly, the degrees of the palindromic numerators are in all cases equal to the degrees of the denominators minus two\footnote{Observation by A.Thomson, Imperial College}. It can readily be verified that all the generating functions given in Table \ref{table2}, Table \ref{table3} and Appendix 3 through Appendix 5 are consistent both with the usual formulae for group dimensions and also with characters obtained by other methods, such as applying Cartan matrices to highest weights and using the Freudenthal multiplicity formula \cite{Fuchs:1997bb}.

A similar analysis can be carried out for unitary groups and we include for reference in Table \ref{table3a} the elements of the character generating functions for unitary groups $U(1)$ through $U(4)$. There is a clear relationship between the $P^G$ functions for the $U(N)$ and the $SU(N)$ series, which can be seen by dropping the last Dynkin labels from the irreps within $P^G$ for a given $U(N)$ series and assigning the corresponding polynomials to the resulting (truncated) irreps of the $SU(N)$ series.

Before concluding this section it is useful also to demonstrate how the Weyl Character Formula can be recast as a generating function. This provides an alternative method for deriving the character generating functions described above; it also makes it possible to replace complicated character generating functions by finite sums over simpler generating functions, which can be helpful for some calculations.

The Weyl Character Formula  \cite{Fuchs:1997bb} is given by:
\begin{equation}
\label{eq:hwhsWCF}
{{\cal X}}\left( \lambda  \right) = \frac{{\sum\limits_{w \in W}^{} {\det \left[ w \right]{e^{w\left( {\lambda  + \rho } \right)}}} }}{{{e^\rho }\prod\limits_{\alpha  \in \Phi  + }^{} {\left( {1 - {e^{ - \alpha }}} \right)}}}.
\end{equation}
The conventional notation translates to the explicit character constructions developed in this note as follows. The weight $\lambda$ corresponds to the desired irrep with Dynkin labels $[n_1, \ldots n_r]$. The weight $\rho$ corresponds to the weight of the Weyl vector, which is $[1,1, \ldots 1]$ and so the formal exponential $e^{\rho}$ corresponds to the CSA coordinate monomial $x_1 x_2 \ldots x_r \equiv x$. The parameter $\alpha$ ranges over the weights of the positive root space, so that $e^{-\alpha}$ corresponds to the monomial $x_1^{\alpha_1} x_2^{\alpha_2} \ldots x_r^{\alpha_r}$ for positive roots $\alpha, \beta, \ldots$. The matrices $w$ or $w_{ij}$ are elements of the finite Weyl group W of the Lie algebra, which acts upon the weights; their signs are given by their determinants, which are all real and unimodular.

The generating functions that we seek are rational functions that generate series of characters, so we introduce the fugacities $m_i$ for the Dynkin labels and use these to form the generating functions:
\begin{equation}
\label{eq:hwhsWCFgen}
{{g^{G}}\left( {{m_i},{\cal X}} \right)}=\sum\limits_{{\lambda} = {0}}^\infty  {{{\cal X}}\left( \lambda  \right){m^\lambda } \equiv \sum\limits_{{n_i} = 0}^\infty  {{{\cal X}}\left[ {{n_1}, \ldots, {n_r}} \right]m_1^{{n_1}} \ldots m_r^{{n_r}}}}.
\end{equation}
We proceed by combining and rearranging \ref{eq:hwhsWCF} and \ref{eq:hwhsWCFgen} :
\begin{equation}
\label{eq:hwhsWCFgen1}
{{g^{G}}\left( {{m_i},{\cal X}} \right)}=
\frac{1}{{{e^\rho }\prod\limits_{\alpha  \in \Phi  + }^{} {\left( {1 - {e^{ - \alpha }}} \right)} }}\sum\limits_{w \in W}^{} {\det \left[ w \right]{e^{w\left( \rho  \right)}}\sum\limits_{\lambda  = [0,\ldots,0]}^\infty  {{e^{w\left( \lambda  \right)}}{m^\lambda }} }.
\end{equation}
We focus on the right hand term and switch to monomial notation, obtaining the rational expression:
\begin{equation}
\label{eq:hwhsWCFgen2}
{\sum\limits_{\lambda  = 0}^\infty  {{e^{w\left( \lambda  \right)}}{m^\lambda }} }=
\sum\limits_{{n_i} = 0}^\infty  {x_1^{w\left( {{n_1}} \right)} \ldots x_r^{w\left( {{n_r}} \right)}m_1^{{n_1}} \ldots m_r^{{n_r}}}  = 
\prod\limits_j {\frac{1}{{1 - {m_j}\prod\limits_i {{x_i}^{{w_{ij}}}} }}}. 
\end{equation}
We can now combine \ref{eq:hwhsWCFgen1} and \ref{eq:hwhsWCFgen2} into a generating function for the series of characters of any irreps of a group G:
\begin{equation}
\label{eq:hwhsWCFgen3}
{g^G}\left( {{m_i},\cal X} \right) = \frac{1}{{x\prod\limits_{\alpha  \in \Phi  + }^{} {\left( {1 - {x^{ - \alpha }}} \right)} }}\sum\limits_{w \in W}^{} {\det \left[ w \right]\prod\limits_j {\frac{{\prod\limits_i {{x_i}^{{w_{ij}}}} }}{{1 - {m_j}\prod\limits_i {{x_i}^{{w_{ij}}}} }}} }.
\end{equation}
The generating function \ref{eq:hwhsWCFgen3} has advantages over the Weyl Character Formula, since the summation over the Weyl group needs only to be carried once, for any number of characters; it is also a finite rational function and can therefore be used in Weyl integration to project out any rational class function into an HWG. The Weyl group matrices necessary for calculations can be obtained from {\it Mathematica} add-on programs such as LieArt \cite{Feger:2012bs}.

\subsection{Transforming a Plethystic Exponential into a Highest Weight Generating Function}

Consider now some other class function, such as the PE of a given set of irreps:

\begin{equation}\label{eq:hwhs2.4}
PE\left[ {\left[ A \right]{t_A} +  \ldots  \left[ D \right]{t_D}} \right] = \sum\limits_{{n_i} = 0}^\infty  {{g^G}_{\left[ {{n_1}, \ldots , {n_r}} \right]}\left( {{t_A}, \ldots , {t_D}} \right)} \left[ {{n_1}, \ldots , {n_r}} \right].
\end{equation}
In this expression, the collective Dynkin labels \{[A],  \ldots ,[D]\} range over some selected irreps of the group and the $\{t_A,  \ldots , {t_D}\}$ are object fugacities. The polynomial coefficients ${g^G}_{\left[ {{n_1} ,\ldots , {n_r}} \right]}\left( {{t_A}, \ldots , {t_D}} \right)$ can be obtained one at a time from  \ref{eq:hwhs2.4} with the help of the character orthonormality and completeness relations and using Weyl integration:

\begin{equation}\label{eq:hwhs2.5}
{g^G}_{\left[ {{n_1} ,\ldots , {n_r}} \right]}\left( {{t_A}, \ldots , {t_D}} \right) = \oint\limits_G {d\mu \left[ {{n_1} ,\ldots , {n_r}} \right]^*PE\left[ {\left[ A \right]{t_A} +  \ldots  \left[ D \right]{t_D}} \right]}. 
\end{equation}
We can aggregate the coefficients ${g^G}_{\left[ {{n_1} ,\ldots , {n_r}} \right]}\left( {{t_A}, \ldots , {t_D}} \right)$ into a polynomial series, which corresponds to an HWG series equivalent to \ref{eq:hwhs2c}:

\begin{equation}\label{eq:hwhs2.6}
{g^G}\left( {{t_A} ,\ldots , {t_D},{m_i}} \right) \equiv \sum\limits_{{n_i} = 0}^\infty  {{g^G}_{\left[ {{n_1}, \ldots , {n_r}} \right]}\left( {{t_A}, \ldots , {t_D}} \right)} {m_1}^{{n_1}} \ldots  {m_r}^{{n_r}}.
\end{equation}
This HWG series \ref{eq:hwhs2.6} encodes all the group theoretic information relating the multiplicities of the objects being symmetrised and the irreps in which they transform. We can extract this information from the HWG in various different ways. For example:
\begin{enumerate}
\item 
Given some multiplicities $\left\{ {{k_A}, \ldots , {k_D}} \right\}$  of the objects represented by the monomials $t_A^{{k_A}} \ldots  t_D^{{k_D}}$, we can identify the linked exponents of highest weight variables $m_1^{{n_1}} \ldots  m_r^{{n_r}}$ amongst the monomials of the HWG series and the integer coefficients of these monomial terms. These enumerate the irreps $[n_1,\ldots , n_r]$ in which the objects transform along with their multiplicities.
\item
Conversely, given some set of irreps $[n_1,\ldots , n_r]$ with their associated monomials $m_1^{{n_1}} \ldots  m_r^{{n_r}}$, we can identify the linked monomials $t_A^{{k_A}} \ldots  t_D^{{k_D}}$ amongst the terms of the HWG. The exponents $\left\{ {{k_A}, \ldots , {k_D}} \right\}$ correspond to the combinations of basic objects that transform in the given irreps.
\item
We can replace the monomials in Dynkin label fugacities $m_1^{{n_1}} \ldots  m_r^{{n_r}}$ within the HWG by the corresponding irrep dimensions and obtain information about the number of combinations of basic objects at any given multiplicity. These unrefined Hilbert series take the form:
\begin{equation}\label{eq:hwhs2.7}
{g^G}\left( {{t_A}, \ldots , {t_D}} \right) \equiv \sum\limits_{{n_i} = 0}^\infty  {{g^G}_{\left[ {{n_1}, \ldots , {n_r}} \right]}\left( {{t_A}, \ldots , {t_D}} \right)} Dim[ n_1, \ldots , n_r].
\end{equation}
\item
We can map distinct object fugacities in a Hilbert series onto a single fugacity $\{ t_A, \ldots , t_D\} \to t$ to obtain an unrefined Hilbert series $g^G(t)$.

\end{enumerate}
The Hilbert series of a theory can thus be presented in various ways. Importantly, the HWG captures all the group theoretic properties of the class functions of the theory; if we encode the information as a HWG series, we can always extract an unrefined HS, but not vice versa. 

Clearly \ref{eq:hwhs2.6} and \ref{eq:hwhs2.7} are in the form of infinite series and it is desirable to obtain these from rational polynomial generating functions. It can often be a non-trivial exercise to find the polynomial generating function for the HWG series $g^G( t_A,m_i)$ for the PE of some given set of objects. This calculation is, however, facilitated for groups for which we have the character generating functions $g^G( m_i,\cal X)$, as described above, available. In these cases, we can, in principle, calculate the HWG by using Weyl integration to project the PE (expressed as a class function in terms of CSA coordinates) onto the complete basis for class functions provided by the character generating function:
\begin{equation}\label{eq:hwhs2.8}
\begin{aligned}
{g^G}\left( {{t_A},{m_i}} \right) &= \sum\limits_{{n_i} = 0}^\infty  {{g^G}_{\left[ {{n_1}, \ldots , {n_r}} \right]}{m_1}^{{n_1}} \ldots  {m_r}^{{n_r}}}
 \\
  &= \sum\limits_{{n_i} = 0}^\infty  {\oint\limits_G {d\mu ~{{{\cal X}}^*}_{\left[ {{n_1} ,\ldots , {n_r}} \right]}{m_1}^{{n_1}} \ldots  {m_r}^{{n_r}}PE\left[ {{{{\cal X}}_{\left[ A \right]}}{t_A} +  \ldots  {{{\cal X}}_{\left[ D \right]}}{t_D}} \right]} } \\
& = \oint\limits_G {d\mu~ {g^G}\left( {{m_i},{{\cal X}}{{^*}}} \right)PE\left[ {{{{\cal X}}_{\left[ A \right]}}{t_A} +  \ldots  {{{\cal X}}_{\left[ D \right]}}{t_D}} \right]} \\
 &= \oint\limits_G {d\mu~ {P^G}\left( {{m_i},{{\cal X}}{{^*}}} \right)PE\left[ {\sum\limits_{i = 1}^r {\left[ {0, \ldots , 1_i, \ldots , 0} \right]^*{m_i}} } \right]PE\left[ {{{{\cal X}}_{\left[ A \right]}}{t_A} +  \ldots  {{{\cal X}}_{\left[ D \right]}}{t_D}} \right]}.
 \end{aligned}
\end{equation}

It will be clear from the foregoing that knowledge of the $P^G( m_i,\cal X)$ function is necessary to construct such an HWG from first principles, although sometimes the HWG can be found by inspection, starting from a finite number of the ${g^G}_{\left[ {{n_1}, \ldots , {n_r}} \right]}\left( {{t_A}, \ldots , {t_D}} \right)$ terms in the series expansion. There may also be situations where it is not necessary to use the full character generating function, for example, if the PE is known to generate only irreps with certain symmetry properties.

HWGs can be used to analyse the combinatorics of group representations quite generally and it is instructive to review some archetypal situations involving (a) finding coefficients for the decomposition of tensor products of irreps and (b) symmetrisations and antisymmetrisations of basic irreps of groups.


\subsection{Decomposing Tensor products using HWGs}

Given two irreps of a group with composite Dynkin labels [A] and [B], we can ask how their tensor product decomposes into a sum of irreps [C] with integer multiplicities $a_{AB}^C$. These integer coefficients are similar to Clebsch-Gordan coefficients, although they relate irreps rather than states within irreps. For $U(N)$ they correspond to Littlewood Richardson coefficients:
\begin{equation}
\label{eq:hwhs3.1}
\left[ A \right] \otimes \left[ B \right] = \sum\limits_C^{} {a_{AB}^C\left[ C \right]}.
\end{equation}
We can transform the relationship \ref{eq:hwhs3.1} using Weyl integration and character completeness/orthonormality to obtain an expression for the individual coefficients:

\begin{equation}
\label{eq:hwhs3.1a}
\begin{aligned}
a_{AB}^C & = \oint\limits_G {d\mu } \left[ C \right]^*\left[ A \right]\left[ B \right].
\end{aligned}
\end{equation}

We now rearrange the problem by introducing three sets of fugacities $\{m_i,n_i,t_i\}$ and by using three character generating functions ${g^G}\left( {{m_i},{{\cal X}}} \right)$, etc.:

\begin{equation}
\label{eq:hwhs3.1b}
\begin{aligned}
\sum\limits_{A,B,C} {a_{AB}^C{m^A}{n^B}{t^C}}  &  = \oint\limits_G {d\mu \sum\limits_{A,B,C} {\left[ C \right]^*\left[ A \right]\left[ B \right]{m^A}{n^B}{t^C}} } \\
 &  = \oint\limits_G {d\mu ~g\left( {{t^C},{{\cal X}}^{\rm{*}}} \right)g\left( {{m^A},{{\cal X}}} \right)} g\left( {{n^B},{{\cal X}}} \right)\\
 &\equiv {g^G}\left( {{m_i},{n_j},{t_k}} \right).
\end{aligned}
\end{equation}
We have used the shorthand notation 
${m^A} \equiv \prod\limits_{i = 1}^r {{{\left( {{m_i}} \right)}^{{A_i}}}}$
 for the product of fugacities relating to irrep [A], and so on \cite{David-Cox:2007fk}. Once calculated explicitly, the resulting HWG generating function ${g^G}\left( {{m_i},{n_j},{t_k}} \right)$ allows the values of $a_{AB}^C$ to be read off from the integer coefficients of the monomials ${m^A}{n^B}{t^C}$  . For $SU(2)$ through $SU(4)$ the tensor product generating functions are shown in Table \ref{table4}.
 
 \begin{table}[htdp]
\caption{HWG for $SU(N)$ Tensor Products}
\begin{center}
\begin{tabular}{|c|c|}

\hline
{}& ${\sum\limits_{A,B,C} {c_{A,B}^C{m^A}{n^B}{t^C}} }$ \\
\hline
${{A_1}}$ & 
${\footnotesize 
{\frac{1}{{\left( {1 - mn} \right)\left( {1 - mt} \right)\left( {1 - nt} \right)}}}}$ \\
\hline
$ {{A_2}}$ &
${\footnotesize 
{\frac{{1 - {m_1}{m_2}{n_1}{n_2}{t_1}{t_2}}}{{\left( {1 - {m_2}{n_1}} \right)\left( {1 - {m_1}{n_2}} \right)\left( {1 - {m_1}{t_1}} \right)\left( {1 - {n_1}{t_1}} \right)\left( {1 - {m_2}{t_2}} \right)\left( {1 - {n_2}{t_2}} \right)\left( {1 - {m_2}{n_2}{t_1}} \right)\left( {1 - {m_1}{n_1}{t_2}} \right)}}}}$ \\
\hline
$ {{A_3}}$ 
& 
$ {\footnotesize 
{\frac{1+\ldots  \textit{152 monomial terms}\ldots +{m_1^3 m_2^4m_3^3n_1^3n_2^4n_3^3t_1^3t_2^4t_3^3}}
{{\left(\begin{array}{c}
( {1 - {m_3}{n_1}})( {1 - {m_2}{n_2}})( {1 - {m_1}{n_3}})( {1 - {m_1}{t_1}})( {1 - {m_2}{t_2}})( {1 - {m_3}{t_3}})\\
 \times ( {1 - {n_1}{t_1}})( {1 - {n_2}{t_2}})( {1 - {n_3}{t_3}})\\
 \times ( {1 - {m_3}{n_2}{t_1}})( {1 - {m_2}{n_3}{t_1}})( {1 - {m_1}{n_1}{t_2}}) \\
 \times \left( {1 - {m_3}{n_3}{t_2}} \right)\left( {1 - {m_2}{n_1}{t_3}} \right)\left( {1 - {m_1}{n_2}{t_3}} \right)\\
 \times \left( {1 - {m_1}{m_3}{n_2}{t_2}} \right)\left( {1 - {m_2}{n_1}{n_3}{t_2}} \right)( {1 - {m_2}{n_2}{t_1}{t_3}} )\end{array}\right)}}}}$\\
 
\hline

\end{tabular}
\end{center}
\label{table4}
\end{table}

For example, to obtain the decomposition of $\left[ {1,0} \right] \otimes \left[ {0,2} \right]$, we identify the terms ${m_1}n_2^2\left( {{t_1}t_2^2 + {t_2}} \right)$ in the Taylor series expansion for the $A_2$ generating function and obtain $\left[ {1,2} \right] \oplus \left[ {0,1} \right]$.

If we examine the individual denominator terms, we find that they obey the rule that the sum of the subscripts on the $m_i$ and $n_i$ less the sum of the subscripts on the $t_i$ variables modulo the dimension of the fundamental is zero. This occurs since (i) the HWG is formed from singlets, (ii) the indices count antisymmetrisations (under the labelling system adopted) and, (iii) antisymmetrisation of the fundamental by the epsilon tensor, which has dimension equal to the fundamental, leads back to a singlet. Put another way, the sum of central charges counting the degree of antisymmetrisation of each irrep is a conserved quantity under a tensor product.

Also, since \ref{eq:hwhs3.1b} is symmetric between the irreps A,B and C*, and the irreps of $SU(N)$ are conjugate under reversal of Dynkin labels, we have the consquence that the HWGs in Table \ref{table4} are symmetric under interchange of any two fugacities $\{m_i,n_i,t_{N-i}\}$.

The numerators of these HWGs for tensor products of A series groups are all palindromic. We can also observe that in all three A series cases, the degree of each variable in the denominator exceeds that in the numerator by two. The palindromic numerator for $A_3$ is somewhat lengthy containing 154 monomial terms in total and we do not present it here.

This approach to calculating the coefficients for decomposing tensor products of irreps from Dynkin label fugacties can, in principle, be generalised to unitary groups, yielding generators for Littlewood Richardson Coefficients, as well as to other classical and exceptional groups.
\subsection{HWGs for Symmetrisation of Basic Irreps of $SU(N)$ Groups}

We can also use the character generating functions ${g^G}( {{m_i},{{\cal X}}})$ to obtain HWGs for symmetrisations of objects transforming in the basic irreps of a $SU(N)$ group by evaluating Weyl integrals:
\begin{equation}
\label{hwhs4.1}
\begin{aligned}
{g^G}\left( {{t_i},{m_j}} \right) &= \oint\limits_G {d\mu } ~{g^G}\left( {{m_j},{{\cal X}}{\rm{^*}}} \right)PE\left[ {\sum\limits_{i = 1}^r {\left[ {0, \ldots , {1_i}, \ldots , 0} \right]{t_i}} } \right]\\
 &= \oint\limits_G {d\mu }~ {P^G}\left( {{m_j},{{\cal X}}^*} \right)PE\left[ {\sum\limits_{j = 1}^r {\left[ {0, \ldots , {1_j}, \ldots , 0} \right]^*{m_j}} } \right]PE\left[ {\sum\limits_{i = 1}^r {\left[ {0, \ldots , {1_i}, \ldots , 0} \right]{t_i}} } \right].\\
\end{aligned}
\end{equation}
The $t_i$ count the objects being symmetrised, while the $m_i$ count the Dynkin labels of the resulting irreps. Explicit evaluations of $g^G(t_i,m_j)$ using generating functions based on Table \ref{table2} are set out in Table \ref{table5} for $SU(2)$ through $SU(4)$. These can alternatively be written more compactly by taking their PLs with respect to the fugacities, as shown in Table \ref{table6}.
\begin{table}
  \centering 
  \caption{HWG of $SU(N)$ basic irreps}\label{table5}
  \begin{tabular}{|c|c|}
\hline
$A_r$   & $HWG ~of ~PE\left[ {\sum\limits_{i = 1}^r {\left[ {0, \ldots , {1_i}, \ldots , 0} \right]{t_i}} } \right]$   \\
\hline
$A_1$   & $\frac{1}{(1-mt)}$  \\
\hline
$A_2$   & ${\frac{1}{{\left( {1 - {t_1}{t_2}} \right)\left( {1 - {m_1}{t_1}} \right)\left( {1 - {m_2}{t_2}} \right)}}}$  \\
\hline
$A_3$   & ${\frac{{1}}{{\left( {1 - {t_1}{t_3}} \right)\left( {1 - t_2^2} \right)\left( {1 - {m_1}{t_1}} \right)\left( {1 - {m_2}{t_2}} \right)\left( {1 - {m_3}{t_3}} \right)\left( {1 - {m_3}{t_1}{t_2}} \right)\left( {1 - {m_1}{t_2}{t_3}} \right)\left( {1 - {m_2}{t_1}{t_2}{t_3}} \right)}}}$  \\
\hline
\end{tabular}
\end{table}
\begin{table}
  \centering 
  \caption{PLs of HWG of $SU(N)$ basic irreps}\label{table6}
   \begin{tabular}{|c|c|}
   \hline
$A_r$   & ${PL\left[ {HWG} \right]}$   \\
\hline
$A_1$   & ${mt}$   \\
\hline
$A_2$   & ${{t_1}{t_2} + {m_1}{t_1} + {m_2}{t_2}}$   \\
\hline
$A_3$   & ${{t_1}{t_3} + t_2^2 + {m_1}{t_1} + {m_2}{t_2} + {m_3}{t_3} + {m_3}{t_1}{t_2} + {m_1}{t_2}{t_3} + {m_2}{t_1}{t_2}{t_3}}$   \\
\hline
 \end{tabular}
\end{table}
In all the cases calculated, the HWGs of the basic irreps lead to finite PLs with all positive terms and are said to be \textit{freely generated}. If we choose to restrict the series by selecting a subset of the basic irreps for symmetrisation (by setting some of the $t_i$ to zero) this simply has the effect of dropping monomials from the PL, and so the subseries of the HWGs are also \textit{freely generated}.

These polynomial generating functions $g^G(t_i,m_j)$ encode, through the exponents of the fugacities, the combinations of irreps formed by the symmetrising action of the PE on the basic irreps. We can in principle carry out similar procedures for other classical or exceptional Lie algebras and also obtain polynomial generating functions that encode the antisymmetrising action of the PEF. The combinatorics of subsets of the basic irreps, when the $t_i$ do not span the rank of the group, can be read off from such HWGs. Alternatively, HWG series can in principle be obtained for any chosen subset of fugacities by following similar procedures to those set out above.
\subsection{HWGs for Invariant Tensors}

Any group has various invariant tensors, such as delta or epsilon tensors, structure constants and intertwiners, and these can be combined in many ways. For example, the invariant tensors can be contracted with vector or spinor fields to yield fields transforming in other irreps. Also, the invariant tensors of the adjoint representation can be combined with the generators of group transformations to create scalar Casimir operators whose eigenvalues identify the irreps in which fields transform \cite{Cvitanovic:2008mw}. The totally symmetric and totally antisymmetric invariant tensors for the defining representations of various Classical and Exceptional Lie Groups are summarised in Table \ref{tableX}. 

\begin{sidewaystable}
\small
  \caption{Invariants of Defining Representations of Classical and Exceptional Lie Groups}
   \begin{center}

\begin{tabular}{|c|c|c|c|c|c|}
\hline
$Group$
&
$ \begin{array}{c}Defining\\Representation\\ \cal X\end{array}$
& 
$ {PL\left[ {\oint {d\mu PE\left[ { {\cal{X}}~ t} \right]} } \right]}$
&
$ {\oint {d\mu PEF\left[ { {\cal{ X}}~ t} \right]} }$
&
$ \begin{array}{c}Degrees\\of~ primitive\\Sym~tensors\end{array}$
&
$ \begin{array}{c}Degrees\\of ~primitive \\AS~tensors\end{array}$ \\
\hline
$ {{A_r}}$&$ {\left[ {1,0, \ldots , 0} \right]}$&$ 0$&$ {1 + {t^{r + 1}}}$&$ - $&$ {r + 1}$ \\
\hline
$ {{B_r}}$&$ {\left\{ \begin{array}{l}
\left[ 2 \right] for~ r = 1\\
\left[ {1,0, \ldots , 0} \right] for~ r > 1
\end{array} \right.}$&$ {{t^2}}$&$ {1 + {t^{2r + 1}}}$&$ 2$&$ {2r + 1}$ \\
\hline
$ {{C_r}}$&$ {\left[ {1,0, \ldots , 0} \right]}$&$ 0$&$ {1 + {t^2} + {t^4} + \ldots , {t^{2r}}}$&$ - $&$ {2,4, \ldots , 2r~ (*)}$ \\
\hline
$ {{D_r}}$&$ {\left\{ \begin{array}{l}
\left[ {1,1} \right] for~ r = 2\\
\left[ {1,0, \ldots , 0} \right] for~ r > 2
\end{array} \right.}$&$ {{t^2}}$&$ {1 + {t^{2r}}}$&$ 2$&$ {2r}$ \\
\hline
$ {{G_2}}$&$ {[0,1]}$&$ {{t^2}}$&$ {\left( {1 + {t^3}} \right)\left( {1 + {t^4}} \right)}$&$ 2$&$ {3,4}$ \\
\hline
$ {{F_4}}$&$ {[0,0,0,1]}$&$ {{t^2} + {t^3}}$&$ {\left( {1 + {t^9}} \right)\left( {1 + {t^{17}}} \right)}$&$ {2,3}$&$ {9,17}$ \\
\hline
$ {{E_6}}$&$ {[0,0,0,0,1,0]}$&$ {{t^3}}$&$ {1 + {t^{27}}}$&$ 3$&$ {27}$ \\
\hline
$ {{E_7}}$&$ {[0,0,0,0,0,1,0]}$&$ {{t^4}}$&$ {1 + {t^2} + {t^4} \ldots , + {t^{56}}}$&$ 4$&$ {2,4, \ldots , 56~ (*)}$ \\
\hline
$ {{E_8}}$&$ {[0,0,0,0,0,0,1,0]}$&$ \begin{array}{l}
{t^2} + {t^8} + {t^{12}} + {t^{14}}\\
 + {t^{18}} + {t^{20}} + {t^{24}} + {t^{30}}
\end{array}$&$ \begin{array}{l}
\left( {1 + {t^3}} \right)\left( {1 + {t^{15}}} \right)\left( {1 + {t^{23}}} \right)\left( {1 + {t^{27}}} \right)\\
 \times \left( {1 + {t^{35}}} \right)\left( {1 + {t^{39}}} \right)\left( {1 + {t^{47}}} \right)\left( {1 + {t^{59}}} \right)
\end{array}$&$ \begin{array}{l}
2,8,12,14,\\
18,20,24,30
\end{array}$&$ \begin{array}{l}
3,15,23,27,\\
35,39,47,59
\end{array}$ \\
\hline
\end{tabular}
\end{center}
(*):Full list of AS tensors rather than primitives.
 \label{tableX}
\end{sidewaystable}

As can be seen from Table \ref{tableX}, each classical and exceptional group has a unique signature in terms of the invariant tensors of its defining representation \cite{Cvitanovic:2008mw}. Within these, there is a minimal set of tensors in terms of which the other invariant tensors can be expressed, termed primitive tensors. If these primitive tensors are symmetric, they can be symmetrised into symmetric tensors of higher degree; if they are antisymmetric, they are forms over the co-cycles of the group manifold \cite{deAzcarraga:1997ya}, and can only be antisymmetrised up to the length of the overall volume form (or epsilon tensor), as determined by the dimension of the defining representation. The symmetric and antisymmetric tensors can also be combined into tensors of mixed symmetry and the number of such possible combinations compounds with increasing rank and dimension; the exceptional groups in particular posses a very complicated set of defining representation invariants when invariant tensors of mixed symmetry are included \cite{Pouliot:2001iw}. 

Corresponding tables can also be generated for other representations, such as the adjoint, in which cases the tensors correspond to the Casimirs of the group. There are also many invariant tensors that can be formed from combinations of representations.

It is therefore interesting to consider how all the invariant tensors of some representation(s) might most effectively be enumerated. This question is closely related to the problem of identification of GIOs in SQCD, which is the subject of the next section. In the case of symmetric and antisymmetric tensors, respectively, these can be enumerated in a straightforward manner by using Weyl integration to project out the singlets of characters that have been symetrised using the PE, or antisymmetrised using the PEF. Thus we find the degrees of primitive symmetric tensors are given by $t^{d_i^S}$ from the Hilbert series:

\begin{equation}
\label{hwhsIT1}
\begin{aligned}
\sum\limits_i^{} {{t^{d_i^S}}}  = PL\left[ {\oint {d\mu~PE\left[ {\cal X}~t \right]} } \right],
\end{aligned}
\end{equation}
and the degrees of antisymmetric tensors given by $t^{d_i^A}$ from the series:
\begin{equation}
\label{hwhsIT2}
\begin{aligned}
1 + \sum\limits_i^{} {{t^{d_i^A}}}  = \oint {d\mu~PEF\left[ {\cal X}~t \right]}. 
\end{aligned}
\end{equation}
The identification of invariant tensors of mixed symmetry is not so straightforward; in particular, it is necessary to have some way of describing the symmetry properties of each tensor, in addition to tracking the number of its indices. One solution is provided by encoding the symmetry properties of the tensor as a HWG of a unitary representation of sufficiently high dimension. This method makes use of the pattern of symmetrisations and antisymmetrisations that arises when a product group is symmetrised, as will be explained.

Consider the symmetrisation of an irrep transforming in a product group $A \otimes B$:
\begin{equation}
\label{hwhsIT2}
\begin{aligned}
Sym^2(A \otimes B) &= Sy{m^2}A \otimes Sy{m^2}B + {\Lambda ^2}A \otimes {\Lambda ^2}B.
\end{aligned}
\end{equation}
The resulting irreps are either symmetrisations of both constituent groups, or antisymmetrisations of both constituent groups. This contrasts with the situation under overall antisymmetrisation, when the constituent group irreps are of opposite symmetry:

\begin{equation}
\label{hwhsIT3}
\begin{aligned}
{\Lambda ^2}({A \otimes B} ) &= Sy{m^2}A \otimes {\Lambda ^2}B + {\Lambda ^2}A \otimes Sy{m^2}B.
\end{aligned}
\end{equation}

This behaviour of product groups under symmetrisation generalises to any order of symmetrisation $n$ to give all possible combinations, such that the symmetry properties of $Sym^n(A \otimes B)$, which can be described in terms of all the possible Young tableaux at order $n$, match between $A$ and $B$ \cite{Pouliot:2001iw}. Whenever the symmetry properties of $B$ correspond to one of its invariant tensors or singlets, this yields a GIO of $B$ that transforms in some representation of $A$; and we can use a HWG to encode the symmetry properties of this representation of $A$.

As a first step we express the series of GIOs of $B$ as a class function in terms of the characters of $U(A)$ using the fugacity $t$:

\begin{equation}
\label{eq:hwhsIT4}
\begin{aligned}
\sum\limits_{{n_i} = 0}^\infty  {{a_{{n_1}, \ldots ,{n_A}}}\left( t \right)\left[ {{n_1}, \ldots ,{n_A}} \right]}  &= \oint\limits_B {d\mu } ~PE \left[ {{\cal X}\left[ { U\left( A \right)  \otimes B} \right] t} \right].
\end{aligned}
\end{equation}
As a second step we use a generating function for the characters of $U(A)$ with the Dynkin label fugacities $m_i$ to transform the class function  \ref{eq:hwhsIT4} into an HWG :

\begin{equation}
\label{hwhsIT5}
\begin{aligned}
{g^{U\left( A \right)}}\left( {{m_i},t} \right) = \oint\limits_{U(A)} {d\mu } ~{g^{U\left( A \right)}}\left( {{m_i},{\cal X}^*} \right)\oint\limits_B {d\mu} ~PE\left[ {{\cal X} \left[ {U(A) \otimes B} \right]t} \right].
\end{aligned}
\end{equation}

The HWG series \ref{hwhsIT5} can be presented in terms of Young tableaux for $U(N)$, with the total number of boxes being given by the exponents of the fugacities $t$ in a partition described by the fugacities $m_i$. Similar results also can be obtained by decomposition of tensor products \cite{Pouliot:2001iw}, however, the HWG approach has the potential advantage of generating the complete infinite series of GIOs, thereby resolving uncertainties about the multiplicities of distinct invariants and/or their appearance at higher orders. It can be noted that equivalent series can also be obtained by using the $SU(N)$  group in place of $U(N)$, since the rightmost Dynkin label of a $U(N)$ representation can be calculated from the Dynkin labels of the corresponding $SU(N)$ representation, together with the total number of Young boxes given by the fugacity $t$. This provides an alternative calculation schema, which can be more efficient.

\FloatBarrier

Thus, HWGs neatly illustrate how the invariant tensors of the group representations comprised within a product group theory determine the Hilbert series of GIOs. The HWG monomials identify both the order at which such GIOs are formed and the representations in which these objects transform. The above series for product groups incorporating $U(N)$ are closely related to the series for SQCD, which are the subject of detailed examples in the next section.

\section{HWGs of Gauge Invariant Objects of SQCD}
We now apply the tools and methods developed in the previous sections to study the HWGs of product groups associated with SQCD quiver theories. These are theories with no superpotential, i.e. $W \equiv 0$, whose Hilbert series have been calculated within the Plethystics Program \cite{Gray:2008yu, Hanany:2008kn, Hanany:2008sb}. We now wish to construct HWG's for such theories. While we do not aim to give an exhaustive treatment, we set out the principles and provide representative examples.


\subsection{GIOs of $SU(N_f) \times SU(N_c)$}
It is helpful to start with an explicit statement of the symmetry transformation properties of the objects within the theory and these are set out in Table \ref{table7}, based on \cite{Gray:2008yu}:

 \begin{table}[htdp]
\caption{SQCD Charge Assignments for $SU(N_f) \times SU(N_c)$}
\begin{tabular}{|c|c|c c|c c c c|}
\hline
{}& {Gauge/Colour}& {Flavour}&{Flavour}& {}&{}&{}&{} \\
\hline
{}&$ {SU( {{N_c}})}$&$ {SU{{( {{N_f}} )}_L}}$&${SU{{( {{N_f}})}_R}}$&$ {U{{( 1 )}_Q}}$&${U{{( 1)}_{\bar Q}}}$&${U{{( 1)}_B}}$&${U{{( 1)}_R}} $\\
\hline
${Q_a^i}$&$ {[ {0, \ldots , 1}]}$&$ {[ {1,0 \ldots }]}$&$1$&$ 1$&$0$&$1$&${( {{N_f} - {N_c}} )/{N_f}} $\\
${\bar Q_i^a}$& ${[ {1,0 \ldots }]}$& 1&${[ {0, \ldots , 1}]}$& 0&{ - 1}&{ - 1}&${( {{N_f} - {N_c}})/{N_f}}$\\ 
\hline
\end{tabular}
\label{table7}
\end{table}
  
The theory consists of chiral quarks $Q_a^i$ and antiquarks $\bar Q_i^a$ transforming in the $SU(N_f)_L \times SU(N_f)_R \times SU(N_c)$ product group, where the indices $i$ range over the fundamentals/antifundamentals of the $SU(N_f)_L$ and $SU(N_f)_R$ flavour groups and the indices $a$ range over the $SU(N_c)$ colour group. The GIOs of the theory consist of $SU(N_c)$ colour singlets composed of quarks and/or antiquarks, with some combination of paired and unpaired $SU{\left( {{N_f}} \right)_L} \times SU{\left( {{N_f}} \right)_R}$ indices determining the flavour representation in which they transform. The fields also carry $U(1)$ charges. The GIOs are restricted to symmetrised combinations of the chiral quarks and antiquarks.

To obtain HWGs for these theories, we first form the characters of the fields for a given $SU(N_f)_L \times SU(N_f)_R \times SU(N_c)$ product group. The characters are given by:

\begin{equation}
\label{hwhs5.2}
\begin{aligned}
{{{\cal X}}_Q} &= {\left[ {1,0, \ldots ,0} \right]_{{{\left( {{N_f}} \right)}_L}}}{\left[ {0, \ldots , 0,1} \right]_{{N_c}}},\\
{{{\cal X}}_{\bar Q}} &= {\left[ {1,0, \ldots ,0} \right]_{{N_c}}}{\left[ {0, \ldots , 0,1} \right]_{{{\left( {{N_f}} \right)}_R}}}.\\
\end{aligned}
\end{equation}

While the colour group CSA coordinates are shared by the quarks and antiquarks, it is necessary to use different CSA coordinates to distinguish the L and R flavour groups. We introduce the fugacities $t_1$ and $t_2$ for the quarks and anti-quarks respectively, reflecting their different $U(1)$ charges, and symmetrise their characters by applying the PE. A first Weyl integration over the colour group projects out the $SU(N_c)$ singlets. We then introduce the character generating functions $g^{(N_f)_L}(l_i,{\cal X}_L^*)$ and $g^{(N_f)_R}(r_i,{\cal X}_R^*)$, where the $l_i$ and $r_i$ Dynkin label fugacities apply to the L and R flavour groups, respectively. We adopt a particular labelling system for the Dynkin label slots by counting $[l_1, \ldots , l_r]$   from the left and $[r_r, \ldots , r_1]$  from the right. Finally, we apply second and third Weyl integrations over the L and R flavour groups. This procedure, which yields the HWG generating functions that enumerate the GIOs formed from the quarks and antiquarks, can be encapsulated as:

\begin{equation}
\label{hwhs5.3}
\begin{aligned}
{g^{\left( {{\left({N_f} \right)}_L,{{\left( {{N_f}} \right)}_R},N_c} \right)}}\left( {{l_i},{r_i},{t_i}} \right) = \oint\limits_{SU\left( {{N_f}} \right)_L} {\oint\limits_{SU{{\left( {{N_f}} \right)}_R}} {\oint\limits_{SU{\left( {{N_c}} \right)}} {d\mu } } }~&{g^{SU{{\left( {{N_f}} \right)}_L}}}\left( {{l_i},{{{\cal X}}_L}^*} \right){g^{SU{{\left( {{N_f}} \right)}_R}}}\left( {{r_i},{{{\cal X}}_R}^*} \right)\\
\times &PE\left[ {{{{\cal X}}_Q}{t_1} + {{{\cal X}}_{\bar Q}}{t_2}} \right].\\
\end{aligned}
\end{equation}

The results of the calculation of \ref{hwhs5.3} are summarised as HWGs in Table \ref{table8}. For comparison with established results \cite{Gray:2008yu} (and for reasons to be explained shortly) this table also includes cases where the symmetry group is $SU(2N_f) \times SU(2)_c$, with just one flavour group and a single fugacity labelled by $t$.

\begin{table}[htdp]
\caption{HWGs of ${SU(N_f)_L} \times {SU(N_f)_R} \times  SU(N_c)$ colour singlets}
\begin{tabular}{|c|c|}
\hline
 $Theory$ & $HWG$ \\
\hline
$SU(2)_f \times SU(2)_c$ & $1/( 1 - {t^2}) $\\
$SU(2N_f \geq 4) \times SU(2)_c$ & $1/( 1 -l_2 {t^2}) $\\
\hline
$U(1)_{f,L} \times \bar{U}(1)_{f,R} \times SU(2)_c$ & $1/(1 -{t_1}{t_2})$\\
$SU(2)_{f,L} \times SU(2)_{f,R} \times SU(2)_c$ & $ 1/( 1 - {t_1}^2) (1 - {t_2}^2) (1 - {l_1}{r_1}{t_1}{t_2})$ \\
$SU(N_f \geq 3)_{L} \times SU(N_f \geq 3)_{R} \times SU(2)_c$ & $1/(1 - {l_2}{t_1}^2) (1 - {r_2}{t_2}^2) (1 - {l_1}{r_1}{t_1}{t_2}) $\\
\hline
$U(1)_{f,L} \times \bar{U}(1)_{f,R} \times SU( 3)_c$ & $1/(1 -{t_1}{t_2})$\\
$SU(2)_{f,L} \times SU(2)_{f,R} \times SU( 3)_c$ & $1/(1 - {l_1}{r_1}{t_1}{t_2})(1 - {{t_1}^2}{t_2}^2)$\\
$SU(3)_{f,L} \times SU(3)_{f,R} \times SU( 3)_c$ & $1/(1 - {t_1}^3)(1 - {t_2}^3)(1 - {l_1}{r_1}{t_1}{t_2})(1 - {l_2}{r_2}{{t_1}^2}{{t_2}^2})$\\
$SU(N_f \geq 4)_{L} \times SU(N_f \geq 4)_{R} \times SU(3)_c$ & $1/(1 - {l_3}{t_1}^3)(1 - {r_3}{t_2}^3)(1 - {l_1}{r_1}{t_1}{t_2})(1 - {l_2}{r_2}{{t_1}^2}{{t_2}^2})$ \\
\hline
\end{tabular}
\label{table8}
\end{table}

The terms in the HWG encode the structure of the colour singlets of the theory and the flavour group representations to which they belong. Thus, for example, the monomials $\{ {{t_1}^3,{t_2}^3,{l_1}{r_1}{t_1}{t_2},{l_2}{r_2}{t_1}^2{t_2}^2}\}$   for $SU(3)_{f,L} \times SU(3)_{f,R} \times SU(3)_c$ correspond, respectively, to three quarks and three antiquarks transforming as flavour singlets, a quark antiquark pair (meson) transforming in ${\left[ {1,0} \right]_L} \otimes {\left[ {0,1} \right]_R}$ and a combination of two quarks and two antiquarks (tetraquark) transforming in ${\left[ {0,1} \right]_L} \otimes {\left[ {1,0} \right]_R}$.

Interestingly, all the HWGs constructed are the same for all flavour groups of fundamental dimension exceeding the fundamental dimension of the colour group. This arises because the antisymmetrisations of the fundamental of the flavour group generated by the PE are limited by the length of the colour group epsilon tensor, which is the most antisymmetric invariant tensor available in the defining representation. The differences in the unrefined HS as the rank of the flavour group is increased are simply due to the different dimensions of the flavour group irreps. Thus, the HWGs of this SQCD theory are the same for all $SU(N)$ flavour groups of fundamental dimension exceeding that of the $SU(N)$ colour group.

A further interesting observation \cite{Gray:2008yu} is that with a $SU(2)$ colour group, the left and right flavour groups can be combined into a $SU(2N_f)$ global symmetry. This particular feature arises for a $SU(2)$ colour group because $SU(2)$ quarks and antiquarks share the same character. Thus, we can set up a fugacity map in which the CSA flavour coordinates plus fugacity degrees of freedom match between $SU(N_f)_L \times SU(N_f)_R$ and $SU(2N_f)_L$, equalling $2N_f$ on both sides.

The only proviso is that it is necessary to choose all the fugacities to differ only by a $U(1)$ phase. Such alternative ways of analysing the same problem give rise to correspondences between the various Hilbert series as tabulated. Thus, for example, the unrefined Hilbert series for $SU(2) \times SU(2) \times SU( 2)$ in Table  \ref{table9}  is the same as the Hilbert series for $SU(4) \times SU(2)$, (which we can obtain by taking the Hilbert series with distinct counting  for $SU(4) \times SU(4) \times SU(2)$  and applying the mapping ${t_1} \to t,{t_2} \to 0$).


\begin{sidewaystable}[htdp]
\caption{Unrefined HS of $SU(N_f)_L \times SU(N_f)_R \times SU(2)_c$ colour singlets}
\begin{center}
\begin{tabular}{|c|c|c|}
\hline
{\small $Theory$ }
& 
{\small $Unrefined~Hilbert~Series~(t_1,t_2)$ } 
& 
{\small $Unrefined~Hilbert~Series~(t)$ } 
\\
\hline
{\small${SU(2)_f \times SU(2)_c}$} 
& ${\frac{1}{{( {1 - {t^2}} )}}}$ 
& ${\frac{1}{{( {1 - {t^2}} )}}}$\\
\hline
{\small${SU(2)_{f,L} \times SU(2)_{f,R} \times SU(2)_c}$ }
& ${\frac{{( {1 + {t_1}{t_2}} )}}{{( {1 - {t_1}^2} )( {1 - {t_2}^2} ){{( {1 - {t_1}{t_2}} )}^3}}}}$ 
& ${\frac{{1 + {t^2}}}{{{{(1 - {t^2})}^5}}}}$\\
\hline
{\small${SU(3)_{f,L} \times SU(3)_{f,R} \times SU(2)_c}$ }
& 
$\frac{{(1 + {t_1}{t_2})(1 + 3{t_1}{t_2} - 3t_1^3{t_2} - 2t_1^2t_2^2 - 3{t_1}t_2^3 + 3t_1^3t_2^3 + t_1^4t_2^4)}}{{{{(1 - t_1^2)}^3}{{(1 - t_2^2)}^3}{{(1 - {t_1}{t_2})}^5}}}$
& 
$\frac{{(1 + {t^2})(1 + 5{t^2} + {t^4})}}{{{{(1 - {t^2})}^9}}}$\\
\hline
{\small${SU(4)_{f,L} \times SU(4)_{f,R} \times SU(2)_c}$ }
&
{\scriptsize
$\frac{{(1 + {t_1}{t_2})\left( \begin{array}{l}
1 + t_1^2 + 8{t_1}{t_2} - 8t_1^3{t_2} + t_2^2\\
 + 2t_1^2t_2^2 - 21t_1^4t_2^2 + 10t_1^6t_2^2 - 8{t_1}t_2^3 - 8t_1^3t_2^3\\
 + 16t_1^5t_2^3 - 21t_1^2t_2^4 + 54t_1^4t_2^4 - 21t_1^6t_2^4 + 16t_1^3t_2^5\\
 - 8t_1^5t_2^5 - 8t_1^7t_2^5 + 10t_1^2t_2^6 - 21t_1^4t_2^6 + 2t_1^6t_2^6\\
 + t_1^8t_2^6 - 8t_1^5t_2^7 + 8t_1^7t_2^7 + t_1^6t_2^8 + t_1^8t_2^8
\end{array} \right)}}{{{{(1 - t_1^2)}^5}{{(1 - t_2^2)}^5}{{(1 - {t_1}{t_2})}^7}}}$
}
& 
$\frac{{(1 + {t^2})(1 + 14{t^2} + 36{t^4} + 14{t^6} + {t^8})}}{{{{(1 - {t^2})}^{13}}}}$
\\
\hline
{\small${SU(5)_{f,L} \times SU(5)_{f,R} \times SU(2)_c}$ }
&
{\scriptsize
 ${\frac{{( {1 + {t_1}{t_2}} )
\left( \begin{array}{l}
1 + 3t_1^2 + t_1^4 + 15{t_1}{t_2} - 5t_1^3{t_2}\\
 - 10t_1^5{t_2} + 3t_2^2 + 30t_1^2t_2^2 - 109t_1^4t_2^2 + 46t_1^6t_2^2\\
 - 5{t_1}t_2^3 - 70t_1^3t_2^3 - 5t_1^5t_2^3 + 115t_1^7t_2^3 - 35t_1^9t_2^3\\
 + t_2^4 - 109t_1^2t_2^4 + 171t_1^4t_2^4 + 106t_1^6t_2^4 - 94t_1^8t_2^4\\
 - 10{t_1}t_2^5 - 5t_1^3t_2^5 + 405t_1^5t_2^5 - 505t_1^7t_2^5 + 115t_1^9t_2^5\\
 + 46t_1^2t_2^6 + 106t_1^4t_2^6 - 404t_1^6t_2^6 + 106t_1^8t_2^6 + 46t_1^{10}t_2^6\\
 + 115t_1^3t_2^7 - 505t_1^5t_2^7 + 405t_1^7t_2^7 - 5t_1^9t_2^7 - 10t_1^{11}t_2^7\\
 - 94t_1^4t_2^8 + 106t_1^6t_2^8 + 171t_1^8t_2^8 - 109t_1^{10}t_2^8 + t_1^{12}t_2^8\\
 - 35t_1^3t_2^9 + 115t_1^5t_2^9 - 5t_1^7t_2^9 - 70t_1^9t_2^9 - 5t_1^{11}t_2^9\\
 + 46t_1^6t_2^{10} - 109t_1^8t_2^{10} + 30t_1^{10}t_2^{10} + 3t_1^{12}t_2^{10} - 10t_1^7t_2^{11}\\
 - 5t_1^9t_2^{11} + 15t_1^{11}t_2^{11} + t_1^8t_2^{12} + 3t_1^{10}t_2^{12} + t_1^{12}t_2^{12}
\end{array} \right)
}}{{{{( {1 - t_1^2} )}^7}{{( {1 - t_2^2} )}^7}{{( {1 - {t_1}{t_2}} )}^9}}}}$
 }
& 
$\frac{{(1 + {t^2})(1 + 27{t^2} + 169{t^4} + 321{t^6} + 169{t^8} + 27{t^{10}} + {t^{12}})}}{{{{(1 - {t^2})}^{17}}}}$
\\
\hline
\end{tabular}
\end{center}
\label{table9}
\end{sidewaystable}

\begin{sidewaystable}[htdp]
\caption{Unrefined Hilbert Series of $SU(N_f)_L \times SU(N_f)_R \times SU(3)_c$ colour singlets}
\small
\begin{center}
\begin{tabular}{|c|c|c|}
\hline
\textit {Theory} & \textit {Unrefined Hilbert Series $(t_1,t_2)$} & \textit {Unrefined Hilbert Series $(t)$}\\
\hline
$U(1)_{f,L} \times \bar{U}(1)_{f,R} \times SU( 3)_c$ & ${\frac{1}{{{{\left( {1 - {t_1}{t_2}} \right)}}}}}$  & ${\frac{1}{\left(1 - {t^2} \right)}}$ \\
\hline
${SU(2)_{f,L} \times SU(2)_{f,R} \times SU(3)_c}$ 
& ${\frac{1}{{{{\left( {1 - {t_1}{t_2}} \right)}^4}}}}$ 
& ${\frac{1}{{{{\left( {1 - {t^2}} \right)}^4}}}}$\\

\hline
${SU(3)_{f,L} \times SU(3)_{f,R} \times SU(3)_c}$ 
& ${\frac{{\left( {1 - t_1^3t_2^3} \right)}}{{\left( {1 - t_1^3} \right)\left( {1 - t_2^3} \right){{\left( {1 - {t_1}{t_2}} \right)}^9}}}}$ 
& ${\frac{{{t^2} - t + 1}}{{\left( {1 - t} \right){{\left( {1 - {t^2}} \right)}^8}\left( {1 - {t^3}} \right)}}}$\\

\hline
${SU(4)_{f,L} \times SU(4)_{f,R} \times SU(3)_c}$ 
& 
{\tiny
$\frac{{\left( \begin{array}{l}
1 + 4{t_1}{t_2} - 4t_1^4{t_2} + 10t_1^2t_2^2 - 16t_1^5t_2^2 + 6t_1^8t_2^2 + 4t_1^3t_2^3\\
 - 16t_1^6t_2^3 + 8t_1^9t_2^3 - 4{t_1}t_2^4 + 2t_1^4t_2^4 - 4t_1^7t_2^4 + 6t_1^{10}t_2^4 - 16t_1^2t_2^5\\
 + 20t_1^5t_2^5 - 4t_1^8t_2^5 - 16t_1^3t_2^6 + 38t_1^6t_2^6 - 16t_1^9t_2^6 - 4t_1^4t_2^7 + 20t_1^7t_2^7\\
 - 16t_1^{10}t_2^7 + 6t_1^2t_2^8 - 4t_1^5t_2^8 + 2t_1^8t_2^8 - 4t_1^{11}t_2^8 + 8t_1^3t_2^9 - 16t_1^6t_2^9\\
 + 4t_1^9t_2^9 + 6t_1^4t_2^{10} - 16t_1^7t_2^{10} + 10t_1^{10}t_2^{10} - 4t_1^8t_2^{11} + 4t_1^{11}t_2^{11} + t_1^{12}t_2^{12}
\end{array} \right)}}{{{{\left( {1 - t_1^3} \right)}^4}{{\left( {1 - t_2^3} \right)}^4}{{\left( {1 - {t_1}{t_2}} \right)}^{12}}}}$
}
& 
{\scriptsize
${\frac{{\left( {1 + {t^2}} \right)
\left( {1 + 3{t^2} + 4{t^3} + 7{t^4} + 4{t^5} + 7{t^6} + 4{t^7} + 3{t^8} + {t^{10}}} \right)
}}{{{{\left( {1 - {t^2}} \right)}^{12}}{{\left( {1 - {t^3}} \right)}^4}}}}$
}
\\
\hline
${SU(5)_{f,L} \times SU(5)_{f,R} \times SU(3)_c}$ 
& 
{\tiny
${\frac{{
\left({ \begin{array}{l}
1 + 3t_1^3 + t_1^6 + 9{t_1}{t_2} + 2t_1^4{t_2} - 16t_1^7{t_2} + 45t_1^2t_2^2 - 75t_1^5t_2^2\\
 + 15t_1^8t_2^2 + 15t_1^{11}t_2^2 + 3t_2^3 + 74t_1^3t_2^3 - 292t_1^6t_2^3 + 245t_1^9t_2^3\\
 - 65t_1^{12}t_2^3 + 2{t_1}t_2^4 + t_1^4t_2^4 - 313t_1^7t_2^4 + 595t_1^{10}t_2^4 - 300t_1^{13}t_2^4\\
 + 50t_1^{16}t_2^4 - 75t_1^2t_2^5 - 45t_1^5t_2^5 + 150t_1^8t_2^5 + 210t_1^{11}t_2^5 - 315t_1^{14}t_2^5\\
 + 75t_1^{17}t_2^5 + t_2^6 - 292t_1^3t_2^6 + 731t_1^6t_2^6 - 210t_1^9t_2^6 - 140t_1^{12}t_2^6\\
 - 35t_1^{15}t_2^6 + 50t_1^{18}t_2^6 - 16{t_1}t_2^7 - 313t_1^4t_2^7 + 1634t_1^7t_2^7 - 2090t_1^{10}t_2^7\\
 + 715t_1^{13}t_2^7 - 35t_1^{16}t_2^7 + 15t_1^2t_2^8 + 150t_1^5t_2^8 + 675t_1^8t_2^8 - 2175t_1^{11}t_2^8\\
 + 1650t_1^{14}t_2^8 - 315t_1^{17}t_2^8 + 245t_1^3t_2^9 - 210t_1^6t_2^9 - 725t_1^9t_2^9 + 100t_1^{12}t_2^9\\
 + 715t_1^{15}t_2^9 - 300t_1^{18}t_2^9 + 595t_1^4t_2^{10} - 2090t_1^7t_2^{10} + 1775t_1^{10}t_2^{10}\\
 + 100t_1^{13}t_2^{10} - 140t_1^{16}t_2^{10} - 65t_1^{19}t_2^{10} + 15t_1^2t_2^{11} + 210t_1^5t_2^{11} \\
  - 2175t_1^8t_2^{11}+ 3900t_1^{11}t_2^{11} + palindrome \ldots  + t_1^{22}t_2^{22}
\end{array}} \right)
}}{{{{\left( {1 - t_1^3} \right)}^7}{{\left( {1 - t_2^3} \right)}^7}{{\left( {1 - {t_1}{t_2}} \right)}^{16}}}}}$ 
}
& 
{\tiny
${\frac{{\left( {1 - t} \right)\left( \begin{array}{l}
1 + t + 10{t^2} + 23{t^3} + 68{t^4} + 135{t^5} + 281{t^6}\\
 + 446{t^7} + 695{t^8} + 895{t^9} + 1090{t^{10}}\\
  + 1115{t^{11}}+ 1090{t^{12}} + 895{t^{13}} + 695{t^{14}} \\
  + 446{t^{15}}+ 281{t^{16}} + 135{t^{17}} + 68{t^{18}} \\
  + 23{t^{19}} + 10{t^{20}} + {t^{21}} + {t^{22}}\\
\end{array} \right)}}{{{{(1 - {t^2})}^{16}}{{(1 - {t^3})}^7}}}}$
}\\
\hline
\end{tabular}
\end{center}
\label{table10}
\end{sidewaystable}


If we are primarily interested in counting dimensions of flavour irreps, the HWGs can be reduced to unrefined Hilbert series by replacing the monomial terms in $\left\{ {{l_i},{r_j}} \right\}$ by the dimensions of the irreps to which they refer. This is equivalent to replacing the coordinates within the characters of the flavour group by unity. We then have:

\begin{equation}
\label{hwhs5.4}
{g^{\left( {{{\left( {{N_f}} \right)}_L},{{\left( {{N_f}} \right)}_R},{N_c}} \right)}}\left( {{t_i}} \right) = \oint\limits_{SU\left( {{N_c}} \right)} {d\mu }~ P{E^{{N_f}}}\left[ {{{\left[ {0, \ldots ,0,1} \right]}_{SU\left( {{N_c}} \right)}}{t_1} + {{\left[ {1,0, \ldots ,0} \right]}_{SU\left( {{N_c}} \right)}}{t_2}} \right].
\end{equation}

\FloatBarrier

These unrefined Hilbert series, shown in Tables \ref{table9} and  \ref{table10}, can be further simplified, as shown, by equating the quark and antiquark fugacities. The results match those given in \cite{Gray:2008yu}. We defer a comparison of the descriptions of the vacuum moduli spaces given by the HWGs with those provided by Hilbert series to the concluding section.

\subsection{GIOs of $SU(N_f) \times SO(N_c)$}

We can apply a similar methodology to the analysis of the SQCD gauge theories that arise when the colour group is a member of the B, C or D series of classical groups. The unrefined Hilbert series of these theories have been studied extensively \cite{Hanany:2008kn}. We can revisit these analyses using HWG methodology.

In the case of the $SU(N_f) \times SO(N_c)$ product groups, the symmetrisations are carried out on an object transforming in the fundamental representation of the flavour group and in the vector representation of the colour group, with charge assignments as in Table \ref{table11}.

\begin{table}[htdp]
\caption{SQCD Charge Assignments: $SU(N_f) \times SO(N_c)$}
\begin{center}
\begin{tabular}{|c|c|c|c|c|}

\hline
{}& {Gauge/Colour}& {Flavour}& {}&{} \\
\hline
{}& $SO(N_c)$& $SU(N_f)$ & $U( 1)_B$ & $U( 1)_R$ \\
\hline
$Q_a^i$ & {vector} & {fundamental} & 1 & $(N_f + 2 - N_c)/N_f$\\ 
\hline

\end{tabular}
\end{center}
\label{table11}
\end{table}
 
The theory consists of quarks $Q_a^i$ transforming in the $SU(N_f) \times SO(N_c)$ product group, where the indices $i$ and $a$ range over the fundamental dimension of the $SU(N_f)$ flavour groups and the vector representation of the $SO(N_c)$ colour group, respectively. The GIOs of the theory consist of colour singlets composed of quarks, with some combination of paired and unpaired $SU(N_f)$ indices determining the flavour irrep in which they transform. The fields also carry $U(1)$ charges, although these are not central to the analysis. The GIOs are restricted to symmetrised combinations of the quarks.

The characters of the quarks are given by:

\begin{equation}
\label{hwhs5.6}
{{{\cal X}}_Q} = \left\{ \begin{array}{l}
\left[ {1,0 \ldots , 0} \right]_{SU}{\left[ 2 \right]_{SO}}~for~{N_c} = 3\\
\left[ {1,0 \ldots , 0} \right]_{SU}{\left[ {1,1} \right]_{SO}}~for~{N_c} = 4\\
\left[ {1,0 \ldots , 0} \right]_{SU}{\left[ {1,0, \ldots , 0} \right]_{SO}}~for~{N_c} > 4.\\
\end{array} \right.
\end{equation}
Proceeding as before, the HWG generating function is given by:

\begin{equation}
\label{hwhs5.7}
{g^{\left( {{N_f},{N_c}} \right)}}\left( {{m_i},t} \right) = \oint\limits_{SU\left( {{N_f}} \right)} {\oint\limits_{SO\left( {{N_c}} \right)} {d\mu } } ~{{{g}}^{{N_f}}}\left( {{m_i},{{\cal X}}^*} \right)PE\left[ {{{{\cal X}}_Q}t} \right],
\end{equation}
where we use the Dynkin label fugacities $m_i$. The unrefined Hilbert series generating function is given by:

\begin{equation}
\label{hwhs5.8}
{g^{\left( {{N_f},{N_c}} \right)}}\left( t \right) = \oint\limits_{SO\left( {{N_c}} \right)} {d\mu }~PE^{{N_f}}{\left[ {{{{\cal X}}_{SO(N_c)}}t} \right]}.
\end{equation}

The HWGs and unrefined Hilbert series for the $SU(N_f) \times SO(N_c)$ product groups are set out in Table \ref{table12}, Table \ref{table13} and Table \ref{table14}. The unrefined HS replicate established results \cite{Hanany:2008kn}.


\begin{table}[htdp]
\caption{Hilbert Series of $SU(N_f) \times SO(3)_c$ product groups}
\begin{center}
\begin{tabular}{|c|c|c|}
\hline
{\small $Theory$ }
&  
{\small $HWG$}
&  
{\small $Unrefined~Hilbert~Series$}
  \\
\hline
{\small $SU(2) \times SO(3)$}
& 
{\small $ {\frac{1}{{\left( {1 - {t^2}{m^2}} \right)\left( {1 - {t^4}} \right)}}}$}
&  
${\frac{1}{{{{\left( {1 - {t^2}} \right)}^3}}}} $ \\
\hline
{\small $SU(3) \times SO(3)$}
& 
$ {\frac{1}{{\left( {1 - {t^2}{m_1}^2} \right)\left( {1 - {t^3}} \right)\left( {1 - {t^4}{m_2}^2} \right)}}}$
& 
$ {\frac{{1 + {t^3}}}{{{{\left( {1 - {t^2}} \right)}^6}}}} $ \\
\hline
{\small  $ {SU(4) \times SO(3)}$}
& $ {\frac{1}{{\left( {1 - {t^2}{m_1}^2} \right)\left( {1 - {t^3}{m_3}} \right)\left( {1 - {t^4}{m_2}^2} \right)}}}$
& $ {\frac{{1 + {t^2} + 4{t^3} + {t^4} + {t^6}}}{{{{\left( {1 - {t^2}} \right)}^9}}}} $ \\
\hline
{\small $  {SU(5) \times SO(3)}$}
& 
{\small $as~above$}
&$  {\frac{{1 + 3{t^2} + 10{t^3} + 6{t^4} + 6{t^5} + 10{t^6} + 3{t^7} + {t^9}}}{{{{\left( {1 - {t^2}} \right)}^{12}}}}} $ \\
\hline
{\small $  {SU(6) \times SO(3)}$}
& 
{\small $as~above$}
& $ {\frac{{1 + 6{t^2} + 20{t^3} + 21{t^4} + 36{t^5} + 56{t^6} + 36{t^7} + 21{t^8} + 20{t^9} + 6{t^{10}} + {t^{12}}}}{{{{\left( {1 - {t^2}} \right)}^{15}}}}}  $\\
\hline
{\small  $ {SU(7) \times SO(3)}$}
&  
{\small $as~above$}
& 
{\scriptsize
$ {\frac{{\left( \begin{array}{l}
1+10 t^2+35 t^3+55 t^4+126 t^5+220 t^6+225 t^7\\
+225 t^8+220 t^9+126 t^{10}+55 t^{11}+35 t^{12}+10 t^{13}+t^{15}
\end{array} \right)}}{{{{\left( {1 - {t^2}} \right)}^{18}}}}} $
}
\\
\hline
\end{tabular}
\end{center}
\label{table12}
\end{table}
\begin{table}[htdp]
\caption{Hilbert Series of $SU(N_f) \times SO(4)_c$ product groups}
\begin{center}
\begin{tabular}{|c|c|c|}
\hline
 {\small ${Theory}$}
& {\small ${HWG}$}
& {\small ${Unrefined~Hilbert~Series} $} \\
\hline
 {\small$ {SU(2) \times SO(4)}$}
& $ {\frac{1}{{\left( {1 - {t^2}{m^2}} \right)\left( {1 - {t^4}} \right)}}}$
&  ${\frac{1}{{{{\left( {1 - {t^2}} \right)}^3}}}} $\\
\hline
 {\small ${SU(3) \times SO(4)}$}
& $ {\frac{1}{{\left( {1 - {t^2}{m_1}^2} \right)\left( {1 - {t^4}m_2^2} \right)\left( {1 - {t^6}} \right)}}}$
& $ {\frac{1}{{{{\left( {1 - {t^2}} \right)}^6}}}}  $\\
\hline
 {\small ${SU(4) \times SO(4)}$}
& $ {\frac{1}{{\left( {1 - {t^2}{m_1}^2} \right)\left( {1 - {t^4}m_2^2} \right)\left( {1 - {t^4}} \right)\left( {1 - {t^6}m_3^2} \right)}}}$
&  ${\frac{{1 + {t^4}}}{{{{\left( {1 - {t^2}} \right)}^{10}}}}} $ \\
\hline
 {\small ${SU(5) \times SO(4)}$}
& 
$ {\frac{1}{{\left( {1 - {t^2}{m_1}^2} \right)\left( {1 - {t^4}m_2^2} \right)\left( {1 - {t^4}{m_4}} \right)\left( {1 - {t^6}m_3^2} \right)}}}$
& 
$ {\frac{{1 + {t^2} + 6{t^4} + {t^6} + {t^8}}}{{{{\left( {1 - {t^2}} \right)}^{14}}}}} $ \\
\hline
  {\small${SU(6) \times SO(4)}$}
& 
{\small $as~above$}
& $ {\frac{{1 + 3{t^2} + 21{t^4} + 20{t^6} + 21{t^8} + 3{t^{10}} + {t^{12}}}}{{{{\left( {1 - {t^2}} \right)}^{18}}}}} $ \\
\hline
 {\small ${SU(7) \times SO(4)}$}
& 
{\small $as~above$}
&  ${\frac{{1 + 6{t^2} + 56{t^4} + 126{t^6} + 210{t^8} + 126{t^{10}} + 56{t^{12}} + 6{t^{14}} + {t^{16}}}}{{{{\left( {1 - {t^2}} \right)}^{22}}}}} $ \\
\hline
 {\small ${SU(8) \times SO(4)}$}
& 
{\small $as~above$}
& 
{\scriptsize $ {\frac{{\left( \begin{array}{l}
1 + 10{t^2} + 125{t^4} + 500{t^6} + 1310{t^8} + 1652{t^{10}}\\
 + 1310{t^{12}} + 500{t^{14}} + 125{t^{16}} + 10{t^{18}} + {t^{20}}
\end{array} \right)}}{{{{\left( {1 - {t^2}} \right)}^{26}}}}} $}\\
\hline
\end{tabular}
\end{center}
\label{table13}
\end{table}

\begin{sidewaystable}[htdp]
\small
\caption{Unrefined Hilbert Series of $SU(N_f) \times SO(5)_c$ colour singlets  }
\begin{center}
\begin{tabular}{|c|c|c|}

\hline
{\small$Theory$}
&
{\small$ HWG$}
&
{\small$ Unrefined~Hilbert~Series$} \\
\hline
{\small$ SU(2) \times SO(5)$}
&
$ {\frac{1}{{( {1 - {t^2}{m^2}} )( {1 - {t^4}} )}}}$
&
$ {\frac{1}{{{{( {1 - {t^2}} )}^3}}}}$ \\
\hline
{\small$ {SU(3) \times SO(5)}$}
&
$ {\frac{1}{{( {1 - {t^2}{m_1}^2} )( {1 - {t^4}{m_2}^2} )( {1 - {t^6}} )}}}$
&
$ {\frac{1}{{{{( {1 - {t^2}} )}^6}}}}$ \\
\hline
{\small$ {SU(4) \times SO(5)}$}
&
$ {\frac{1}{{( {1 - {t^2}{m_1}^2} )( {1 - {t^4}{m_2}^2} )( {1 - {t^6}{m_3}^2} )( {1 - {t^8}} )}}}$
&
$ {\frac{1}{{{{( {1 - {t^2}} )}^{10}}}}}$ \\
\hline
{\small$ {SU(5) \times SO(5)}$}
&
$ {\frac{1}{{( {1 - {t^2}{m_1}^2} )( {1 - {t^4}{m_2}^2} )( {1 - {t^5}} )( {1 - {t^6}{m_3}^2} )( {1 - {t^8}{m_4}^2} )}}}$
&
$ {\frac{{1 + {t^5}}}{{{{( {1 - {t^2}} )}^{15}}}}}$ \\
\hline
{\small$ {SU(6) \times SO(5)}$}
&
$ {\frac{1}{{( {1 - {t^2}{m_1}^2} )( {1 - {t^4}{m_2}^2} )( {1 - {t^5}{m_5}} )( {1 - {t^6}{m_3}^2} )( {1 - {t^8}{m_4}^2} )}}}$
&
$ {\frac{{1 + {t^2} + {t^4} + 6{t^5} + {t^6} + {t^8} + {t^{10}}}}{{{{( {1 - {t^2}} )}^{20}}}}}$ \\
\hline
{\small$ {SU(7) \times SO(5)}$}
&
{\small $as~above$}
&
{\small$ {\frac{{\left( \begin{array}{l}
1 + 3{t^2} + 6{t^4} + 21{t^5} + 10{t^6} + 15{t^7}\\
 + palindrome~.~.~.~ + {t^{15}}
\end{array} \right)}}{{{{( {1 - {t^2}} )}^{25}}}}}$} \\
\hline
{\small$ {SU(8) \times SO(5)}$}
&
{\small $as~above$}
&
{\small$ {\frac{{{\left( \begin{array}{l}
 
1+6 t^2+21 t^4+56 t^5+56 t^6+120 t^7\\
 +126 t^8+160 t^9+252 t^{10}+160 t^{11}+126 t^{12}\\
 +120 t^{13}+56 t^{14}+56 t^{15}+21 t^{16}+6 t^{18}+t^{20}
  
\end{array} \right)}}}
{{{{( {1 - {t^2}} )}^{30}}}}}$ }\\
\hline
$ {SU(9) \times SO(5)}$
&
{\small $as~above$}
&
{\small$ {\frac{{\left( \begin{array}{l}
1 + 10{t^2} + 55{t^4} + 126{t^5} + 220{t^6} + 540{t^7}\\
 + 715{t^8} + 1270{t^9} + 2002{t^{10}} + 2080{t^{11}} + 2485{t^{12}}\\
 + palindrome~.~.~.~+ {t^{25}}
\end{array} \right)}}{{{{( {1 - {t^2}} )}^{35}}}}}$} \\
\hline

\end{tabular}
\end{center}

\label{table14}

\end{sidewaystable}
\FloatBarrier
As before, the HWGs of this SQCD theory are the same for all flavour $SU(N_f)$ groups of fundamental dimension exceeding the vector dimension of the $SO(N_c)$ colour group.
\FloatBarrier
\subsection{GIOs of $SU(N_f) \times USp(2n_c)$}

In the case of $SU(N_f) \times USp(2n_c)$ product groups, the symmetrisations are carried out on an object transforming in the fundamental representation of the flavour group and in the defining $2n$ dimensional fundamental representation of the symplectic colour group. We use the charge assignments in Table \ref{table15}  \cite{Hanany:2008kn}.

\begin{table}[htdp]
\caption{SQCD Charge Assignments: $SU(2N_f) \times USp(2n_c)$}
\begin{center}
\begin{tabular}{|c|c|c|c|c|}
\hline
{}& {Gauge/Colour} & {Flavour} & {} & {} \\
\hline
{}& $USp(2n_c)$ & $SU(2N_f)$ & $U( 1)_B$ & $U( 1)_R$ \\
\hline
$Q_a^i$ & {vector} & {fundamental} & 1 & $(N_f -1 - n)/N_f$\\ 
\hline
\end{tabular}
\end{center}
\label{table15}
\end{table}
 
The theory consists of quarks $Q_a^i$ transforming in the $SU(2N_f) \times {USp(2n_c)} $ product group, where the indices i and a range over the fundamental dimension of the flavour group and the $2n$ dimensional fundamental representation of the colour group, respectively. The GIOs of the theory are colour singlets composed of quarks, with some combination of paired and unpaired flavour indices determining the flavour irrep in which they transform. The fields also carry $U(1)$ charges, although these are not central to the analysis. The GIOs are restricted to symmetrised combinations of quarks. The further restriction in Table \ref{table15} to flavour groups of even fundamental dimension follows \cite{Witten:1982fp}. For greater generality, we derive below the HWGs for any flavour group $SU(N_f)$; these results can readily be specialised to flavour groups of even dimension, as desired.

The characters of the quarks are given by:

\begin{equation}
\label{hwhs5.9}
{\cal X}_Q = [1, \ldots , 0]_{SU} [1, \ldots , 0]_{USp}.
\end{equation}
Proceeding as before, the HWG generating function is given by:

\begin{equation}
\label{hwhs5.10}
{g^{\left( {{n_c},{N_f}} \right)}}(m_i,t) = \oint \limits_{SU(N_f)} \oint\limits_{USp(2n)} {d\mu } ~ {g^{{N_f}}}\left( {{m_i},{{\cal X}}^*} \right)PE\left[ {{{{\cal X}}_Q}t} \right],
\end{equation}
where we use the Dynkin label fugacities $m_i$, and the unrefined Hilbert series generating function is given by:

\begin{equation}
\label{hwhs5.11}
{g^{\left( {{n_c},{N_f}} \right)}}\left( t \right) = \oint\limits_{USp\left( {{2n}} \right)} {d\mu }~PE^{{N_f}}{\left[ {{{\left[ {1, \ldots , 0} \right]}_{USp}}t} \right]}.
\end{equation}
The HWGs and unrefined Hilbert series for the $SU(N_f) \times USp(2n_c)$ product groups are set out in Table \ref{table16} and Table \ref{table17}. The unrefined Hilbert series for flavour groups of even fundamental dimension match established results \cite{Gray:2008yu, Hanany:2008kn}. 

\begin{table}[htdp]
\caption{Hilbert Series of $SU(N_f) \times USp(2)_c$ colour singlets}
\begin{center}
\begin{tabular}{|c|c|c|}

\hline
  $Theory $ & $  {HWG} $ & $  {Unrefined~Hilbert~series} $ \\
\hline
  ${SU(2) \times USp(2)} $ & $  {\frac{1}{{1 - {t^2}}}} $ & $  {\frac{1}{{1 - {t^2}}}}$  \\
  ${SU(3) \times {USp(2)}} $ & $  {\frac{1}{{1 - {t^2}{m_2}}}} $ & $  {\frac{1}{{{{\left( {1 - {t^2}} \right)}^3}}}} $ \\
  ${SU(4) \times {USp(2)}} $ & 
{\small $as~above$}
  & $  {\frac{{1 + {t^2}}}{{{{\left( {1 - {t^2}} \right)}^5}}}} $ \\
  ${SU(5) \times {USp(2)}} $ & 
 {\small $as~above$}
  & $  {\frac{{1 + 3{t^2} + {t^4}}}{{{{\left( {1 - {t^2}} \right)}^7}}}}  $\\
  ${SU(6) \times {USp(2)}} $ & 
  {\small $as~above$} 
  & $  {\frac{{(1 + {t^2})(1 + 5{t^2} + {t^4})}}{{{{\left( {1 - {t^2}} \right)}^9}}}}  $\\
\hline

\end{tabular}
\end{center}
\label{table16}
\end{table}%

\begin{table}[htdp]
\small
\caption{Hilbert Series of $SU(N_f) \times USp(4)_c$ colour singlets}
\begin{center}
\begin{tabular}{|c|c|c|}

\hline
 $Theory$ & $  {HWG} $ & $  {Unrefined~Hilbert~Series} $ \\
\hline
 $ {SU(2) \times {USp(4)}} $ & $  {\frac{1}{{\left( {1 - {t^2}} \right)}}} $ & $  {\frac{1}{{\left( {1 - {t^2}} \right)}}}  $\\
\hline
  ${SU(3) \times {USp(4)}} $ & $  {\frac{1}{{\left( {1 - {t^2}{m_2}} \right)}}} $ & $  {\frac{1}{{{{\left( {1 - {t^2}} \right)}^3}}}}  $\\
\hline
  ${SU(4) \times {USp(4)}} $ & $  {\frac{1}{{\left( {1 - {t^2}{m_2}} \right)\left( {1 - {t^4}} \right)}}} $ & $  {\frac{1}{{{{\left( {1 - {t^2}} \right)}^6}}}} $ \\
\hline
  ${SU(5) \times {USp(4)}} $ & $  {\frac{1}{{\left( {1 - {t^2}{m_2}} \right)\left( {1 - {t^4}{m_4}} \right)}}} $ & $  {\frac{1}{{{{\left( {1 - {t^2}} \right)}^{10}}}}}  $\\
\hline
  ${SU(6) \times {USp(4)}} $ & 
 {\small $as~above$}
  & $  {\frac{{1 - {t^6}}}{{{{\left( {1 - {t^2}} \right)}^{15}}}}}  $\\
\hline
  ${SU(7) \times {USp(4)}} $ & 
  {\small $as~above$}
  & $  {\frac{{1 + 3{t^2} + 6{t^4} + 3{t^6} + {t^8}}}{{{{\left( {1 - {t^2}} \right)}^{18}}}}} $ \\
\hline
  ${SU(8) \times {USp(4)}} $ & 
  {\small $as~above$}
  & $  {\frac{{1 + 6{t^2} + 21{t^4} + 28{t^6} + 21{t^8} + 6{t^{10}} + {t^{12}}}}{{{{\left( {1 - {t^2}} \right)}^{22}}}}}  $\\
\hline
  ${SU(9) \times {USp(4)}} $ & 
  {\small $as~above$}
& 
{\scriptsize
$  {\frac{{\left( \begin{array}{l}
1 + 10{t^2} + 55{t^4} + 136{t^6} + 190{t^8}\\
 + 136{t^{10}} + 55{t^{12}} + 10{t^{14}} + {t^{16}}
\end{array} \right)}}{{{{\left( {1 - {t^2}} \right)}^{26}}}}} $
} \\
\hline
  ${SU(10) \times {USp(4)}} $ &
  {\small $as~above$} 
&
{\scriptsize
 $  {\frac{{\left( \begin{array}{l}
1 + 15{t^2} + 120{t^4} + 470{t^6} + 1065{t^8} + 1377{t^{10}} + \\
1065{t^{12}} + 470{t^{14}} + 120{t^{16}} + 15{t^{18}} + {t^{20}}
\end{array} \right)}}{{{{\left( {1 - {t^2}} \right)}^{30}}}}}  $
}\\

\hline

\end{tabular}
\end{center}
\label{table17}
\end{table}
 
Once again, the HWGs are the same for all flavour $SU(N_f)$ groups of fundamental dimension exceeding that of the colour group.
\FloatBarrier

\subsection{GIOs of $SU(N_f) \times G_{2}$}
Finally, it is interesting to examine the situation where the SQCD colour group is taken as an exceptional group, of which $G_2$ is the lowest rank example. We use the charge assignments in Table \ref{table18}.

\begin{table}[htdp]
\caption{SQCD Charge Assignments: $SU(N_f) \times {G_{2}}$}
\begin{center}
\begin{tabular}{|c|c|c|}

\hline
{}& {Gauge/Colour} & {Flavour} \\
\hline
{}& ${G_2}$ & $SU(N_f)$ \\
\hline
$Q_a^i$ & $[0,1]$ & $[1,0,\ldots ,0]$\\ 
\hline

\end{tabular}
\end{center}
\label{table18}
\end{table}
The theory consists of quarks $Q_a^i$ transforming in the $SU(N_f) \times {G_{2}}$ product group, where the indices $i$ and $a$ range over the fundamental representation $[1,0,\ldots,0]$ of the $SU(N_f)$ flavour group and fundamental representation $[0,1]$  of the $G_2$ colour group, respectively. The GIOs of the theory consist of colour singlets composed of quarks, with some combination of paired and unpaired flavour indices determining the irrep in which they transform. $U(1)$ charges are not shown. The GIOs are restricted to symmetrised combinations of quarks.

The characters of the quarks are given by:

\begin{equation}
{{{\cal X}}_Q} = {\left[ {1,0, \ldots , 0} \right]_{SU\left( {{N_f}} \right)}}{\left[ {0,1} \right]_{{G_2}}}.
\label{hwhs5.12}
\end{equation}
The HWG generating function is given by:

\begin{equation}
{g^{\left( {SU({N_f}),{G_2}} \right)}}\left( {{m_i},t} \right) ={\oint\limits_{SU\left( {{N_f}} \right)}  \oint\limits_{{G_2}} {d\mu } } ~{g^{{N_f}}}\left( {{m_i},{{\cal X}}^*} \right)PE\left[ {{{{\cal X}}_Q}t} \right],
\label{hwhs5.13}
\end{equation}
where we use the $SU(N_f)$ Dynkin label fugacities $m_i$. The unrefined Hilbert series generating function is given by:

\begin{equation}
{g^{\left( {SU\left( {{N_f}} \right),{G_2}} \right)}}\left( t \right) = \oint\limits_{{G_2}} {d\mu }~PE^{{N_f}}{\left[ {{{\left[ {0,1} \right]}_{{G_2}}}t} \right]}.
\label{hwhs5.14}
\end{equation}
The HWGs and unrefined Hilbert series for the $SU(N_f) \times G_{2}$ product groups are set out in Tables \ref{table19} and  \ref{table20}.

\begin{table}[htdp]
\caption{HWGs of $SU(N_f) \times G_2$ colour singlets}
\begin{center}
\begin{tabular}{|c|c|c|}
\hline
$  Theory $ & $  {HWG}  $\\
\hline
 $ {SU(2) \times {G_2}} $ & $  {\frac{1}{{\left( {1 - {m^2}{t^2}} \right)\left( {1 - {t^4}} \right)}}} $ \\
\hline
$  {SU(3) \times {G_2}} $ & $ \frac{1}{{\left( {1 - m_1^2{t^2}} \right)\left( {1 - {t^3}} \right)\left( {1 - m_2^2{t^4}} \right)\left( {1 - {t^6}} \right)}} $\\
\hline
 $ {SU(4) \times {G_2}} $ & $ \frac{{\left( {1 + {m_1}{m_2}{t^7}} \right)\left( {1 + {m_2}{m_3}{t^9}} \right)}}{{\left( {1 - m_1^2{t^2}} \right)\left( {1 - {m_3}{t^3}} \right)\left( {1 - {t^4}} \right)\left( {1 - m_2^2{t^4}} \right)\left( {1 - {m_1}{t^5}} \right)\left( {1 - m_3^2{t^6}} \right)\left( {1 - {t^8}} \right)\left( {1 - m_2^2{t^{12}}} \right)}} $\\
\hline
\end{tabular}
\end{center}
\label{table19}
\end{table}

\begin{table}[htdp]
\caption{Unrefined Hilbert series of $SU(N_f) \times G_2$ colour singlets}
\begin{center}
\begin{tabular}{|c|c|c|}
\hline
  $Theory $ & $  {Unrefined~Hilbert~Series} $ \\
\hline
$  {SU(2) \times {G_2}} $ & $  {\frac{1}{{{{\left( {1 - {t^2}} \right)}^3}}}} $ \\
\hline
 $ {SU(3) \times {G_2}} $ & $  {\frac{1}{{\left( {1 - {t^3}} \right){{\left( {1 - {t^2}} \right)}^6}}}} $ \\
\hline
$  {SU(4) \times {G_2}} $ & $  {\frac{{1 + {t^4}}}{{{{\left( {1 - {t^3}} \right)}^4}{{\left( {1 - {t^2}} \right)}^{10}}}}} $ \\
\hline
 $ {SU(5) \times {G_2}} $ & $ 
\frac{\left(1+t^2+3 t^3+6 t^4+3 t^5+7 t^6+8 t^7+7 t^8+3 t^9+6 t^{10}+3 t^{11}+t^{12}+t^{14}\right)}{\left(1-t^2\right)^{14} \left(1-t^3\right)^7}
  $\\
\hline
\end{tabular}
\end{center}
\label{table20}
\end{table}

We can identify the roles of the $G_2$ symmetric invariant tensors of order 2, 4, 6 and of the antisymmetric invariant tensors of order 3 and 4, which appear in the HWGs for $SU(2)$, $SU(3)$ and $SU(4)$ flavour groups. There are also other invariant tensors that appear. In the case of an SU(5) flavour group, proceeding as before, we can obtain the refined Hilbert series in the form of a class function:
\begin{equation}
{g^{\left( {SU\left( 5 \right),{G_2}} \right)}}\left( {\cal X},t \right) = \left( {1 - {t^5}} \right)\left( \begin{array}{r}
\left( {1 + {t^5} - {t^{15}} - {t^{20}}} \right)\left[ {0,0,0,0} \right]\\
 + {t^4}\left( {1 + {t^5} + {t^{10}}} \right)\left[ {0,0,0,1} \right]\\
 + {t^{13}}\left[ {0,0,1,0} \right]\\
 - {t^7}\left[ {0,1,0,0} \right]\\
 - {t^6}\left( {1 + {t^5} + {t^{10}}} \right)\left[ {1,0,0,0} \right]
\end{array} \right) PE\left[ {\left[ {2,0,0,0} \right] {t^2}} \right]PE\left[ {\left[ {0,0,1,0} \right] {t^3}} \right].
\label{su5g2a}
\end{equation}
To illustrate the rich nature of the series of GIOs arising in $G_2$ gauge theories, we give below the terms of the refined Hilbert series expansion up to order 15 in the fugacity t, using HWG notation. The exact rational HWG corresponding to this series, however, remains to be found\footnote{Due to computing constraints}:

\begin{equation}
\begin{aligned}
{g^{\left( {SU\left( 5 \right),{G_2}} \right)}}\left( {{m_i},t} \right) & =
1 + m_1^2{t^2} + {m_3}{t^3} + m_1^4{t^4} + m_2^2{t^4} + {m_4}{t^4} + m_1^2{m_3}{t^5} + {m_1}{m_4}{t^5} \\
&+ {m_1}{t^6} + m_1^6{t^6} + m_1^2m_2^2{t^6} + 2m_3^2{t^6} + m_1^2{m_4}{t^6}\\
 &+ {m_2}{t^7} + m_1^4{m_3}{t^7} + m_2^2{m_3}{t^7} + m_1^3{m_4}{t^7} + {m_1}{m_2}{m_4}{t^7} + {m_3}{m_4}{t^7}\\
 &+ m_1^3{t^8} + m_1^8{t^8} + {m_1}{m_2}{t^8} + m_1^4m_2^2{t^8} + m_2^4{t^8} + 2m_1^2m_3^2{t^8} \\
&+ m_1^4{m_4}{t^8} + m_2^2{m_4}{t^8} + {m_1}{m_3}{m_4}{t^8} + 2m_4^2{t^8}\\
 &+ m_1^2{m_2}{t^9} + 2{m_1}{m_3}{t^9} + m_1^6{m_3}{t^9} + m_1^2m_2^2{m_3}{t^9} + 2m_3^3{t^9}\\
 &+ m_1^5{m_4}{t^9} + m_1^3{m_2}{m_4}{t^9} + {m_1}m_2^2{m_4}{t^9} + m_1^2{m_3}{m_4}{t^9} \\
  &+{m_2}{m_3}{m_4}{t^9} + {m_1}m_4^2{t^9}+ {t^{10}} + m_1^5{t^{10}} + m_1^{10}{t^{10}} \\
  &+ m_1^3{m_2}{t^{10}} + {m_1}m_2^2{t^{10}} + m_1^6m_2^2{t^{10}} + m_1^2m_2^4{t^{10}} \\
  &+ 2{m_2}{m_3}{t^{10}} + 2m_1^4m_3^2{t^{10}}+ 2m_2^2m_3^2{t^{10}} + {m_1}{m_4}{t^{10}} \\
  &+ m_1^6{m_4}{t^{10}} + m_1^2m_2^2{m_4}{t^{10}} + m_1^3{m_3}{m_4}{t^{10}} \\
  &+ {m_1}{m_2}{m_3}{m_4}{t^{10}} + 2m_3^2{m_4}{t^{10}} + 3m_1^2m_4^2{t^{10}} +  \ldots 
  \end{aligned}
\label{su5g2b}
\end{equation}

Moreover, it is clear that the HWGs calculated have not yet converged in the manner noted earlier for classical colour groups (of defining dimension lower than $G_2$). Noting the seven dimensional epsilon tensor of $G_2$, we can expect that such convergence of the HWG should occur for flavour groups $SU(8)$ and above.
\section{HWGs of Instanton Moduli Spaces}
Instantons continue to attract considerable interest since their discovery as self-dual solutions of Yang-Mills field equations in 1975 \cite{Tong:2005un}. Many studies have been carried out on different aspects of instantons, such as \cite{Nekrasov:2004vw}, which gives an ADHM construction for the instantons of Yang-Mills fields transforming in Classical gauge groups. This paper focuses just on the analysis of the moduli spaces of instantons, in terms of their Hilbert series and character expansions. In this context, the Hilbert series for the moduli spaces of one and two instantons of Classical gauge groups have been constructed from SUSY quiver theories \cite{Benvenuti:2010pq, Hanany:2012dm}. Efforts have been made to construct the moduli spaces of two instantons of Exceptional gauge groups \cite{Keller:2012da}. Our aim here is to construct the HWGs for the moduli spaces of some low rank $SU(N)$ instantons on $\mathbb{C}^2$ and to show how HWGs can be used to study the structure of these moduli spaces.

We set out in Table \ref{table21} the field content of a quiver theory for the moduli space of $k$ $SU(N)$ instantons on $\mathbb{C}^2$, following \cite{Benvenuti:2010pq}. The instanton moduli space is identified with the Higgs branch of the quiver theory \cite{Douglas:1995bn}. The fields in the quiver theory transform in some representation of a quiver gauge group determined by the number of instantons, in addition to transforming under the Yang-Mills gauge group (also referred to in the literature as a {\it flavour} group). The instanton moduli spaces are given by field combinations that are singlets of the quiver gauge group.

Since the representations of unitary groups do not contain antiparticles \cite{Cvitanovic:2008mw}, it is helpful to decompose the unitary representations within the quiver into special unitary representations, by extracting overall $U(1)$ charges. Then the $X_{12}$ and $X_{21}$ bi-fundamental fields transform in conjugate representations with respect to the $SU(k)$ quiver gauge and $SU(N)$ Yang-Mills gauge groups and also carry conjugate $U(1)$ charges. The fields $\{\Phi ,\phi ^{(a)}\}$  transform in the $U(k)$ adjoint, formed from the product of the fundamental $U(k)$ representation with its conjugate, which can be decomposed as a $SU(k)$ adjoint plus a singlet. The $U(1)$ symmetry associated with the Yang-Mills gauge group has been absorbed into the local $U(1)$ symmetry of the quiver gauge group. We use a fugacity $t$, corresponding to the global $U(1)$ charge\footnote{Technically this global $U(1)$ counts the highest weights of an $SU(2)$ R-symmetry}, to count the fields.

\begin{table}[htdp]
\small
\caption{Field Content of Quiver Theory for Moduli Space of k $SU(N)$ Instantons on $\mathbb{C}^2$}
\begin{center}
\begin{tabular}{|c|c c|c|c|c|}

\hline
 $ Brane~perspective $ & $  k~D3~branes $ & ${} $ & $  N~D7~branes $ & $  \mathbb{C}^2 $ & ${}$ \\
\hline
  ${} $ 
& $  \begin{array}{c}Quiver\\Gauge\\Group\end{array}: $  
& ${U(k)} $ 
& $  \begin{array}{c}Yang-Mills\\Gauge\\Group\end{array} $ 
& $   \begin{array}{c}Global\\Symmetry\end{array}:U(2) $ 
& $  \begin{array}{c}Fugacity\end{array} $ \\
 $ {Field} $ 
& $  {\begin{array}{c}{SU(k)}\end{array}} $
& ${U(1)} $ & $  {SU(N)} $ 
& $  {\begin{array}{*{20}{r}} {SU(2)_g}  & ~~{U(1)_g}\end{array}} $ 
& $  {}  $\\
\hline
  $\Phi  $ 
& $  {\begin{array}{*{20}{c}}{\left[ {1,0, \ldots , 0,1} \right] + 1}\end{array}} $ 
& $0 $ 
& $  {\left[ {0, \ldots ,0} \right]} $ 
& $  {\begin{array}{*{20}{l}} {[0]~~~}  & 0\end{array}} $ 
& $   -  $ \\
$  {{\phi ^{\left(\alpha \right)}}} $ 
& $  {\begin{array}{*{20}{c}}{\left[ {1,0, \ldots , 0,1} \right] + 1}\end{array}} $ 
& $0 $ 
& $  {\left[ {0, \ldots ,0} \right]} $ 
& $  {\begin{array}{*{20}{c}} {\left[ 1 \right]}  &~~ {\left( 1 \right)}\end{array}} $ 
& $  {{t}} $ \\
$  {{X_{12}}} $ 
& $  {\begin{array}{*{20}{c}} {\left[ {1,0, \ldots ,0} \right]} \end{array}} $ 
& ${\left( 1 \right)} $ & $  {\left[ {0, \ldots ,0,1} \right]} $ 
& $  {\begin{array}{*{20}{c}}{\left[ 0 \right]}  &~~ {\left( 1 \right)}\end{array}} $ 
& $  {{t}}  $\\
$  {{X_{21}}} $ 
& $  {\begin{array}{*{20}{c}} {\left[ {0, \ldots , 0,1} \right]}\end{array}} $ 
& ${\left( { - 1} \right)} $ 
& $  {\left[ {1,0 \ldots ,0} \right]} $ 
& $  {\begin{array}{*{20}{c}}{\left[ 0 \right]}  
&~~ {\left( 1 \right)}\end{array}} $ 
& $  {{t}} $ \\
\hline
$  {CSA~Coordinates} $ 
& $  {\begin{array}{*{20}{c}}{{y_1}, \ldots ,{y_{k - 1}}}\end{array}} $ 
& $w $ & $  {{x_1}, \ldots ,{x_{N - 1}}} $ 
& $  {\begin{array}{*{20}{l}}x  
& {~~~~~~}\end{array}} $ 
& $  {}  $\\
\hline

\end{tabular}
\end{center}
\label{table21}
\end{table}
  
The theory is defined not only by its basic fields, but also by its superpotential \cite{Benvenuti:2010pq}:

\begin{equation}
\label{hwhs6.1}
{{\cal W}} = Tr\left( {{X_{21}} \Phi {X_{12}} + {\varepsilon _{\alpha \beta }}{\phi ^{\left( \alpha  \right)}} \Phi {\phi ^{\left( \beta  \right)}}} \right).
\end{equation}
The trace is taken over all unpaired gauge group indices. We apply variational principles, requiring that the superpotential should be extremised with respect to the field $\Phi$ :

\begin{equation}
\label{hwhs6.2}
\delta {{\cal W}}{\rm{ = }}\frac{{\partial {{\cal W}}}}{{\partial \Phi }}\delta \Phi  = 0.
\end{equation}
This in turn leads to the following F-term constraints or selection rules between $SU(N)$ singlets formed from the $X$ fields and the global singlets of the $\phi ^{(\alpha)}$ fields, respectively, where we denote quiver gauge indices by $(a,b,\ldots)$ and Yang-Mills gauge indices by $(i,j,\ldots)$:

\begin{equation}
\label{hwhs6.3}
\left( {{X_{12}}} \right)_a^{~i}\left( {{X_{21}}} \right)_i^{~b} = {\varepsilon _{\alpha \beta }}\left( {{\phi ^{\left( \alpha  \right)}}} \right)_a^{~c}\left( {{\phi ^{\left( \beta  \right)}}} \right)_c^{~b}.
\end{equation}
For the gauge group $U(k)$, there are $k^2$ relations in total. For a $U(1)$ gauge symmetry the commutator of the $\phi ^{(\alpha)}$ fields vanishes and this imposes the F-term constraint that there can be no $SU(N)$ singlets formed from pairs of $X$ fields. For $k>1$, the commutator does not vanish, and so the $SU(N)$ singlets formed from pairs of $X$ fields become identical to global $SU(2)$ singlets formed by contracting pairs of $\phi ^{(\alpha)}$ fields.

If we specialise these relations to quiver gauge singlets, we obtain the general constraint:

\begin{equation}
\label{hwhs6.4}
\left( {{X_{12}}} \right)_a^{~i}\left( {{X_{21}}} \right)_i^{~a} = {\varepsilon _{\alpha \beta }}\left( {{\phi ^{\left( \alpha  \right)}}} \right)_a^{~c}\left( {{\phi ^{\left( \beta  \right)}}} \right)_c^{~a} = 0.
\end{equation}

We are interested in finding symmetrised combinations of the fields $\{ {{\phi ^{\left( a \right)}},{X_{12}},{X_{21}}}\}$ that are also singlets of the $U(k)$ gauge group. We can generate such combinations of the fields using the PE followed by Weyl integration, however, the fields are also subject to the relations \ref{hwhs6.3}. When we construct Hilbert series for the various theories we need to incorporate the appropriate F-term constraints into our generating functions to exclude any disallowed combinations and to avoid over-counting.

Following Plethystic Program methodology, we start by identifying the irreps of the fields with their characters, which transform in the given product groups. Thus:

\begin{equation}
\label{hwhs6.5}
\begin{aligned}
{\cal{X}}({{\phi^{(\alpha)}}})&={[{1,0\ldots ,0,1}]_{SU(k)}}{[1]_{SU(2)}}+{[1]_{SU(2)}},\\
{\cal{X}}({{X_{12}}})&={[{1,0\ldots ,0,0}]_{SU(k)}}{[{0,0\ldots ,1}]_{SU(N)}}{(1)_{U(1)}},\\
{\cal{X}}({{X_{21}}})&={[{0,0\ldots ,0,1}]_{SU(k)}}{[{1,0\ldots ,0}]_{SU(N)}}{(1)_{U(1)}}.
\end{aligned}
\end{equation}
We can choose to incorporate F-term constraints at this point by subtracting the PE of the character of field combinations made redundant by the relations \ref{hwhs6.3}.

Next we extract the quiver gauge group singlets. This projection onto singlets is carried out by Weyl integration over both the $U(1)$ and $SU(k)$ quiver gauge groups. The first integration imposes the constraint that the overall $U(1)$ quiver gauge charge is zero and so the X fields within the generating functions always consist of pairs of $SU(k)$ fundamentals and anti-fundamentals, and, similarly, of pairs of $SU(N)$ fundamentals and anti-fundamentals.

HWGs can be calculated by projecting these  generating functions onto the irreps of the $SU(N)$ Yang-Mills symmetry and the global $SU(2)$ symmetry groups. The HWGs faithfully encode the structure of the representation space and this can facilitate the application of F-term constraints.

Instanton moduli spaces invariably contain a component generated by the fundamental of the global $SU(2)$ symmetry. This component represents the position of the instanton on the on $\mathbb{C}^2$ manifold. For multiple instanton theories, this corresponds to the centre of mass. Instanton moduli spaces can be presented in reduced form by taking a quotient of the full moduli space by this $SU(2)$ symmetry. This can lead to significant simplifications in the HWGs for the moduli spaces, as will be shown.

The Hilbert series of an instanton moduli space can also be summarised in an unrefined form by replacing the characters of the $SU(N)$ and global $SU(2)$ symmetry groups by their dimensions.

We set out in the following sections the results of such calculations for one $SU(3)$ instanton on $\mathbb{C}^2$  and also for two and three $SU(2)$ instantons on $\mathbb{C}^2$.


\subsection{Moduli Space of One $SU(3)$ Instanton}
Using Table \ref{table21} and noting that, for one instanton, the adjoint of $U(k)$ becomes the adjoint of $U(1)$, we obtain the following characters in terms of the CSA coordinates $\left\{ {x,{x_1},{x_2},w} \right\}$:
\begin{equation}
\label{hwhs6.6}
\begin{aligned}
{{\cal X}}({\phi ^{(\alpha )}}) &= 1/x + x,\\
{{\cal X}}({X_{12}}) &= \left( {{x_1} + {x_2} + 1/{x_1}{x_2}} \right)w,\\
{{\cal X}}({X_{21}}) &= \left( {1/{x_1} + 1/{x_2} + {x_1}{x_2}} \right){w^{ - 1}}.\\
\end{aligned}
\end{equation}
In addition to symmetrising the various fields using the PE, we need to correct for the relation \ref{hwhs6.3}. Since the commutator of two $\phi$ fields transforming in the Abelian $U(1)$ gauge group is zero, this relation entails that $SU(3)$ singlets composed of pairs of $X$ fields should vanish and so the resulting F-term constraint is that ${t^2}m_1^0m_2^0 = {t^2} \to 0$. Singlets of the Yang-Mills gauge group composed of pairs of $X$ fields are therefore excluded. We can achieve this by taking a {\it hyper-Kahler quotient}, which incorporates the redundancy of the $t^2$ $SU(3)$ singlets into the PE before the projection onto the quiver gauge singlets as:
\begin{equation}
\label{hwhs6.12}
g^{1,SU( 3 )}(t,x,x_1,x_2) \equiv \oint\limits_{U( 1 )} {d\mu }
~PE[{\cal X}( \phi^{(\alpha)}) t + {\cal X}(X_{12}) t + {\cal X}(X_{21}) t - \underbrace {{\cal X}([0,0;0] )}_1{t^2} ].
\end{equation}
The hyper-Kahler quotient follows from the last term of \ref{hwhs6.12} (with negative sign) and the Weyl integral evaluates as:
\begin{equation}
\begin{aligned}
g^{1,SU( 3)}(t,x,x_1,x_2) &= (1-t^2)^2((1+2t^2+2t^4+2t^6+t^8)[0,0]-t^4 [1,1]) \\
& ~~~~\times  ~PE[[1,1]~t^2]~PE[ [1]~t],
\end{aligned}
\label{hwhs6.8}
\end{equation}
where we are using character notation for the class functions of the $SU(3)$ Yang-Mills gauge and $SU(2)$ global symmetry groups, for brevity.

To obtain a HWG we use the generating functions for the characters of $SU(2)$ and $SU(3)$ as per \ref{eq:hwhs2.3} and Table \ref{table2}, using Dynkin label fugacities $\{m_1, m_2 \}$ for the Yang-Mills symmetry and $m$ for the global symmetry. We then apply Weyl integration over the $SU(3)$ and $SU(2)$ symmetry groups to obtain the generating function:
\begin{equation}
\label{hwhs6.10}
\begin{aligned}
HWG^{1,SU( 3 )}(t,m,m_1,m_2) & \equiv \oint\limits_{SU( 2 )} \oint\limits_{SU( 3 )} {d\mu }  ~(1-m_1 m_2 ) PE[ {{\cal X}^*}[1,0]m_1 +{{\cal X}^*}[0,1]m_2 ]\\
&~~~~~~~~~~\times PE[ {\cal X}[1]m ]~g^{1,SU( 3 )}(t,x,x_1,x_2)\\
& = \frac{1}{{( {1 - {t^2}{m_1}{m_2}} )(1 - {t}m)}}.
\end{aligned}
\end{equation}
The HWG analysis explicates the basic objects that generate the moduli space. We can identify the combinations of the fields within the HWG \ref{hwhs6.10}   as shown in Table \ref{table22}.
\begin{table}[htdp]
\caption{Basic Objects of HWG for One $SU(3)$ Instanton Moduli Space}
\begin{center}
\begin{tabular}{|c|c|c|}
\hline
$ SU(3)_{YM};SU(2)_{global}$ & $ {HWG~Terms} $&$ {Basic~Objects} $\\
\hline
$ {[ {1,1;0} ]}$ & $ {{t^2}{m_1}{m_2}}$ & $ {SU( 3 )~adjoint~from~{X_{12}}~and~{X_{21}}} $\\
\hline
$ {[ {0,0;1} ]}$ & $ {{t}m}$ & $ {SU( 2 )~fundamental~from~{\phi^{(\alpha)}}} $\\
\hline
\end{tabular}
\end{center}
\label{table22}
\end{table}

The Dynkin label fugacities $m_1m_2$ always appear paired and so this is an example of the general result \cite{Benvenuti:2010pq} that, for one instanton, the resulting tensor products between $SU(N)$ particles and antiparticles always transform in a real representation that is a symmetrisation $[n,0,\ldots ,0,n]$ of the $SU(N)$ adjoint representation. We can also see that the $t^2$ singlets have been excluded, as intended, by the hyper-Kahler quotient.

The physical interpretation of the HWG is that the term $t m$ enumerates the representations of the global $SU(2)$ symmetry that describe the position of the instanton on $\mathbb{C}^2$ , while the ${t^2} {m_1}{m_2}$ term enumerates the holomorphic operators in the reduced moduli space. Thus the reduced moduli space for one $SU(3)$ instanton is given by the one dimensional HWG:
\begin{equation}
\label{hwhs6.13}
HWG_{reduced}^{1,SU( 3 )}(t,m_1,m_2)= PE[m_1 m_2 t^2].
\end{equation}
We can unrefine the series given by the HWG \ref{hwhs6.10}, by Taylor expansion followed by replacement of the monomials in the Dynkin label fugacities by the dimensions of the corresponding irreducible representations using the schema:
\begin{equation}
\label{hwhs6.13a}
m_1^{{n_1}}m_2^{{n_2}}{m^n} \to Dim[ {{n_1},{n_2}} ]Dim[ n ]=\frac{1}{2}({n_1} + 1)({n_2} + 1)({n_1} + {n_2} + 2)\left( {n + 1} \right).
\end{equation}
The resulting unrefined Hilbert series matches that given by \cite{Benvenuti:2010pq}:
\begin{equation}
\label{hwhs6.14}
HS_{unrefined}^{1,SU( 3 )}( t ) = \frac{{1 + 4{t^2} + {t^4}}}{{{{( {1 - t} )}^2}{{( {1 - {t^2}} )}^4}}}.
\end{equation}


\subsection{Moduli Space of Two $SU(2)$ Instantons}
The analysis for multiple instantons is complicated by the enlarged gauge group symmetry; the characters combine the three separate non-Abelian product groups; quiver gauge $U(k=2)$, Yang-Mills $SU(N=2)$ and the global $SU(2)$, in addition to their $U(1)$ charges. We use the CSA coordinates $\{ {w,x,{x_1},{y_1}} \}$ and apply Table \ref{table21} to obtain the following characters for the various fields:
\begin{equation}
\label{hwhs6.15}
\begin{aligned}
{{\cal X}}({\phi ^{(\alpha )}}) &= \left( {y_1^2 + 1 + \frac{1}{{y_1^2}}} \right)\left( {x + \frac{1}{x}} \right) + \left( {x + \frac{1}{x}} \right),\\
{{\cal X}}({X_{12}}) &= \left( {{y_1} + \frac{1}{{{y_1}}}} \right)\left( {{x_1} + \frac{1}{{{x_1}}}} \right)w,\\
{{\cal X}}({X_{21}}) &= \left( {{y_1} + \frac{1}{{{y_1}}}} \right)\left( {{x_1} + \frac{1}{{{x_1}}}} \right){w^{ - 1}}.
\end{aligned}
\end{equation}
In addition to symmetrising the fields using the PE,
we need to incorporate the F-term constraints that follow from \ref{hwhs6.3} and \ref{hwhs6.4}, corresponding to the tensor relations:
\begin{equation}
\label{hwhs6.17}
( {{X_{12}}} )_a^{~i}( {{X_{21}}} )_i^{~b} = {\varepsilon _{\alpha \beta }}( {{\phi ^{( \alpha  )}}} )_a^{~c}( {{\phi ^{( \beta  )}}} )_c^{~b},
\end{equation}
and
\begin{equation}
\label{hwhs6.18}
( {{X_{12}}} )_a^{~i}( {{X_{21}}} )_i^{~a} = {\varepsilon _{\alpha \beta }}( {{\phi ^{( \alpha  )}}} )_a^{~c}( {{\phi ^{( \beta  )}}} )_c^{~a} = 0.
\end{equation}
Firstly, \ref{hwhs6.18} entails that singlets under all three symmetries (quiver gauge, Yang-Mills and global), which are composed just of pairs of $\phi$  fields (or pairs of $X$ fields), should vanish. Indeed, since the only Casimir of $SU(2)$ is of degree 2, it follows that all singlets formed from chains of odd numbers of pairs of $\phi$ fields (or $X$ fields) should vanish. Thus:
\begin{equation}
\label{hwhs6.19}
\underbrace {( {{X_{12}}} )_a^{~i}( {{X_{21}}} )_i^{~b} \ldots  \ldots ( {{X_{12}}} )_c^{~k}( {{X_{21}}} )_k^{~a}}_{odd~number~of~{X_{12}}{X_{21}}~pairs} = 0.
\end{equation}
This F-term constraint corresponds to adding the following term to the generating function:
\begin{equation}
\label{hwhs6.20}
PE[ { - [ {0;0;0} ]{t^2}} ] = PE[ { - {t^2}} ].
\end{equation}
Secondly, \ref{hwhs6.17} entails that simple symmetrisation of the characters \ref{hwhs6.15} over-counts combinations of pairs of $\phi$ fields (or pairs of $X$ fields) transforming in the adjoint of the $SU(2)$ quiver gauge group. The relation does not apply generally, but only between pairs that are Yang-Mills gauge singlets and global singlets. We can correct for this duplication by incorporating a generating function term containing the character for the adjoint of the $SU(2)$ quiver gauge group:
\begin{equation}
\label{hwhs6.21}
PE[ - [2;0;0 ]t^2 ] = PE[  - ( y_1^2 + 1 + \frac{1}{{y_1^2}} )t^2].
\end{equation}
We combine the characters  \ref{hwhs6.15} and the F-term constraints  \ref{hwhs6.20} and  \ref{hwhs6.21} into a hyper-Kahler quotient and then apply Weyl integration, first over the $U(1)$ and then over the $SU(2)$ of the quiver gauge group to obtain the generating function:
\begin{equation}
\label{hwhs6.23}
g^{2,SU( 2 )}(t,x,x_1) \equiv \oint\limits_{SU( 2 )} \oint\limits_{U( 1 )} {d\mu }~PE[{\cal X}(\phi^{(\alpha)}) t + {\cal X}(X_{12}) t + {\cal X}(X_{21}) t - t^2 - {\cal X}[2;0;0] t^2 ].
\end{equation}
We evaluate \ref{hwhs6.23} and then rearrange the resulting functions of the $\{x,x_1 \}$ CSA coordinates into characters, with Dynkin labels ordered as [Yang-Mills; global], to give the refined HS generating function: 
\begin{equation}
\label{hwhs6.24a}
\begin{aligned}
g^{2,SU(2)}\left( {t,{\cal X}} \right){\rm{ }} & = \left( {1 - {t^2}} \right)\left( {\begin{array}{*{20}{r}}
{\left( {1 + {t^2} - {t^{14}} - {t^{16}}} \right)[0;0]}\\
{ + {t^3}\left( {1 - {t^{10}}} \right)[0;1]}\\
{ - {t^6}\left( {1 - {t^4}} \right)[2;0]}\\
{ - {t^5}\left( {1 - {t^6}} \right)[2;1]}
\end{array}} \right) \\
& \times PE\left[ {[0;1]t + [2;0]{t^2} + [0;2]{t^2} + [2;1]{t^3} - [0;1]{t^3}} \right].
\end{aligned}
\end{equation}
We now introduce the Dynkin label fugacities $m_1$ and $m$ to track the Yang-Mills $SU(2)$ and global $SU(2)$ irreps respectively and use Weyl integration to project the generating function \ref{hwhs6.24a} onto the irreps of the $SU(2)$ groups to give the HWG:
\begin{equation}
\label{hwhs6.25}
{\small
HWG_{}^{2,SU( 2 )}( {t,m,{m_1}} ) \equiv \oint\limits_{SU( 2 )} {\oint\limits_{SU( 2 )} {d\mu } } ~PE[ [1]{m_1}+[1]_{global}m]g^{2,SU( 2 )}( {t,\cal X} )
}.
\end{equation}
Evaluation yields the result:
\begin{equation}
\label{hwhs6.26}
{\small
HWG_{}^{2,SU(2)}\left( {t,m,{m_1}} \right) = \frac{{Num(t,m,{m_1})}}{{\left( {1 - mt} \right)\left( {1 - {m^2}{t^2}} \right)\left( {1 - m_1^2{t^2}} \right)\left( {1 - mm_1^2{t^3}} \right){{\left( {1 - {t^4}} \right)}^2}{{\left( {1 - m_1^2{t^4}} \right)}}}},
}
\end{equation}
where
\begin{equation}
\label{hwhs6.27}
{\small
Num\left( {t,m,{m_1}} \right) = 1 + m{t^3} + mm_1^2{t^5} + m_1^2{t^6} - {m^2}m_1^2{t^6} - mm_1^2{t^7} - mm_1^4{t^9} - {m^2}m_1^4{t^{12}}
}.
\end{equation}
Interestingly, the polynomial $Num(t,m,m_1)$  is anti-palindromic of degree (12,2,4) in the variables $(t,m,m_1)$. We can identify combinations of the fields giving rise to the HWG denominator terms as shown in Table \ref{table23}.

\begin{table}[htdp]
\small
\caption{Basic Objects of HWG for Two $SU(2)$ Instanton Moduli Space}
\begin{center}
\begin{tabular}{|c|c|c|}
\hline
$SU(2)_{YM}; SU(2)_{global} $&$ HWG~Terms $& $Basic~Objects$\\
\hline
$ [0;0]$ & $ t^4$ & $ Singlets~from~\phi^{(\alpha)}$ \\
\hline
$[0;1]$ & $m t$ & $ Global~SU( 2 )~fundamental~from~\phi^{(\alpha)} $\\
\hline
$[0;2]$ & $m^2 t^2$ & $ Global~SU( 2 )~adjoint~from~\phi^{(\alpha)}$ \\
\hline
$[2;0]$ & $ m_1^2{t^2}$ & $ Yang-Mills~SU( 2 )~adjoint~from~X_{12}~and~X_{21}$\\
$[2;0]$ & $ m_1^2t^4$ & $Yang-Mills~SU( 2 )~adjoint~from~\phi^{(\alpha)},X_{12}~and~X_{21}$\\
\hline
$[2;1]$ & $m m_1^2{t^3}$ & 
$\begin{array}{c}Yang-Mills~SU(2)~adjoint~\&~global~SU(2)~fund.\\
from~\phi ^{(a)}, X_{12}~and~ X_{21}\end{array}$
\\
\hline
\end{tabular}
\end{center}
\label{table23}
\end{table}
We read off the exponents of the fugacities $\{m,m_1\}$, which give the Dynkin labels of the global $SU(2)$ and Yang-Mills $SU(2)$ representations respectively, and identify the building blocks of the theory according to the irreps in which they transform. In particular, while the global $SU(2)$ irreps include the fundamental, the only Yang-Mills $SU(2)$ irreps that occur are the adjoint and symmetrisations of the adjoint. We can identify the term $m m_1^2{t^3}$, which is responsible for the mixing that occurs between the global and Yang-Mills symmetries in multiple instanton theories. $Num( {t,m,{m_1}} )$ also contains the monomial term $mt^3$, which is not just a product of the other building blocks. Some of these basic objects can be identified from the PE in \ref{hwhs6.24a}, however, the HWG gives a complete enumeration.

We can verify that the addition of the terms \ref{hwhs6.20} and \ref{hwhs6.21} has had the desired effect of excluding Yang-Mills gauge singlets formed from pairs of X fields from the Hilbert series. Thus, if we specialise the series \ref{hwhs6.26} to Yang-Mills singlets, by setting $m_1$ to zero, we obtain:
\begin{equation}
\label{hwhs6.28}
HWG^{2,SU( 2 )}(t,m,0) = \frac{{1 + mt^3}}{{{{( {1 - t^4} )}^2}( {1 - m{t}} )( {1 - {m^2}t^2} )}}.
\end{equation}
This series does not contain any $t^2$ monomials, confirming that all Yang-Mills singlet pairs of $X$ or $\phi$ fields have been excluded as intended by applying the F-term constraint terms to the Hilbert series. Also, the only global and Yang-Mills singlets are at orders of $t^4$, confirming also that the singlets only contain even numbers of pairs of $X$ or $\phi$ fields.

Returning to the HWG for the two $SU(2)$ instanton moduli space given by \ref{hwhs6.26} and \ref{hwhs6.27}, we can see that the global symmetry only appears in its generating monomials as $m$ or $m^2$, corresponding to the [1] and [2] global irreps after Taylor expansion. This appears to be part of a more general pattern, where the global symmetry appears amongst the HWG generating monomials at orders up to $m^k$, where k is the instanton number and equals the maximum degree of the Casimirs of the $U(k)$ quiver gauge group.

Importantly, the [1] irrep of the global $SU(2)$ arises from the $PE[[0;1]t]$ term within the refined HS generating function  \ref{hwhs6.24a} and we can simplify the HWG considerably by factoring out the tensor products that result from this term. This gives us the {\it reduced} moduli space. Physically, the [1] irrep of the global $SU(2)$ corresponds to the centre of mass of a system of instantons and so working with the reduced moduli space corresponds to an analysis in the instanton rest frame.

If we reduce the HS generating function \ref{hwhs6.24a} by taking a quotient by this PE term, but otherwise proceed as before, the HWG \ref{hwhs6.26} simplifies to the reduced two $SU(2)$ instanton moduli space:
\begin{equation}
\label{hwhs6.29a}
HWG_{reduced}^{2,SU(2)}\left( {t,m,{m_1}} \right) = \frac{{1 + m m_1^2{t^5}}}{{\left( {1 - {m^2}{t^2}} \right)\left( {1 - m_1^2{t^2}} \right)\left( {1 - m m_1^2{t^3}} \right)\left( {1 - {t^4}} \right)}}.
\end{equation}
This result is equivalent to a series expansion of the HWG presented in \cite{Hanany:2012dm}, but uses the more concise HWG notation, which clarifies the structure of the basic objects in the moduli space. Unlike the HWG \ref{hwhs6.26} for the full moduli space, the HWG \ref{hwhs6.29a} for the reduced moduli space constitutes a complete intersection. We can simplify an expansion of \ref{hwhs6.29a} to give an unrefined series by replacing monomial terms in the $\{ m,m_1 \}$ fugacities by the corresponding irrep dimensions using the mapping:
\begin{equation}
\label{hwhs6.29}
{m^n}{m_1}^{{n_1}} \to Dim[ n ]Dim[n_1]=(n+1)(n_1+1).
\end{equation}
We obtain:
\begin{equation}
\label{hwhs6.30}
HS_{reduced,unrefined}^{2,SU(2)}(t) = \frac{{\left( {1 - t} \right)\left( {1 + t + 3{t^2} + 6{t^3} + 8{t^4} + 6{t^5} + 8{t^6} + 6{t^7} + 3{t^8} + {t^9} + {t^{10}}} \right)}}{{{{\left( {1 - {t^2}} \right)}^4}{{\left( {1 - {t^3}} \right)}^3}}}.
\end{equation}
The numerator of the generating function \ref{hwhs6.30} is a palindromic polynomial, and the result matches that given in \cite{Hanany:2012dm}.

\subsection{Moduli Space of Three $SU(2)$ Instantons}
The presence of three instantons gives rise to a $U(3)$ quiver gauge symmetry, and the field characters combine three separate non-Abelian product groups: quiver gauge $SU(k=3)$, Yang-Mills $SU(N=2)$ and the global $SU(2)$, in addition to the local and global $U(1)$ symmetries. We express the characters using the CSA coordinates $\{ {{y_1},{y_2},x,{x_1},w}\}$  in accordance with Table \ref{table21}:
\begin{equation}
\label{hwhs6.31}
\begin{aligned}
{{\cal X}}( {{\phi^{(\alpha)}}} ) &= ( {{y_2}y_1^2 + y_2^2{y_1} + \frac{{{y_1}}}{{{y_2}}} + \frac{1}{{y_1^2{y_2}}} + \frac{1}{{{y_1}y_2^2}} + \frac{{{y_2}}}{{{y_1}}} + 2} )( {x + \frac{1}{x}} ) + ( {x + \frac{1}{x}} ),\\
{{\cal X}}( {{X_{12}}} ) &= ( {{y_1} + {y_2} + \frac{1}{{{y_1}{y_2}}}} )w( {{x_1} + \frac{1}{{{x_1}}}} ),\\
{{\cal X}}( {{X_{21}}} ) &= ( {{y_1}{y_2} + \frac{1}{{{y_1}}} + \frac{1}{{{y_2}}}} ){w^{ - 1}}( {{x_1} + \frac{1}{{{x_1}}}} ).\\
\end{aligned}
\end{equation}
We also need to identify the F-term constraints that follow from \ref{hwhs6.3} and \ref{hwhs6.4}. As before, \ref{hwhs6.4} entails that singlets under all three symmetries (quiver gauge, Yang-Mills and global), which are composed just of pairs of $\phi$ fields (or pairs of $X$ fields), should vanish, and this constraint corresponds to the generating function term:
\begin{equation}
\label{hwhs6.34}
PE[ - [ {0,0;0;0} ]t^2] = PE[-t^2].
\end{equation}
Secondly, \ref{hwhs6.3} entails that to avoid duplication of pairs of fields we should also incorporate a generating function term to eliminate symmetrised characters from the adjoint of $SU(3)$:
\begin{equation}
\label{hwhs6.35}
PE[ { - [ {1,1;0;0} ]{t^2}} ] = PE[ { - ( {{y_2}y_1^2 + y_2^2{y_1} + \frac{{{y_1}}}{{{y_2}}} + \frac{1}{{y_1^2{y_2}}} + \frac{1}{{{y_1}y_2^2}} + \frac{{{y_2}}}{{{y_1}}} + 2} ){t^2}} ].
\end{equation}
Combining the PEs of the characters \ref{hwhs6.31} with the hyper-Kahler quotient terms \ref{hwhs6.34} and \ref{hwhs6.35}, we obtain a refined generating function by Weyl integration over the $\{ {w,{y_1},{y_2}} \}$ CSA coordinates:
\begin{equation}
\label{hwhs6.36}
g^{3,SU(2)}(t,x,x_1) = \oint\limits_{SU(3)} \oint\limits_{U(1)} {d\mu} ~PE[ {\cal X}(\phi^{(\alpha)}) t + {\cal X}(X_{12}) t + {\cal X}(X_{21})t - t^2 - [1,1;0;0] t^2].
\end{equation}
We shall not give the explicit evaluation of the generating function in this form since it is extremely unwieldy. Instead we simplify the analysis by taking a quotient by the global symmetry and working with the reduced three instanton moduli space:

\begin{equation}
\label{hwhs6.36a}
g_{reduced}^{3,SU(2)}(t,x,{x_1}) = {g^{3,SU(2)}}(t,x,{x_1})~PE[ - [0,0;0;1] t].
\end{equation}

As before, we introduce character generating functions and the Dynkin label fugacities $m_1$ and $m$ to track the Yang-Mills $SU(2)$ and global $SU(2)$ irreps respectively and apply Weyl integration over the remaining $\{ {{x_1},{x_2}} \}$ CSA coordinates to obtain an HWG:
\begin{equation}
\label{hwhs6.37}
HWG_{reduced}^{3,SU(2)}({m_1},m,t) \equiv \oint\limits_{SU(2)} {\oint\limits_{SU(2)} {d\mu } }~PE\left[ {{{[1]}_{YM}}{m_1} + {{[1]}_{global}}m} \right]g_{reduced}^{3,SU(2)}(t,x,{x_1}).
\end{equation}
Evaluation yields:
\begin{equation}
\label{hwhs6.38}
HWG_{reduced}^{3,SU(2)}({m_1},m,t) = \frac{{Num( {{m_1},m,t} )}}{{Den( {{m_1},m,t} )}},
\end{equation}
where the denominator is given by:
\begin{equation}
\label{hwhs6.39}
\begin{aligned}
Den( {{m_1},m,t} ) &=
{\left( {1 - {t^4}} \right)^2}\left( {1 - {t^6}} \right)\left( {1 - {t^8}} \right)\\
&~~~ \times \left( {1 - {m^2}{t^2}} \right)\left( {1 - m{t^3}} \right)\left( {1 - {m^3}{t^3}} \right)\\
&~~~ \times \left( {1 - m_1^2{t^2}} \right)\left( {1 - m_1^2{t^4}} \right)\left( {1 - m_1^6{t^{10}}} \right)\\
&~~~ \times \left( {1 - mm_1^2{t^3}} \right)\left( {1 - {m^2}m_1^2{t^4}} \right),\\
\end{aligned}
\end{equation}
and the numerator $Num( {{m_1},m,t})$ consists of unity followed by 248 monomial terms, being palindromic of degree (12,7,43) in the variables $({m_1},m,t)$:
\begin{equation}
\label{hwhs6.39a}
{\tiny
\left(
 \begin{array}{l}
1 - m{t^3} + m{t^5} + {m^3}{t^5} - {t^6} + {m^2}{t^6} + m{t^7} - {m^4}{t^8} + m{t^9} - {m^3}{t^{11}} + {t^{12}} - {m^3}{t^{13}} - {m^2}{t^{14}} + {m^4}{t^{14}} - m{t^{15}} - {m^3}{t^{15}} + \\
{m^3}{t^{17}} - {m^4}{t^{20}} - m_1^2{t^4} + 2mm_1^2{t^5} + m_1^2{t^6} + 2{m^2}m_1^2{t^6} + 2mm_1^2{t^7} + {m^3}m_1^2{t^7} + m_1^2{t^8} - {m^2}m_1^2{t^8} - {m^4}m_1^2{t^8} - \\
mm_1^2{t^9} - 3{m^3}m_1^2{t^9} - {m^5}m_1^2{t^9} + 2m_1^2{t^{10}} - {m^2}m_1^2{t^{10}} - {m^4}m_1^2{t^{10}} - 2mm_1^2{t^{11}} - {m^3}m_1^2{t^{11}} + {m^5}m_1^2{t^{11}} - 2{m^2}m_1^2{t^{12}} + {m^4}m_1^2{t^{12}} + {m^6}m_1^2{t^{12}} - \\
2mm_1^2{t^{13}} - 2{m^3}m_1^2{t^{13}} - m_1^2{t^{14}} - {m^2}m_1^2{t^{14}} - mm_1^2{t^{15}} + {m^3}m_1^2{t^{15}} + {m^5}m_1^2{t^{15}} - m_1^2{t^{16}} - {m^2}m_1^2{t^{16}} + mm_1^2{t^{17}} + 2{m^3}m_1^2{t^{17}} + \\
2{m^2}m_1^2{t^{18}} - {m^6}m_1^2{t^{18}} + mm_1^2{t^{19}} + {m^3}m_1^2{t^{19}} + {m^5}m_1^2{t^{19}} - {m^4}m_1^2{t^{20}} - {m^3}m_1^2{t^{21}} - {m^5}m_1^2{t^{21}} + {m^4}m_1^2{t^{22}} + {m^5}m_1^2{t^{23}} + {m^4}m_1^2{t^{24}} + {m^6}m_1^2{t^{24}} + \\
mm_1^4{t^7} + m_1^4{t^8} - 2{m^3}m_1^4{t^9} - 3{m^2}m_1^4{t^{10}} - {m^4}m_1^4{t^{10}} - 3{m^3}m_1^4{t^{11}} + {m^4}m_1^4{t^{12}} + {m^6}m_1^4{t^{12}} - mm_1^4{t^{13}} + {m^3}m_1^4{t^{13}} + 2{m^5}m_1^4{t^{13}} - \\
2{m^2}m_1^4{t^{14}} - mm_1^4{t^{15}} + {m^3}m_1^4{t^{15}} - {m^5}m_1^4{t^{15}} - {m^7}m_1^4{t^{15}} + {m^2}m_1^4{t^{16}} - {m^6}m_1^4{t^{16}} - mm_1^4{t^{17}} + 2{m^3}m_1^4{t^{17}} + 2{m^5}m_1^4{t^{17}} + \\
{m^2}m_1^4{t^{18}} + {m^3}m_1^4{t^{19}} - m_1^4{t^{20}} + {m^2}m_1^4{t^{20}} + {m^4}m_1^4{t^{20}} - {m^3}m_1^4{t^{21}} + {m^7}m_1^4{t^{21}} + {m^6}m_1^4{t^{22}} + mm_1^4{t^{23}} + {m^3}m_1^4{t^{23}} + \\
2{m^4}m_1^4{t^{24}} + {m^6}m_1^4{t^{24}} - {m^4}m_1^4{t^{26}} - {m^6}m_1^4{t^{26}} - {m^5}m_1^4{t^{27}} - {m^7}m_1^4{t^{27}} - {m^6}m_1^4{t^{28}} - mm_1^6{t^{11}} + m_1^6{t^{12}} - {m^2}m_1^6{t^{12}} + \\
{m^3}m_1^6{t^{13}} + m_1^6{t^{14}} + {m^4}m_1^6{t^{14}} - mm_1^6{t^{15}} - {m^3}m_1^6{t^{15}} + {m^5}m_1^6{t^{15}} - 2{m^2}m_1^6{t^{16}} - {m^4}m_1^6{t^{16}} - {m^6}m_1^6{t^{16}} - {m^3}m_1^6{t^{17}} - \\
2m_1^6{t^{18}} + 2{m^4}m_1^6{t^{18}} + mm_1^6{t^{19}} - {m^3}m_1^6{t^{19}} + 2{m^5}m_1^6{t^{19}} + {m^7}m_1^6{t^{19}} - m_1^6{t^{20}} + {m^4}m_1^6{t^{20}} + mm_1^6{t^{21}} + {m^3}m_1^6{t^{21}} - {m^7}m_1^6{t^{21}} - \\
m_1^6{t^{22}} + {m^4}m_1^6{t^{22}} + {m^6}m_1^6{t^{22}} + {m^3}m_1^6{t^{23}} - {m^7}m_1^6{t^{23}} + m_1^6{t^{24}} + 2{m^2}m_1^6{t^{24}} - {m^4}m_1^6{t^{24}} + {m^6}m_1^6{t^{24}} + 2{m^3}m_1^6{t^{25}} - 2{m^7}m_1^6{t^{25}} - {m^4}m_1^6{t^{26}} - \\
mm_1^6{t^{27}} - {m^3}m_1^6{t^{27}} - 2{m^5}m_1^6{t^{27}} + {m^2}m_1^6{t^{28}} - {m^4}m_1^6{t^{28}} - {m^6}m_1^6{t^{28}} + {m^3}m_1^6{t^{29}} + {m^7}m_1^6{t^{29}} + {m^4}m_1^6{t^{30}} - {m^5}m_1^6{t^{31}} + {m^7}m_1^6{t^{31}} - \\
{m^6}m_1^6{t^{32}} - mm_1^8{t^{15}} - m_1^8{t^{16}} - {m^2}m_1^8{t^{16}} - mm_1^8{t^{17}} - {m^3}m_1^8{t^{17}} + mm_1^8{t^{19}} + 2{m^3}m_1^8{t^{19}} + {m^4}m_1^8{t^{20}} + {m^6}m_1^8{t^{20}} + mm_1^8{t^{21}} + \\
m_1^8{t^{22}} - {m^4}m_1^8{t^{22}} + {m^3}m_1^8{t^{23}} + {m^5}m_1^8{t^{23}} - {m^7}m_1^8{t^{23}} + {m^4}m_1^8{t^{24}} + {m^5}m_1^8{t^{25}} + 2{m^2}m_1^8{t^{26}} + 2{m^4}m_1^8{t^{26}} - {m^6}m_1^8{t^{26}} - \\
mm_1^8{t^{27}} + {m^5}m_1^8{t^{27}} - m_1^8{t^{28}} - {m^2}m_1^8{t^{28}} + {m^4}m_1^8{t^{28}} - {m^6}m_1^8{t^{28}} - 2{m^5}m_1^8{t^{29}} + 2{m^2}m_1^8{t^{30}} + {m^4}m_1^8{t^{30}} - {m^6}m_1^8{t^{30}} + mm_1^8{t^{31}} + {m^3}m_1^8{t^{31}} - \\
3{m^4}m_1^8{t^{32}} - {m^3}m_1^8{t^{33}} - 3{m^5}m_1^8{t^{33}} - 2{m^4}m_1^8{t^{34}} + {m^7}m_1^8{t^{35}} + {m^6}m_1^8{t^{36}} + mm_1^{10}{t^{19}} + {m^3}m_1^{10}{t^{19}} + {m^2}m_1^{10}{t^{20}} + {m^3}m_1^{10}{t^{21}} - \\
{m^2}m_1^{10}{t^{22}} - {m^4}m_1^{10}{t^{22}} - {m^3}m_1^{10}{t^{23}} + {m^2}m_1^{10}{t^{24}} + {m^4}m_1^{10}{t^{24}} + {m^6}m_1^{10}{t^{24}} - mm_1^{10}{t^{25}} + 2{m^5}m_1^{10}{t^{25}} + 2{m^4}m_1^{10}{t^{26}} + {m^6}m_1^{10}{t^{26}} - \\
{m^5}m_1^{10}{t^{27}} - {m^7}m_1^{10}{t^{27}} + {m^2}m_1^{10}{t^{28}} + {m^4}m_1^{10}{t^{28}} - {m^6}m_1^{10}{t^{28}} - {m^5}m_1^{10}{t^{29}} - {m^7}m_1^{10}{t^{29}} - 2{m^4}m_1^{10}{t^{30}} - 2{m^6}m_1^{10}{t^{30}} + \\
mm_1^{10}{t^{31}} + {m^3}m_1^{10}{t^{31}} - 2{m^5}m_1^{10}{t^{31}} + {m^2}m_1^{10}{t^{32}} - {m^4}m_1^{10}{t^{32}} - 2{m^6}m_1^{10}{t^{32}} - {m^3}m_1^{10}{t^{33}} - {m^5}m_1^{10}{t^{33}} + 2{m^7}m_1^{10}{t^{33}} - \\
{m^2}m_1^{10}{t^{34}} - 3{m^4}m_1^{10}{t^{34}} - {m^6}m_1^{10}{t^{34}} - {m^3}m_1^{10}{t^{35}} - {m^5}m_1^{10}{t^{35}} + {m^7}m_1^{10}{t^{35}} + {m^4}m_1^{10}{t^{36}} + 2{m^6}m_1^{10}{t^{36}} + 2{m^5}m_1^{10}{t^{37}} + {m^7}m_1^{10}{t^{37}} + \\
2{m^6}m_1^{10}{t^{38}} - {m^7}m_1^{10}{t^{39}} - {m^3}m_1^{12}{t^{23}} + {m^4}m_1^{12}{t^{26}} - {m^4}m_1^{12}{t^{28}} - {m^6}m_1^{12}{t^{28}} + {m^3}m_1^{12}{t^{29}} - {m^5}m_1^{12}{t^{29}} - {m^4}m_1^{12}{t^{30}} + {m^7}m_1^{12}{t^{31}} - \\
{m^4}m_1^{12}{t^{32}} + {m^6}m_1^{12}{t^{34}} - {m^3}m_1^{12}{t^{35}} + {m^6}m_1^{12}{t^{36}} + {m^5}m_1^{12}{t^{37}} - {m^7}m_1^{12}{t^{37}} + {m^4}m_1^{12}{t^{38}} + {m^6}m_1^{12}{t^{38}} - {m^6}m_1^{12}{t^{40}} + {m^7}m_1^{12}{t^{43}}
\end{array}
 \right)
 .
}
\end{equation}

We identify in Table \ref{table24} the combinations of the fields in the denominator that contribute to the HWG.
\begin{table}[htdp]
\caption{Basic Objects of HWG for Three $SU(2)$ Instanton Reduced Moduli Space}
\begin{center}
\begin{tabular}{|c|c|c|}
\hline
$SU(2)_{YM}; SU(2)_{global}$ & ${HWG}~Terms$ & $Basic~Objects$\\
\hline
$ {[ {0;0} ]}$ & $ {t^4,t^6,t^8}$ & $
 Singlets~from~\phi^{(\alpha)}
 $\\
\hline
$ {[ {0;1} ]}$ & $ {mt^3}$ & $Global~SU(2)$\\
$ {[ {0;2} ]}$ & $ {m^2}t^2$ & $irreps$\\
$ {[ {0;3} ]}$ & $ {m^3}{t^3}$ & $from~\phi ^{(\alpha )}$\\
\hline
$ {[ {2;1} ]}$ & $ {m_1^2m{t^3}}$ & $
SU(2)~adjoint ~\& ~global~SU( 2 )~irreps
$\\
$ {[ {2;2} ]}$ & $ {m_1^2{m^2}{t^4}}$ & $
from~\phi^{(\alpha)},X_{12}~and~X_{21}
$\\
\hline
$ {[ {2;0} ]}$ & $ {m_1^2{t^2}}$ & $ {SU( 2 )~adjoint~from~{X_{12}}~and~{X_{21}}} $\\
$ {[ {2;0} ]}$ & $ {m_1^2{t^4}}$ & $ {SU( 2 )~adjoint~from~{\phi^{(\alpha)}},{X_{12}}~and~{X_{21}}} $\\
\hline
$ {[ {6;0} ]}$ & $ {m_1^6 t^{10}}$ & $ SU(2)~symmetrised~adjoint~from~\phi^{(\alpha)},X_{12}~and~X_{21} $\\
\hline
\end{tabular}
\end{center}
\label{table24}
\end{table}
We read off the exponents of the Dynkin label fugacities $m_1$ and $m$, which label the Yang-Mills $SU(2)$ and global $SU(2)$ representations respectively, and identify the building blocks of the theory according to the irreps in which they transform. In particular, while the global $SU(2)$ irreps include the fundamental, the only Yang-Mills $SU(2)$ irreps that occur in the PL are the adjoint and symmetrisations of the adjoint. The objects do not include Yang-Mills singlets comprised of $X$ fields, so we can verify that the F-term constraints have been implemented as intended by the terms \ref{hwhs6.34} and \ref{hwhs6.35}.

As before, we can also simplify the HWG into an unrefined version by either (a) setting the Yang-Mills gauge and global $SU(2)$ CSA coordinates in \ref{hwhs6.36a} to unity or (b) replacing monomial terms in the $m$ and $m_1$ Dynkin label fugacities in a Taylor expansion of \ref{hwhs6.38} by the corresponding irrep dimensions using the mapping:
\begin{equation}
\label{hwhs6.41}
{m^n}{m_1}^{{n_1}} \to Dim[ n ]Dim[n_1]=(n+1)(n_1+1).
\end{equation}
In either case we obtain the quotient of palindromic polynomials \footnote{This is consistent with the series obtained by using instanton counting methods set out in \cite{Nakajima:2003pg}.}:
\begin{equation}
\label{hwhs6.42}
HS_{reduced, unrefined}^{3,SU(2)}(t) = \frac{{\left( \begin{array}{l}
1 + 3{t^2} + 6{t^3} + 12{t^4} + 16{t^5} + 31{t^6} + 36{t^7} + 55{t^8} + 54{t^9} + 60{t^{10}}\\
 + 54{t^{11}} + 55{t^{12}} + 36{t^{13}} + 31{t^{14}} + 16{t^{15}} + 12{t^{16}} + 6{t^{17}} + 3{t^{18}} + {t^{20}}
\end{array} \right)}}{{{{(1 - {t^2})}^3}{{(1 - {t^3})}^4}{{(1 - {t^4})}^3}}}.
\end{equation}

\section{Discussion and Conclusions}

We have shown how the HWG methodology takes the often complicated plethystic class functions of refined Hilbert series and transforms them into corresponding generating functions for the coefficients of the irreps in the series, identified by their Dynkin label fugacities. Both refined and unrefined Hilbert series can be recovered in a straightforward manner from these HWGs.  It is instructive, therefore, to compare the properties of HWGs with those of HS\footnote{It is worth commenting that both palindromic properties and the use of unrefined Hilbert series can be extremely useful as a check on the overall counting of dimensions during the calculation of HWGs.}. The geometric properties of refined Hilbert series largely follow from those of unrefined Hilbert series and so we do not comment on them separately.

We use PLs for the purpose of comparison since these encode the geometric nature of the series most concisely. Tables \ref{table25} through  \ref{table27} set out the PLs of the HWG and unrefined HS for the colour singlets of the various SQCD theories examined herein.

\begin{table}[htdp]
\small
\caption{PLs of HWGs and Hilbert Series of $SU(N_f)_L \times SU(N_f)_R \times SU(N_c)$ colour singlets}
\begin{center}
\begin{tabular}{|c|c|c|}
\hline
$Theory$ &$ {PL[ {HWG} ]}$& $PL[Unrefined~HS] $\\
\hline
 ${SU(2)_f \times SU(2)_c}$ & $ {{t^2}}$ &$ {{t^2}} $ \\
  ${SU(4)_f \times SU(2)_c}$ & $ l_2{t^2}$ & $ 6 {t^2}-{t^4} $\\
  ${SU(2N_f \ge 6)_f \times SU(2)_c}$ & $ l_2{t^2}$ & $ {infinite~series} $\\
  \hline
 ${SU(2)_{f,L} \times SU(2)_{f,R} \times SU(2)_c}$ & $ {{t_1}^2 + {t_2}^2 + {l}{r}{t_1}{t_2}}$ & $ {{t_1}^2 + {t_2}^2 + 4{t_1}{t_2} - t_1^2t_2^2} $\\
 ${SU({N_f \ge 3} )_L \times SU( { \ge 3} )_R \times SU(2)_c}$ & $ {{l_2}{t_1}^2 + {r_2}{t_2}^2 + {l_1}{r_1}{t_1}{t_2}}$ & $ {infinite~series} $\\
\hline
 ${SU(2)_{f,L}  \times SU(2)_{f,R} \times SU(3)_c}$ & $ {{l}{r}{t_1}{t_2} + {t_1}^2{t_2}^2}$ & $ {4{t_1}{t_2}}$ \\
 ${SU(3)_{f,L} \times SU(3)_{f,R} \times SU(3)_c}$ & $ {{t_1}^3 + {t_2}^3 + {l_1}{r_1}{t_1}{t_2} + {l_2}{r_2}{t_1}^2{t_2}^2}$ & $ {9{t_1}{t_2} + {t_1}^3 + {t_2}^3 - {t_1}^3{t_2}^3} $\\
$ {SU( {N_f \ge 4} )_L \times SU( {N_f \ge 4} )_R \times SU(3)_c}$ & $ {{l_3}{t_1}^3 + {r_3}{t_2}^3 + {l_1}{r_1}{t_1}{t_2} + {l_2}{r_2}{t_1}^2{t_2}^2}$ & $ {infinite~series} $\\
\hline
\end{tabular}
\end{center}
\label{table25}
\end{table}

\begin{table}[htdp]
\caption{PLs of HWGs and Hilbert Series of $SU(N_f) \times SO(N_c)$ colour singlets}
\begin{center}
\begin{tabular}{|c|c|c|}
\hline
$Theory$ &$ {PL[ {HWG} ]}$& $PL[Unrefined~HS] $\\
\hline
 ${SU(2) \times SO(3)}$ & $ {{t^2}{m^2} + {t^4}}$ & $ {3{t^2}}$ \\
 ${SU(3) \times SO(3)}$ & $ {{t^2}{m_1}^2 + {t^3} + {t^4}{m_2}^2}$ & $ {6{t^2} + {t^3} - {t^6}} $\\
 ${SU( \ge 4) \times SO(3)}$ & $ {{t^2}{m_1}^2 + {t^3}{m_3} + {t^4}{m_2}^2}$ & $ {infinite~series} $\\
\hline
 ${SU(2) \times SO(4)}$ & $ {{t^2}{m^2} + {t^4}}$ & $ {3{t^2}}$ \\
 ${SU(3) \times SO(4)}$ & $ {{t^2}{m_1}^2 + {t^4}{m_2}^2 + {t^6}}$ & $ {6{t^2}} $\\
 ${SU(4) \times SO(4)}$ & $ {{t^2}{m_1}^2 + {t^4}{m_2}^2 + {t^6}m_3^2 + {t^4}}$ & $ {10{t^2} + {t^4} - {t^8}} $\\
 ${SU( \ge 5) \times SO(4)}$ & $ {{t^2}{m_1}^2 + {t^4}{m_2}^2 + {t^4}{m_4} + {t^6}m_3^2}$ & $ {infinite~series} $\\
\hline
 ${SU(2) \times SO(5)}$ & $ {{t^2}{m^2} + {t^4}}$ & $ {3{t^2}}$ \\
 ${SU(3) \times SO(5)}$ & $ {{t^2}{m_1}^2 + {t^4}{m_2}^2 + {t^6}}$ & $ {6{t^2}} $\\
 ${SU(4) \times SO(5)}$ & $ {{t^2}{m_1}^2 + {t^4}{m_2}^2 + {t^6}m_3^2 + {t^8}}$ & $ {10{t^2}} $\\
 ${SU(5) \times SO(5)}$ & $ {{t^2}{m_1}^2 + {t^4}{m_2}^2 + {t^5} + {t^6}m_3^2 + {t^8}m_4^2}$ & $ {15{t^2} + {t^5} - {t^{10}}} $\\
 ${SU( \ge 6) \times SO(5)}$ & $ {{t^2}{m_1}^2 + {t^4}{m_2}^2 + {t^5}{m_5} + {t^6}m_3^2 + {t^8}m_4^2}$ & $ {infinite~series} $\\
\hline   
\end{tabular}
\end{center}
\label{table26}
\end{table}

 \begin{table}[htdp]
\caption{PLs of HWGs and Hilbert Series of $SU(N_f) \times USp(2n_c)$ colour singlets}
\begin{center}
\begin{tabular}{|c|c|c|}
\hline
$Theory$&${PL[{HWG}]}$&${PL[{Unrefined~Hilbert~Series}]}$\\
\hline
${SU(2)\times{USp(2)}}$&${{t^2}}$&${{t^2}}$\\
${SU(3)\times{USp(2)}}$&${{t^2}{m_2}}$&${3{t^2}}$\\
${SU(4)\times{USp(2)}}$&${{t^2}{m_2}}$&${6{t^2}-{t^4}}$\\
${SU(\ge5)\times{USp(2)}}$&${{t^2}{m_2}}$&${infinite~series}$\\
\hline
${SU(2)\times{USp(4)}}$&${{t^2}}$&${{t^2}}$\\
${SU(3)\times{USp(4)}}$&${{t^2}{m_2}}$&${3{t^2}}$\\
${SU(4)\times{USp(4)}}$&${{t^2}{m_2}+{t^4}}$&${6{t^2}}$\\
${SU(5)\times{USp(4)}}$&${{t^2}{m_2}+{t^4}{m_4}}$&${10{t^2}}$\\
${SU(6)\times{USp(4)}}$&${{t^2}{m_2}+{t^4}{m_4}}$&${15{t^2}-{t^6}}$\\
${SU(\ge7)\times{USp(4)}}$&${{t^2}{m_2}+{t^4}{m_4}}$&${infinite~series}$\\
\hline
\end{tabular}
\end{center}
\label{table27}
\end{table}

The PLs of the unrefined Hilbert series for SQCD theories correspond to established results \cite{Gray:2008yu, Hanany:2008kn}. The PLs of the HWGs differ from the unrefined Hilbert series for all but the simplest series and explicate the structure of the GIOs of the theory. Thus, for example, the PL ${t^2}{m_2} + {t^4}$ in Table \ref{table27} indicates that the highest weight basis for all GIO colour singlets formed from quarks transforming in a $SU(4) \times {USp(4)}$  product group consists of a contraction of two quarks transforming in the [0, 2, 0] irrep of the $SU(4)$ flavour group and a contraction of 4 quarks transforming as a [0, 0, 0] flavour singlet. This HWG has all positive terms in the PL and is a \textit{freely generated} series. Negative terms in a PL indicate that a series is a \textit{complete intersection} or the quotient of two freely generated series. When a PL does not terminate, this shows that the GIOs of the theory cannot be reduced to symmetrisations of a finite basis set of GIOs, or their quotients, and the series is termed a \textit{non-complete intersection} \cite{Gray:2008yu}. In the case of SQCD, Tables \ref{table25} through \ref{table27} illustrate how the HWGs can have moduli spaces that are freely generated or complete intersections, when the unrefined HS do not have finite PLs. 

As noted in Section 3, providing one adopts a labelling system that reflects group symmetries, the HWG generating functions are identical for all the $SU(N)$ flavour groups once their fundamental dimension exceeds the defining space dimension of the colour group. This arises because the antisymmetrisations of the fundamental of the flavour group generated by the PE are limited by the length of the colour group epsilon tensor. The differences in the unrefined Hilbert series as the rank of the flavour group is increased are simply due to the different dimensions of the flavour group irreps. Thus, we can reason that the HWGs of SQCD are the same for all $SU(N)$ flavour groups of fundamental dimension exceeding that of the colour group. This important insight makes it possible to calculate the Hilbert series for GIOs of SQCD theories with high rank flavour groups, for which a direct calculation using Weyl integration might not be feasible. The result corresponds to observations within \cite{Gray:2008yu, Hanany:2008kn}.

Following similar tensorial reasoning, it is possible to relate all the HWGs for classical SQCD to the primitive invariant tensors of the classical groups, as set out in Table \ref{tableX}, with each HWG monomial corresponding to a contraction of delta and epsilon tensors with the numbers of quarks identified by the fugacities. Using this knowledge of the structure of the primitive invariant tensors, we can write down expressions for the HWGs describing the SQCD theory for a given classical product group as in Table \ref{table28a} .

\begin{table}[htdp]
\caption{Generalised PLs of HWGs for SQCD with Classical Colour Groups}
\begin{center}
\begin{tabular}{|c|c|}
\hline
 $ {Theory} $ & $ {PL\left[ {HWG} \right]} $ \\
\hline
 $ {SU\left( {{N_f}} \right) \times SU\left( {{N_c}} \right)} $ 
 &
  $ {{m_{N_c}}{t^{N_c}}} $ \\
\hline
$ {SU{{\left( {{N_f}} \right)}_L} \times SU{{\left( {{N_f}} \right)}_R} \times SU\left( {{N_c}} \right)} $ 
& 
${\sum\limits_{i = 1}^{\min ({\rm{N_f}},{\rm{N_c}}) - 1} {{l_i}} {r_i}{\rm{t}}_1^i{\rm{t}}_2^i + \left\{ {\begin{array}{*{20}{c}}
{if ~{N_f} < {N_c}:}&{t_1^{{N_f}}t_2^{{N_f}}}\\
{if ~{N_f} = {N_c}:}&{t_1^{N_c} + t_2^{N_c}}\\
{if ~{N_f} > {N_c}:}&{l_{N_c}^{}t_1^{N_c} + r_{N_c}^{}t_2^{N_c}}
\end{array}} \right.}$\\
\hline
 $ {SU\left( {{N_f}} \right) \times SO\left( {{N_c}} \right)} $ 
 &
 ${\sum\limits_{i = 1}^{\min ({\rm{N_f}},{\rm{N_c}}) - 1} {m_i^2} {t^{2i}} + \left\{ {\begin{array}{*{20}{c}}
{if~{N_f} < {N_c}:}&{{t^{2N_f}}}\\
{if~{N_f} = {N_c}:}&{{t^{N_f}}}\\
{if~{N_f} > {N_c}:}&{m_{N_c}^{}{t^{N_c}}}
\end{array}} \right.}$\\
\hline
 ${SU\left( {{N_f}} \right) \times USp\left( {{N_c}} \right)} $ 
 &
${\sum\limits_{i = 1}^{\min \left( {\left\lfloor {\frac{{{\rm{N_f}} - 1}}{2}} \right\rfloor ,\frac{{{\rm{N_c}}}}{2}} \right)} {{m_{2i}}} {t^{2i}} + \left\{ {\begin{array}{*{20}{c}}
{if~{N_f} \le {N_c}~\&~{N_f}~is~even:}&{{t^{N_f}}}\\
{if~{N_f} > {N_c}:}&{m_{N_c}^{}{t^{N_c}}}
\end{array}} \right.}$\\
\hline
\end{tabular}
\end{center}
\label{table28a}
\end{table}
\FloatBarrier

The HWGs typically contain a different number of generators and relations to the HS. The generators can be identified by PL terms with positive coefficients, while the relations between the generators are given by PL terms with negative coefficients. The overall dimension of a moduli space is given by the number of generators less relations \cite{Gray:2008yu}. We summarise the HWG and HS descriptions of the moduli spaces of a selection of classical SQCD theories in Table \ref{table28c}. The dimensions are obtained by summing the PLs in Tables \ref{table25} through \ref{table27} with all coordinate fugacities set to unity, or, in the case of the non-terminating PLs of unrefined Hilbert series, by summing the orders of the poles calculated in Section 3.

\begin{table}[htdp]
\caption{Dimensions of Moduli Spaces of Classical SQCD Theories}
\begin{center}
\begin{tabular}{|c|c|c|c|c|}
\hline
$ Theory $ 
&
$\begin{array}{c}HWG\\Dimension\\(a) \end{array}$
&
$\begin{array}{c}HWG\\Irrep\end{array}$
&
$\begin{array}{c}HWG\\Irrep\\Degree\\(b)\end{array}$
&
$\begin{array}{c}Moduli~Space\\Dimension\\(HS)\\(a)+(b)\end{array}$ \\
\hline
$ {SU(2) \times SU(2)} $ & $ 1 $ & ${[0]} $ & $0 $ & $ 1$ \\
$ {SU(4) \times SU(2)} $ & $ 1 $ & ${[{0,n,0}]} $ & $4 $ & $ 5$ \\
$ {SU(6) \times SU(2)} $ & $ 1 $ & ${[{0,n,0,0,0}]} $ & $8 $ & $ 9$ \\
$ {SU(8) \times SU(2)} $ & $ 1 $ & ${[{0,n,0,0,0,0,0}]} $ & ${12} $ & $ {13}$ \\
$ {SU({10}) \times SU(2)} $ & $ 1 $ & ${[{0,n,0,0,0,0,0,0,0}]} $ & ${16} $ & $ {17}$ \\
\hline
$ {SU(2) \times SU(2) \times SU(3)} $ & $ 2 $ & ${[n][n]} $ & $2 $ & $ 4$ \\
$ {SU(3) \times SU(3) \times SU(3)} $ & $ 4 $ & ${[n_1,n_2][n_1,n_2]} $ & $6 $ & $ {10}$ \\
$ {SU(4) \times SU(4) \times SU(3)} $ & $ 4 $ & ${[{{n_1},{n_2},{n_3}}][{{n_1},{n_2},{n_4}}]} $ & ${12} $ & $ {16}$ \\
\hline
$ {SU(2) \times SO(3)} $ & $ 2 $ & ${[2n]} $ & $1 $ & $ 3$ \\
$ {SU(3) \times SO(3)} $ & $ 3 $ & ${[{{2n_1},{2n_2}}]} $ & $3 $ & $ 6$ \\
$ {SU(4) \times SO(3)} $ & $ 3 $ & ${[{{2n_1},{2n_2},{n_3}}]} $ & $6 $ & $ 9$ \\
\hline
$ {SU(2) \times SO(4)} $ & $ 2 $ & ${[2n]} $ & $1 $ & $ 3$ \\
$ {SU(3) \times SO(4)} $ & $ 3 $ & ${[{{2n_1},{2n_2}}]} $ & $3 $ & $ 6$ \\
$ {SU(4) \times SO(4)} $ & $ 4 $ & ${[{{2n_1},{2n_2},{2n_3}}]} $ & $6 $ & $ {10}$ \\
$ {SU(5) \times SO(4)} $ & $ 4 $ & ${[{{2n_1},{2n_2},{2n_3},{n_4}}]} $ & ${10} $ & $ {14}$ \\
\hline
$ {SU(2) \times SO(5)} $ & $ 2 $ & ${[2n]} $ & $1 $ & $ 3$ \\
$ {SU(3) \times SO(5)} $ & $ 3 $ & ${[{{2n_1},{2n_2}}]} $ & $3 $ & $ 6$ \\
$ {SU(4) \times SO(5)} $ & $ 4 $ & ${[{{2n_1},{2n_2},{2n_3}}]} $ & $6 $ & $ {10}$ \\
$ {SU(5) \times SO(5)} $ & $ 5 $ & ${[{{2n_1},{2n_2},{2n_3},{2n_4}}]} $ & ${10} $ & $ {15}$ \\
$ {SU(6) \times SO(5)} $ & $ 5 $ & ${[{{2n_1},{2n_2},{2n_3},{2n_4},{n_5}}]} $ & ${15} $ & $ {20}$ \\
\hline
$ {SU(2) \times USp(2)} $ & $ 1 $ & ${[0]} $ & $0 $ & $ 1$ \\
$ {SU(3) \times USp(2)} $ & $ 1 $ & ${[{0,n}]} $ & $2 $ & $ 3$ \\
$ {SU(4) \times USp(2)} $ & $ 1 $ & ${[{0,n,0}]} $ & $4 $ & $ 5$ \\
$ {SU(5) \times USp(2)} $ & $ 1 $ & ${[{0,n,0,0}]} $ & $6 $ & $ 7$ \\
\hline
$ {SU(2) \times USp(4)} $ & $ 1 $ & ${[0]} $ & $0 $ & $ 1$ \\
$ {SU(3) \times USp(4)} $ & $ 1 $ & ${[{0,n}]} $ & $2 $ & $ 3$ \\
$ {SU(4) \times USp(4)} $ & $ 2 $ & ${[{0,n,0}]} $ & $4 $ & $ 6$ \\
$ {SU(5) \times USp(4)} $ & $ 2 $ & ${[{0,{n_1},0,{n_2}}]} $ & $8 $ & $ {10}$ \\
$ {SU(6) \times USp(4)} $ & $ 2 $ & ${[{0,{n_1},0,{n_2},0}]} $ & ${12} $ & $ {14}$ \\
$ {SU(7) \times USp(4)} $ & $ 2 $ & ${[{0,{n_1},0,{n_2},0,0}]} $ & ${16} $ & $ {18}$ \\
\hline

\end{tabular}
\end{center}
\label{table28c}
\end{table}

As can be seen, the Hilbert series typically describe moduli spaces with a higher dimension than those of the HWGs. We can develop a systematic account of the relationship between the dimensions of a Hilbert series and those of its underlying HWG by expanding the latter and analysing the dimensional structure of its irreps. For example, the expansion for $SU(4)_{f,L} \times SU(4)_{f,R} \times SU(3)_c$ in Table \ref{table25} takes the form:

\begin{equation}
\label{hwhs5.X1}
PE\left[ {{l_3}t_1^3 + {r_3}t_2^3 + {l_1}{r_1}{t_1}{t_2} + {l_2}{r_2}t_1^2t_2^2} \right] 
\Leftrightarrow 
\sum\limits_{{n_1},{n_2},{n_3},{n_4} = 0}^\infty  {\left[ {{n_1},{n_2},{n_3}} \right]\left[ {{n_1},{n_2},{n_4}} \right]} t_1^{{n_1} + 2{n_2} + 3{n_3}}~t_2^{{n_1} + 2{n_2} + 3{n_4}}.
\end{equation}
The Dynkin labels in this HWG series expansion are described by four different parameters  $\{n_1,n_2,n_3,n_4\}$ corresponding to the four generators 
$\{
{l_1}{r_1}{t_1}{t_2}, 
~{l_2}{r_2}t_1^2t_2^2,
~{l_3}t_1^3 , 
~{r_3}t_2^3
 \}$ respectively. These parameters define the sub-lattice of the group spanned by the irreps of the HWG. The dimensions of the irreps are given by a polynomial function of the parameters and the degree of this polynomial accounts for the extra dimensions of the Hilbert series compared to the HWG.
 
We now define the \textit{HWG Irrep Degree} as the total degree of the polynomial that gives the dimensions of the HWG irreps.\footnote{This definition leads to HWG Irrep Degrees that are consistent with the degrees of dimensional polynomials given in \cite{Hanany:2012dm}.} Example \ref{hwhs5.X1} is built from irreps of $SU(4)$, for which the dimension formula is:
 \begin{equation}
\label{hwhs5.X2}
Dim[n_1,n_2,n_3]=\frac{1}{12} (n_1+1) (n_2+1) (n_3+1) (n_1+n_2+2) (n_2+n_3+2) (n_1+n_2+n_3+3).
\end{equation}
The degree of this polynomial is six and thus the HWG Irrep Degree is 12, being the sum of the degrees for the two $SU(4)$ sub-groups. In HWGs where some Dynkin labels are fixed at zero, the HWG Irrep Degree is reduced. The HWG Irrep Degree, as defined, exactly accounts for the difference between the dimensions of the HWG and the Hilbert series. This analysis can be repeated for all the SQCD theories studied and is summarised in Table \ref{table28c}.
\FloatBarrier

Noting the constant nature of the HWG dimension for $N_f \ge N_c$ we can, by inspection, generalise the dimensions of the moduli spaces for large $N_f$ as in Table \ref{table28b}.

\begin{table}[htdp]
\caption{Dimensions of Moduli Spaces of Classical SQCD Theories for $N_f \ge N_c$}
\begin{center}
\begin{tabular}{|c|c|c|c|}
\hline
$ Theory $ 
&
$\begin{array}{c}HWG\\Dimension\\(a) \end{array}$
&
$\begin{array}{c}HWG\\Irrep~Degree\\(b)\end{array}$
&
$\begin{array}{c}Moduli~Space\\Dimension~(HS)\\(a)+(b)\end{array}$ \\
\hline
$ {SU(2N_f) \times SU(2)} $ & $ 1 $  & $4 N_f-4 $ & $ 4 N_f-3$ \\
\hline
$ {SU(N_f) \times SU(N_f) \times SU(N_c)} $ & $ N_c+1 $  & $2 N_f N_c - N_c^2 - N_c $ & $ 2 N_f N_c - N_c^2 + 1$ 
\\\hline
$ {SU(N_f) \times SO(N_c)} $ & $ N_c $ & $ N_f N_c - N_c(N_c+1)/2$ & $N_f N_c - N_c(N_c-1)/2 $ \\
\hline
$ {SU(N_f) \times USp(N_c)} $ & $ N_c/2 $ & $N_f N_c - N_c(N_c+2)/2 $ & $N_f N_c - N_c(N_c+1)/2$ \\
\hline

\end{tabular}
\end{center}
\label{table28b}
\end{table}
In all these cases, for $N_f \ge N_c$ the gauge group is completely broken, and the dimension of the Hilbert series is given by the dimension of the underlying product group representation less the group dimension of the colour group.  This Hilbert series dimension reduces into the dimension of the HWG and the degree of the dimensional polynomial for its irreps.

\begin{table}[htdp]
\caption{PLs of HWGs and Hilbert Series of $SU(N_f) \times G_2$  colour singlets}
\begin{center}
\begin{tabular}{|c|c|c|}
\hline
$Theory$&${PL[{HWG}]}$&${PL[{Unrefined~Hilbert~Series}]}$\\
\hline
${SU(2)\times{G_2}}$&${{m^2}{t^2}+{t^4}}$&${3{t^2}}$\\
\hline
${SU(3)\times{G_2}}$&${{t^3}+{t^6}+m_1^2{t^2}+m_2^2{t^4}}$&${{t^3}+6{t^2}}$\\
\hline
${SU(4)\times{G_2}}$&$\begin{array}{c}
m_1^2{t^2}+{m_3}{t^3}+{t^4}\\
+m_2^2{t^4}+{m_1}{t^5}+m_3^2{t^6}+{m_1}{m_2}{t^7}+{t^8}\\
+{m_2}{m_3}{t^9}+m_2^2{t^{12}}-m_1^2m_2^2{t^{14}}-m_2^2m_3^2{t^{18}}
\end{array}$&${10{t^2}+4{t^3}+{t^4}-{t^8}}$\\
\hline
${SU(5)\times{G_2}}$&${to~be~calculated}$&${infinite~series}$\\
\hline
\end{tabular}
\end{center}
\label{table28}
\end{table}

The Hilbert series of the GIOs of exceptional gauge groups are considerably more complicated than those of classical gauge groups. This complicated structure can be seen from Table  \ref{table28}, which contains some Hilbert series and HWGs for $G_2$. We can identify basic GIOs built on the $G_2$ primitive symmetric invariant tensor of rank two and the $G_2$ primitive antisymmetric invariant tensors of rank three and four, being $m_1^2 t^2$, $m_3 t^3$ and $t^4$ respectively, corresponding to the [2,0,0], [0,0,1] and [0,0,0] irreps of $SU(4)$. The HWG identifies, in addition, the complicated pattern of GIOs in the many other $SU(N)$ irreps that can be formed from combinations of these basic objects, taking account of the relations amongst them. The dimensions of the Hilbert series and the HWGs are related to each other in a similar manner to those of the classical SQCD theories analysed above, as can be seen from Table \ref{table28d}.

\begin{table}[htdp]
\caption{Dimensions of Moduli Spaces of $SU(N_f) \times G_2$  colour SQCD Theories}
\begin{center}
\begin{tabular}{|c|c|c|c|c|}

\hline
$ Theory $ 
&
$\begin{array}{c}HWG\\Dimension\\(a) \end{array}$
&
$\begin{array}{c}HWG\\Irrep\end{array}$
&
$\begin{array}{c}HWG\\Irrep\\Degree\\(b)\end{array}$
&
$\begin{array}{c}Moduli~Space\\Dimension\\(HS)\\(a)+(b)\end{array}$ \\
\hline
$ {SU(2) \times G_2} $ & $ 2 $ & ${[n]} $ & $1$ & $ 3$ \\
$ {SU(3) \times G_2} $ & $ 4 $ & ${[{n_1,n_2}]} $ & $3 $ & $ 7$ \\
$ {SU(4) \times G_2} $ & $ 8 $ & ${[{n_1,n_2,n_3}]} $ & $6 $ & $ 14$ \\
$ {SU(5) \times G_2} $ & $ t.b.c. $ & ${[{n_1,n_2,n_3,n_4}]} $ & $10 $ & $ 21 $ \\
\hline
\end{tabular}
\end{center}
\label{table28d}
\end{table}

It is argued that the Hilbert series of SQCD are always palindromic, which entails that the moduli spaces of the fugacities are Calabi-Yau\cite{Gray:2008yu}. All the Hilbert series, and also all the HWGs for SQCD calculated herein, are palindromic, considering that the numerators of freely generated series and complete intersections are also simple palindromes, and therefore Calabi-Yau. This palindromic property of many generating functions for Hilbert series is shared with the character generating functions discussed in Section 2 that are used to derive the HWGs and Hilbert series. 

An important demonstration from the HWG analysis is that, in all cases, the unrefined HS are reducible to sums of series associated with individual flavour group irreps. This contrasts with the conjecture regarding SQCD \cite{Gray:2008yu} (Observation 3.11) that:
\begin{quote}
Òwe find in all case studies that ${\cal M}{(N_f ,N_c)}$ is irreducible using primary decomposition and conjecture this to hold in general.Ó
\end{quote}
The group theoretic reducibility of the unrefined Hilbert series follows from the facts that the PE and PEF map class functions into other class functions and that these can always be decomposed in terms of characters. This reducibility corresponds to the precise description of the group structures underlying the Hilbert series in terms of the HWG generating functions.

Turning to instanton moduli spaces, we have set out in Section 4, the HWG for one $SU(3)$ instanton on $\mathbb {C}^2$  and for two and three $SU(2)$ instantons on $\mathbb {C}^2$. We have shown that all the operators in the moduli space of three $SU(2)$ instantons on $\mathbb {C}^2$ transform in some symmetrised irrep of the adjoint, with Dynkin labels $[\underbrace {0,0}_{gauge};\underbrace {{2{n_1}}}_{flavour};\underbrace {n}_{global}]$, for non-negative integers $n$ and $n_1$. This is similar to the established result for one instanton Hilbert series \cite{Benvenuti:2010pq}. It results from the initial choice of gauge group as $U(k)$, since this has no epsilon tensor, and its singlets can only be formed from equal numbers of $U(k)$ representations and conjugate $\bar U(k)$ representations.

The HWGs for instanton moduli spaces are generally considerably more complicated than those of SQCD, since they involve symmetrisations of the adjoint in addition to those of basic irreps. This can be seen from Table \ref{table29}.

\begin{table}[htdp]
\caption{PLs of Moduli Spaces of Selected Instanton Theories}
\begin{center}
\begin{tabular}{|c|c|c|c|}
\hline
${Instanton~Theory}$&${PL[{HWG_{reduced}}]}$&${PL[{HWG}]}$&${PL[HS_{unrefined}]}$\\
\hline
${One~SU(2)}$&$ {m_1}^2{t^2}$&${m{t}+{m_1}^2{t^2}}$&${infinite~series}$\\
\hline
${Two~SU(2)}$&
$\begin{array}{c} {m^2}{t^2} + m_1^2{t^2}  + mm_1^2{t^3} + {t^4}\\+ mm_1^2{t^5} - {m^2}m_1^4{t^{10}} \end{array} $
&${infinite~series}$&${infinite~series}$\\
\hline
${Three~SU(2)}$&${infinite~series}$&${infinite~series}$&${infinite~series}$\\
\hline
\hline
${One~SU(3)}$&${m_1}{m_2}{t^2} $&${m{t}+{m_1}{m_2}{t^2}}$&${infinite~series}$\\
\hline
\end{tabular}
\end{center}
\label{table29}
\end{table}
 
While the generating functions for the HWGs and Hilbert series of instanton moduli spaces are all palindromic, only some of these spaces turn out to be freely generated. In particular, the HWGs for single instanton theories are freely generated. We have also shown that the {\it reduced} moduli space for two $SU(N)$ instantons is a complete intersection, although this is the case neither for the higher multiple instanton theories examined, nor for any of the unrefined Hilbert series.

We should mention also, without giving details, that the moduli spaces of multiple instanton theories all contain the moduli spaces of one instanton theories and so an alternative approach to multiple instanton theories is to study the quotient space between the multiple instanton theory and its underlying one instanton theory. Such quotient spaces lend themselves naturally to analysis in terms of HWGs.

Finally, the dimensions of the instanton moduli spaces decompose into the dimension of their HWGs and the degrees of the dimensional polynomial of the HWG irreps in a similar manner to the SQCD theories, as can be seen from Table \ref{table28e}.

\begin{table}[htdp]
\caption{Dimensions of Moduli Spaces of Selected Instanton Theories}
\begin{center}
\begin{tabular}{|c|c|c|c|c|}

\hline
$ Theory $ 
&
$\begin{array}{c}HWG\\Dimension\\(a)\end{array}$
&
$\begin{array}{c}HWG\\Irrep\end{array}$
&
$\begin{array}{c}HWG\\Irrep\\Degree\\(b)\end{array}$
&
$\begin{array}{c}Moduli~Space\\Dimension\\(HS)\\(a)+(b)\end{array}$ \\
\hline
$ {One~SU(2)~Instanton} $ & $ 2 $ & ${[2n_1,n]} $ & $2$ & $4$ \\
$ {Two~SU(2)~Instantons} $ & $ 6$ & ${[2n_1;n]} $ & $2$ & $8$ \\
$ {Three~SU(2)~Instantons} $ & $10$ & ${[2n_1;n]} $ & $2$ & $12$ \\
\hline
$ {One~SU(3)~Instanton} $ & $ 2 $ & ${[n_1,n_1;n]} $ & $4$ & $6$ \\
\hline
\end{tabular}
\end{center}
\label{table28e}
\end{table}

The instanton moduli spaces calculated in Table \ref{table28e} all include a contribution from global $SU(2)$ symmetries. If this contribution is excluded, we obtain reduced instanton moduli spaces, as discussed earlier. The HWGs of these reduced instanton moduli spaces of one instanton theories based on simple Lie groups are all one dimensional. For example, the HWG of the reduced moduli space for one $SU(3)$ instanton is just $m_1 m_2 t^2$. This leads to an elegant decomposition of the moduli spaces of one instanton theories into one dimensional HWGs and corresponding HWG dimensional polynomials, calculated in the same manner as above, and these are set out in Table \ref{tableXYZ}. All the reduced one instanton moduli spaces have a dimension equal to the sum of the Dual Coxeter labels of the group \cite{Benvenuti:2010pq}.

\begin{table}[htdp]
\caption{Dimensional Analysis of Reduced One Instanton Moduli Spaces}
\begin{center}
\begin{tabular}{|c|c|c|c|c|c|c|}
\hline
$ {Series}$&$ \begin{array}{c}Adjoint\\Representation\end{array}$&$ \begin{array}{c}Instanton\\HWG\end{array}$&$\begin{array}{c}Dim\\HWG\end{array}$&$ \begin{array}{c}Degree~of\\Dimensional\\Polynomial\end{array}$&$ \begin{array}{c}Dim\\Instanton
\end{array}$&$ \begin{array}{c}Sum~Dual\\Coxeter\\Labels \end{array}$\\
$ {}$&$ {}$&$ {}$&$ {(a)}$&$ {(b)}$&$ {(a) + (b)}$&$ {\frac{1}{2}\left( {(a) + (b)} \right)}$ \\
\hline
$ {{A_n}}$&$ \begin{array}{l}{A_1}:\left[ 2 \right]\\{A_2}:\left[ {1,1} \right]\\{A_{ \ge 3}}:\left[ {1,0, \ldots ,1} \right]\end{array}$&$ \begin{array}{c}
{m^2}t\\{m_1}{m_2}{t^2}\\{m_1}{m_n}{t^2}\end{array}$&$ \begin{array}{c}1\\1\\1\end{array}$&$ \begin{array}{c}1\\3\\2n - 1\end{array}$&$ \begin{array}{c}2\\4\\2n\end{array}$&$ \begin{array}{c}1\\2\\n\end{array}$ \\
\hline
$ {{B_n}}$&$ \begin{array}{l}
{B_1}:\left[ 2 \right]\\
{B_2}:\left[ {0,2} \right]\\
{B_{ \ge 3}}:\left[ {0,1, \ldots ,0} \right]
\end{array}$&$ \begin{array}{c}
{m^2}t\\
{m_2}^2t\\
{m_2}t
\end{array}$&$ \begin{array}{c}
1\\
1\\
1
\end{array}$&$ \begin{array}{c}
1\\
3\\
4n - 5
\end{array}$&$ \begin{array}{c}
2\\
4\\
4n - 4
\end{array}$&$ \begin{array}{c}
1\\
2\\
2n - 2
\end{array}$ \\
\hline
$ {{C_n}}$&$ \begin{array}{l}
{C_1}:\left[ 2 \right]\\
{C_2}:\left[ {2,0} \right]\\
{C_{ \ge 3}}:\left[ {2,0, \ldots ,0} \right]
\end{array}$&$ \begin{array}{c}
{m^2}t\\
{m_1}^2t\\
m_1^2t
\end{array}$&$ \begin{array}{c}
1\\
1\\
1
\end{array}$&$ \begin{array}{c}
1\\
3\\
2n - 1
\end{array}$&$ \begin{array}{c}
2\\
4\\
2n
\end{array}$&$ \begin{array}{c}
1\\
2\\
n
\end{array}$ \\
\hline
$ {{D_n}}$&$ \begin{array}{l}
{D_2}:\left[ {2,0} \right] \oplus [0,2]\\
{D_3}:\left[ {0,1,1} \right]\\
{D_{ \ge 4}}:\left[ {0,1, \ldots ,0} \right]
\end{array}$&$ \begin{array}{c}
m_1^2t \oplus m_2^2t\\
{m_2}{m_3}{t^2}\\
{m_2}t
\end{array}$&$ \begin{array}{c}
1 \oplus 1\\
1\\
1
\end{array}$&$ \begin{array}{c}
1 \oplus 1\\
5\\
4n - 7
\end{array}$&$ \begin{array}{c}
2 \oplus 2\\
6\\
4n - 6
\end{array}$&$ \begin{array}{c}
1 \oplus 1\\
3\\
2n - 3
\end{array}$ \\
\hline
$ {{E_6}}$&$ {\left[ {0,0,0,0,0,1} \right]}$&$ {{m_6}t}$&$ 1$&$ {21}$&$ {22}$&$ {11}$ \\
$ {{E_7}}$&$ {\left[ {1,0,0,0,0,0,0} \right]}$&$ {{m_1}t}$&$ 1$&$ {33}$&$ {34}$&$ {17}$ \\
$ {{E_8}}$&$ {\left[ {0,0,0,0,0,0,1,0} \right]}$&$ {{m_7}t}$&$ 1$&$ {57}$&$ {58}$&$ {29}$ \\
\hline
$ {{F_4}}$&$ {\left[ {1,0,0,0} \right]}$&$ {{m_1}t}$&$ 1$&$ {15}$&$ {16}$&$ 8$ \\
\hline
$ {{G_2}}$&$ {\left[ {1,0} \right]}$&$ {{m_1}t}$&$ 1$&$ 5$&$ 6$&$ 3$ \\
\hline
\end{tabular}
\end{center}
\label{tableXYZ}
\end{table}

\paragraph{Conclusion}

In conclusion, the HWG approach provides an efficient means of encoding, calculating and decomposing Hilbert series and opens up many avenues for further investigation. These could include explication of the general relationship between the invariant tensors of groups and the structures of GIOs that arise within product groups. Specific theories that could merit further study include SQCD with fields transforming in various representations of classical and/or exceptional gauge groups and multiple instanton theories generally. Finally, it could prove interesting to understand more fully the geometric nature of the HWG manifolds, which all appear to be palindromic and therefore Calabi-Yau in nature, and to relate this more precisely to the geometry of the Hilbert series manifolds.

It should be added that these investigations make extensive use of contour integration and face the challenge of implementing algorithms in \textit{Mathematica} to combine and simplify large numbers of residues within the computing constraints of memory limits and a practical timescale. The development of more effective algorithms would therefore facilitate the extension of the results herein to a wider range of theories built from other group or product group representations.

\FloatBarrier

\paragraph{Acknowledgements}
Rudolph Kalveks expresses his gratitude to Rak-Kyeong Seong  for his guidance around the key concepts of the Plethystics Program and for practical assistance with Latex and its myriad of helper applications and also to Andrew Thomson, Imperial College for many valuable discussions around SQCD.

\FloatBarrier

\section{Appendices}

\subsection{Appendix 1: Plethystic Exponential and Logarithm}
Consider a function in some variable t, which can be expressed as a power series:
\begin{equation}
\label{hwhs8.1}
f(t) \equiv \sum\limits_{n = 0}^\infty  {{a_n}{t^n}}.
\end{equation}
The Plethystic Exponential (ÒPEÓ) for such a function is defined \cite{Gray:2008yu} as :
\begin{equation}
\label{hwhs8.2}
\begin{aligned}
PE[{f(t),t}]&\equiv\exp({\sum\limits_{k=1}^\infty{\frac{{f({{t^k}})-f(0)}}{k}}})\\
&=\prod\limits_{n=1}^\infty{\frac{1}{{{{({1-{t^n}})}^{{a_n}}}}}}.\\
\end{aligned}
\end{equation}
The PE can be generalised for power series of more than one variable, so that for:
\begin{equation}
\label{hwhs8.3}
f({{t_1},\ldots ,{t_N}})\equiv\sum\limits_{n=0}^\infty{\sum\limits_{i=1}^N{{a_{n_i}}t_i^n}},
\end{equation}
we obtain the PE:
\begin{equation}
\label{hwhs8.4}
\begin{aligned}
PE[{f({{t_1},\ldots ,{t_N}}),({{t_1},\ldots ,{t_N}})}]&\equiv\exp({\sum\limits_{k=1}^\infty{\frac{{f({t_1^k,\ldots , t_N^k})-f({0,\ldots ,0})}}{k}}})\\
&=\prod\limits_{n=1}^\infty{\prod\limits_{i=1}^N{\frac{1}{{{{({1-t_i^n})}^{{a_{ni}}}}}}}}.\\
\end{aligned}
\end{equation}
In order to avoid ambiguities, we shall, where necessary, use the notation:
\begin{equation}
\label{hwhs8.5}
PE[{f({{t_1},\ldots ,{t_N}}),({{t_1},\ldots ,{t_N}})}],
\end{equation}
to clarify the variables with respect to which the PE is being taken (and similarly for the PL).

The Plethystic Logarithm (PL) can be used to invert the PE. The PL makes use of the Mobius function $\mu(k)$  , which is defined as $(-1)^n$   for an integer that is the product of n distinct primes other than unity, and zero otherwise, such that $\mu(1)=1, \mu(2)= \mu(3)= -1 ,\ldots$ etc. For the general case, the PL is defined as:
\begin{equation}
\label{hwhs8.6}
PL[{g({{t_1}\ldots ,{t_N}}),({{t_1}\ldots ,{t_N}})}]\equiv\sum\limits_{k=1}^\infty{\frac{1}{k}\mu(k)\log g({{t^k}_1,\ldots ,{t^k}_N})}.
\end{equation}
If we set $g({{t_1}\ldots ,{t_N}})=PE[{f({{t_1}\ldots ,{t_N}})}]$, we then obtain $f({{t_1}\ldots ,{t_N}})=PL[{g({{t_1},\ldots ,{t_N}}),({{t_1},\ldots ,{t_N}})}]$, as required. The identity can be proved by manipulation of the various series using the properties of the Mobius function \cite{Feng:2007ur}, which include the key simplifying identity:
\begin{equation}
\label{hwhs8.7}
\sum\limits_{l=1}^\infty{\sum\limits_{m=1}^\infty{\frac{{\mu(l)}}{{lm}}{t^{klm}}}=}{t^k}.
\end{equation}
The PE, which is a symmetrising function, has a related antisymmetrising function called the Fermionic Plethystic Exponential (ÒPEFÓ). This is defined as:
\begin{equation}
\label{hwhs8.8}
\begin{aligned}
PEF[{f({{t_1},\ldots ,{t_N}}),({{t_1},\ldots ,{t_N}})}]&\equiv\exp({\sum\limits_{k=1}^\infty{{{({-1})}^{k+1}}\frac{{f({t_1^k,\ldots , t_N^k})-f({0,\ldots ,0})}}{k}}})\\
&=\prod\limits_{n=1}^\infty{\prod\limits_{i=1}^N{{{({1+t_i^n})}^{{a_{ni}}}}}}.\\
\end{aligned}
\end{equation}
Like the PE, the PEF also has an inverse function, which we term the Fermionic Plethystic Logarithm (or ``PLF"), given by:
\begin{equation}
\label{hwhs8.8a}
\begin{aligned}
PLF[g({t_1}, \ldots , {t_N}),({t_1}, \ldots , {t_N})] &= \sum\limits_{m = 0}^\infty  PL[g(t_1^{2^m}, \ldots , t_N^{2^m}),( t_1^{2^m}, \ldots , t_N^{2^m} )] \\
&  = \sum\limits_{m = 0}^\infty  \sum\limits_{k = 1}^\infty  {\frac{1}{k}}\mu ( k )\log g(t_1^{(2^m)k} \ldots , t_N^{(2^m)k}). 
\end{aligned}
\end{equation}
The PE and PEF have the useful properties that:
\begin{equation}
\label{hwhs8.9}
\begin{aligned}
PE[{{f_1}+{f_2}}]&=PE[{{f_1}}]PE[{{f_2}}]\\
PEF[{{f_1}+{f_2}}]&=PEF[{{f_1}}]PEF[{{f_2}}]\\
\end{aligned}
\end{equation}
and the PL and PLF have the related properties that:
\begin{equation}
\label{hwhs8.10}
\begin{aligned}
PL[{{g_1}{g_2}}]&=PL[{{g_1}}]+PL[{{g_2}}]\\
PLF[{{g_1}{g_2}}]&=PLF[{{g_1}}]+PLF[{{g_2}}].
\end{aligned}
\end{equation}
All the above results are exact, providing that the series are convergent.

We can use the Plethystic Exponential to symmetrise the character of an irrep of some group G as follows. Suppose the character $\cal X$ of the irrep is composed of monomials ${A_i}({{x_1},\ldots ,{x_r}})$, where the $x_j$ are CSA coordinates ranging over the rank $r$ of the group and the index $i$ ranges over the dimension $Dim(\cal X)$  of the irrep: 
\begin{equation}
\label{hwhs8.12}
{\cal X} =\sum\limits_{i = 1}^{Dim ( {{\cal X}} )} {{A_i}( {{x_1}, \ldots , {x_r}} )}. 
\end{equation}
We form a generating function ${g^G}(t,{\cal X})$ by taking the PE of the sum of fugacities ${f_i} \equiv t{A_i}$, which are given by the products of each coordinate monomial with a fugacity $t$, where $0< |t| <1$:
\begin{equation}
\label{hwhs8.13}
\begin{aligned}
{g^G}({t,{{\cal X}}})&\equiv PE[{{{\cal X}}t}]\\
&\equiv PE[{\sum\limits_{i=1}^{Dim(\cal X)}{{f_i}},({{f_1},\ldots ,{f_{Dim(\cal X)}}})}]\\
&=\prod\limits_{i=1}^{Dim(\cal X)}{\frac{1}{{({1-t{A_i}})}}}.
\end{aligned}
\end{equation}
The Taylor expansion of ${g^G}(t,{\cal X})$ generates an infinite polynomial in the fugacity $t$, whose coefficients are functions of the coordinate monomials. Importantly, the PE of a character is a class function and the Peter Weyl Theorem \cite{Fuchs:1997bb} entails that the characters of a compact group form a complete basis for its class functions, so this Taylor expansion can be decomposed as a sum of characters of irreps, each of which is related to the initial character by symmetrisation, and each of which has a distinct coefficient in the form of a polynomial in the fugacity $t$.
\begin{equation}
\label{hwhs8.14}
PE[{{{\cal X}}t}]=\sum\limits_{irreps}^{}{{g_{irrep}}(t){{{\cal X}}_{irrep}}({{A_i}})}.
\end{equation}
The symmetrising PE is complemented by the PEF, which we can use in a similar manner to antisymmetrise the character of an irrep.
\begin{equation}
\label{hwhs8.15}
\begin{aligned}
g_F^G({t,{{\cal X}}})\equiv PEF[{{{\cal X}}t}]\\
&\equiv PEF[{\sum\limits_{i=1}^{Dim(X)}{{f_i}},({{f_1},\ldots ,{f_{Dim(X)}}})}]\\
&=\prod\limits_{i=1}^{Dim(X)}{({1+t{A_i}})}.
\end{aligned}
\end{equation}
Following similar reasoning to that above, the PEF can also be expanded as a sum of characters:
\begin{equation}
\label{hwhs8.16}
PEF[{{{\cal X}}t}]=\sum\limits_{irreps}^{}{g_{F~irrep}^G(t){{{\cal X}}_{irrep}}({{A_i}})}.
\end{equation}
In this case, the sum of characters is finite.


\subsection{Appendix 2: Weyl Integration}

Weyl integration (also known as Molien-Weyl integration) provides a helpful method for integrating the class functions of a group $G$, such as characters, which depend only upon the identity of a chosen irrep, over the volume of the group. Normally, group integration of a function $f[ \gamma]$, where $\gamma \in G$, requires taking the integral over all the dimensions $Dim(G)$ of the group:
\begin{equation}
\label{hwhs9.1}
I=\int\limits_{G}{d\mu(\gamma)f(\gamma)},
\end{equation}
where $d\mu(\gamma)$ is the Haar measure. In Weyl integration the integral is simplified to one over the maximal torus of the group (as generated by its Cartan subalgebra), by conjugating the class function with other elements of the Group, such that it is always represented by an element of the maximal torus. This is permissible since the integral is effectively a trace over the group. This conjugation reduces the number of integrations required from the dimension of the group to the rank of the group. To do this consistently, the Haar measure, which is effectively a volume element, has to be modified by scaling to reflect the projection of the entire group onto its maximal torus \cite{Fuchs:1997bb, Fulton:2004id}.

Without digressing further into the technical details, we can usefully tabulate the modified Haar measures for $U(r)$ and the classical groups \cite{Hughes:2006ww} as in Table \ref{table30}. In this table, the rank of a group is always labelled by $r$ and $\{ {{\theta _1}, \ldots , {\theta _r}}\}$  are coordinates of periodicity $ 2\pi$ on its maximal torus. The Haar measure is normalised such that it integrates to unity: $\int\limits_{{\theta _i} = 0}^{2\pi } {d\mu  = 1} $.

\begin{table}[htdp]
\small
\caption{Modified Haar Measures on the Maximal Torus}
\begin{center}
\begin{tabular}{|c|c c|}
\hline
 ${Group}$&$Haar~Measure~on~Maximal~Torus$&$~$ \\
\hline
$ {U(r)}$&$ {\frac{1}{{{{( {2\pi } )}^r}r!}} {\int\limits_{{\theta _i} = 0}^{2\pi }}~{{\prod\limits_{1 \le j < k \le r} {| {{e^{i{\theta _j}}} - {e^{i{\theta _k}}}} |} }^2}d{\theta _1} \ldots , d{\theta _r}}$&$~ $\\
\hline
$ {SU(r + 1)}$&$ {\frac{1}{{{{( {2\pi } )}^r}( {r + 1} )!}} {\int\limits_{{\theta _i} = 0}^{2\pi }}~ {{\prod\limits_{1 \le j < k \le r + 1} {| {{e^{i{\theta _j}}} - {e^{i{\theta _k}}}} |} }^2}d{\theta _1} \ldots , d{\theta _r}}$&${where~{\theta _{r + 1}} \equiv  - \sum\limits_{j = 1}^r {{\theta _j}} } $\\
\hline
$ {SO( {2r + 1} )}$&$ {\frac{{{2^{{r^2}}}}}{{{{( {2\pi } )}^r}r!}} {\int\limits_{{\theta _i} = 0}^{2\pi }} ~{{\prod\limits_{1 \le j < k \le r} {( {\cos ( {{\theta _j}} ) - \cos ( {{\theta _k}} )} )} }^2}\prod\limits_{n = 1}^r {{{\sin }^2}( {\frac{1}{2}{\theta _n}} )} d{\theta _1} \ldots , d{\theta _r}}$&$~$ \\
\hline
$ {USp( 2r )}$&$ {\frac{{{2^{{r^2}}}}}{{{{( {2\pi } )}^r}r!}} {\int\limits_{{\theta _i} = 0}^{2\pi }}~ {{\prod\limits_{1 \le j < k \le r} {( {\cos ( {{\theta _j}} ) - \cos ( {{\theta _k}} )} )} }^2}\prod\limits_{n = 1}^r {{{\sin }^2}( {{\theta _n}} )} d{\theta _1} \ldots , d{\theta _r}}$&$~ $\\
\hline
$ {SO( {2r} )}$&$ {\frac{{{2^{(r - 1)^2}}}}{{{{( {2\pi } )}^r}r!}} {\int\limits_{{\theta _i} = 0}^{2\pi }}~ {{\prod\limits_{1 \le j < k \le r} {( {\cos ( {{\theta _j}} ) - \cos ( {{\theta _k}} )} )} }^2}d{\theta _1} \ldots , d{\theta _r}}$&$~ $\\
\hline
\end{tabular}
\end{center}
\label{table30}
\end{table}
 
\begin{table}[htdp]
\caption{Modified Haar Measures on the Maximal Torus: Contour Integrals}
\small
\begin{center}
\begin{tabular}{|c|c c|}
\hline
${Group}$&${Haar~Measure~on~Maximal~Torus}$&$~$\\
\hline
${U(r)}$&${\frac{1}{{{{({2 \pi i})}^r}r!}}\oint \limits_{|{{x_i}}|=1}{\frac{{d{x_i}}}{{{x_i}}}{{\prod \limits_{1\le j<k \le r}{|{{x_j}-{x_k}}|}}^2}}}$&$~$\\
\hline
${SU(r+1)}$&${\frac{1}{{{{({2 \pi i})}^r}({r+1})!}}\prod \limits_{i=1}^r{\oint \limits_{|{{x_i}}|=1}{\frac{{d{x_i}}}{{{x_i}}}}{{\prod \limits_{1 \le j<k \le r+1}{|{{x_j}-{x_k}}|}}^2}}}$&${{x_{r+1}}\equiv \prod \limits_{j=1}^r{\frac{1}{{{x_j}}}}}$\\
\hline
${SO({2r+1})}$&${\frac{1}{{{2^r}{{({2\pi i})}^r}r!}}\prod\limits_{i=1}^r{\oint\limits_{|{{x_i}}|=1}{\frac{{d{x_i}}}{{{x_i}}}}}\prod\limits_{1\le j<k\le r}{\frac{{{{({{x_j}-{x_k}})}^2}{{({1-{x_j}{x_k}})}^2}}}{{{x^2}_j{x^2}_k}}}\prod\limits_{m=1}^r{\frac{{({1-{x_m}})({{x_m}-1})}}{{{x_m}}}}}$&$~$\\
\hline
${USp(2r)}$&${\frac{1}{{{2^r}{{({2\pi i})}^r}r!}}\prod\limits_{i=1}^r{\oint\limits_{|{{x_i}}|=1}{\frac{{d{x_i}}}{{{x_i}}}}}\prod\limits_{1\le j<k\le r}{\frac{{{{({{x_j}-{x_k}})}^2}{{({1-{x_j}{x_k}})}^2}}}{{{x^2}_j{x^2}_k}}}\prod\limits_{m=1}^r{\frac{{({1-{x^2}_m})({{x^2}_m-1})}}{{{x^2}_m}}}}$&$~$\\
\hline
${SO({2r})}$&${\frac{1}{{{2^{r-1}}{{({2\pi i})}^r}r!}}\prod\limits_{i=1}^r{\oint\limits_{|{{x_i}}|=1}{\frac{{d{x_i}}}{{{x_i}}}}}\prod\limits_{1\le j<k\le r}{\frac{{{{({{x_j}-{x_k}})}^2}{{({1-{x_j}{x_k}})}^2}}}{{{x^2}_j{x^2}_k}}}}$&$~$\\
\hline
\end{tabular}
\end{center}
\label{table31}
\end{table}

It can be helpful to express the Weyl integral and Haar measures as unimodular contour integrals. If we make the coordinate substitution:
\begin{equation}
\label{hwhs9.2}
{x_j} = {e^{i{\theta _j}}},
\end{equation}
we can then rewrite the Weyl integrals in Table \ref{table30} as in Table \ref{table31}. This form of the Weyl integral readily lends itself to application of the residue theorem.

Importantly, the characters of irreps are orthonormal under the Weyl integral \cite{Fuchs:1997bb}); if we consider the characters of two irreps of a group $G$ labelled by ${\cal X}_{[A]}$ and ${\cal X}_{[B]}$, their inner product (appropriately normalised) is given by:
\begin{equation}
\label{hwhs9.3}
\int\limits_G {d\mu{{\cal X}}_{[A]}^*} {{{\cal X}}_{[B]}} = {\delta _{AB}}.
\end{equation}
Thus, the Weyl integral can be used to form an inner product that projects out the singlet content of products of characters (or functions of characters) and, as a corollary, the Weyl integral of a single character is zero for any irrep other than the singlet itself.

We can obtain the characters of irreps by the usual methods from Cartan matrices and by using the Freudenthal recursion formula to find the correct multiplicities of weights \cite{Fuchs:1997bb, Slansky:1981yr}. The modified Haar measures do however depend on the CSA coordinate system used. The modified Haar measures in Table \ref{table31} are correct for the characters of defining representations as set out in Table \ref{table32}. If a different choice of weights is used for the coordinate monomials of the defining representations, then the modified Haar measures must be transformed to the new coordinate system to ensure that the orthonormality relations \ref{hwhs9.3} remain satisfied.
\begin{table}[htdp]
\caption{Characters used by Haar Measures}
\begin{center}
\begin{tabular}{|c|c|c|}
\hline
$ \begin{array}{l}Group\\Series\end{array}$&$ \begin{array}{c}Defining\\Representation\end{array}$&$ \begin{array}{c}Defining~Character\\used~by\\Haar~Measure\end{array}$ \\
\hline
$ {{A_r}}$&$ {[ 1, \ldots , 0 ]}$&$ {\sum\limits_{i = 1}^r {{x_i}}  +  \prod\limits_{i = 1}^r {x_i^{ - 1}} } $\\
\hline
 ${{B_r}}$&
 $\begin{array}{*{20}{c}}{[2]}\\
{[1, \ldots ,0]}\end{array}
 \begin{array}{*{20}{c}}{r = 1}\\
{r > 1}
\end{array}$
 &$ {\sum\limits_{i = 1}^r {{x_i}}  + 1 + \sum\limits_{i = 1}^r {x_i^{ - 1}} } $\\
\hline
$ {{C_r}}$&$ {[ {1, \ldots , 0} ]}$&$ {\sum\limits_{i = 1}^r {{x_i}}  + \sum\limits_{i = 1}^r {x_i^{ - 1}} } $\\
\hline
 ${{D_r}}$&
$ \begin{array}{*{20}{c}}
{[1,1]}\\
{[1, \ldots ,0]}
\end{array}\begin{array}{*{20}{c}}
{r = 2}\\
{r > 2}
\end{array}$
 &$ {\sum\limits_{i = 1}^r {{x_i}}  + \sum\limits_{i = 1}^r {x_i^{ - 1}} } $\\
\hline
\end{tabular}
\end{center}
\label{table32}
\end{table}

A simpler form of the Haar measure is noted in \cite{Hanany:2008sb}, which is based on characters whose coordinate monomials carry canonical weights derived from the Cartan matrix and gives the Haar measure in terms of the positive root space only. This generalises to exceptional groups and can produce simpler expressions that can be evaluated more quickly.
\begin{equation}
\label{hwhs9.4}
\oint\limits_G {d\mu= \frac{1}{{{{( {2\pi i} )}^r}}}} \oint\limits_{| {{x_i}} | = 1} {\prod\limits_{i = 1}^r {\frac{{d{x_i}}}{{{x_i}}}} } \prod\limits_{\alpha\in \Phi+ } {( {1 - {A_\alpha }( {{x_1}, \ldots , {x_r}} )} )}.
\end{equation}

\subsection{Appendix 3: Numerator of Generating Function for $G_2$ Characters}

\includegraphics[scale=0.9]{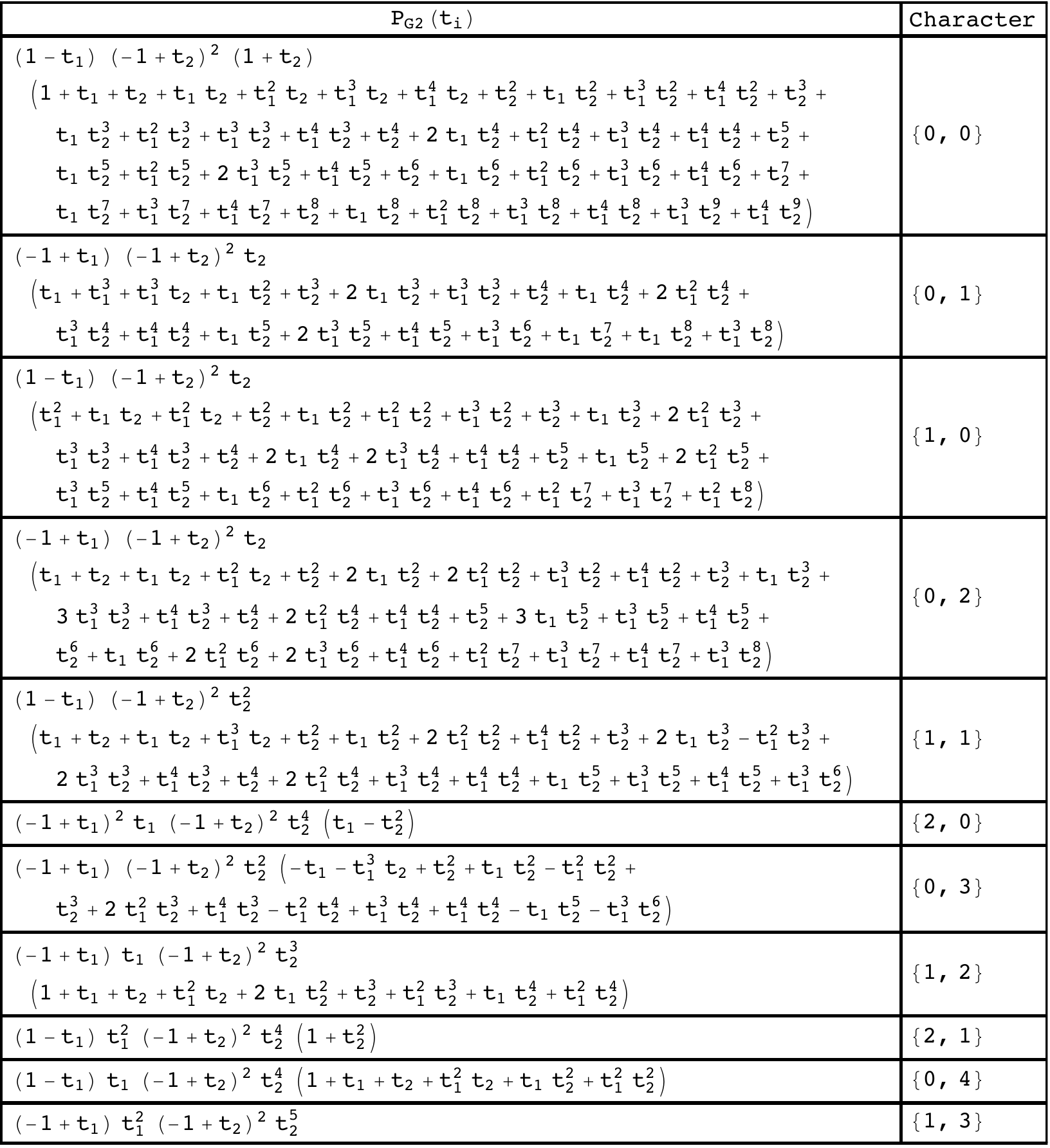}
\clearpage

\subsection{Appendix 4: Numerator of Generating Function for $A_4$ Characters}

\includegraphics[scale=0.85]{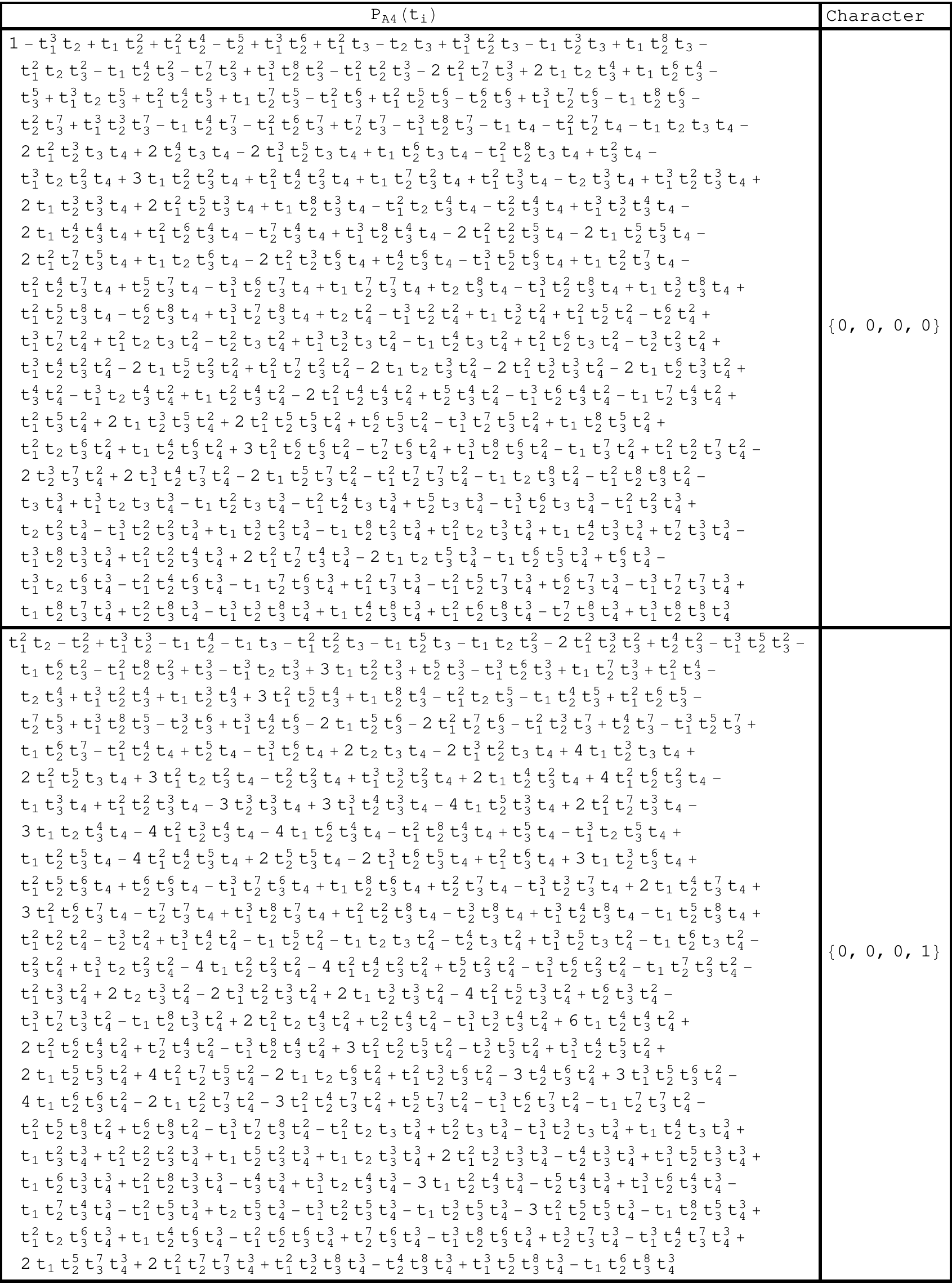}
\clearpage
\pagebreak
\includegraphics[scale=0.85]{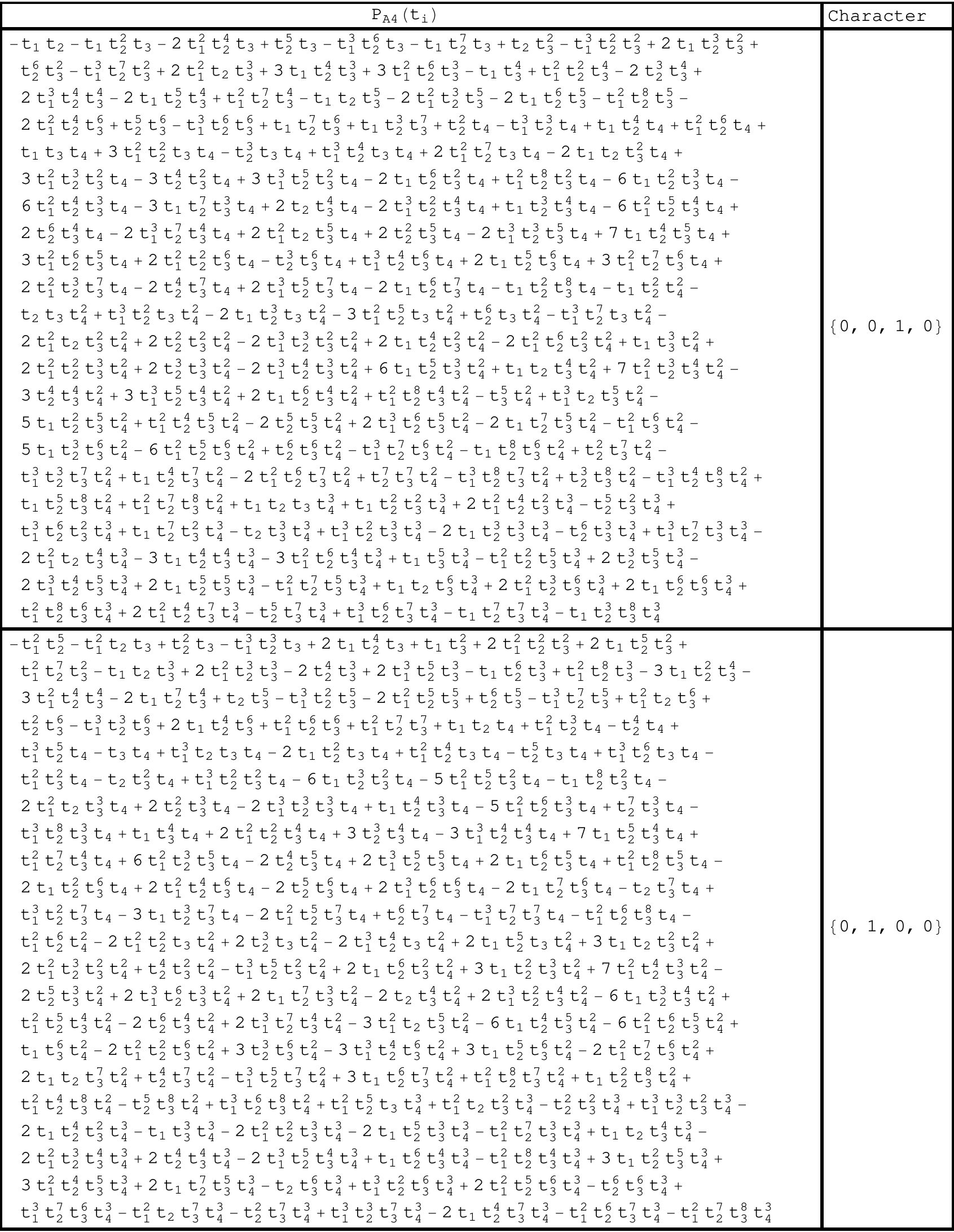}
\clearpage
\pagebreak
\includegraphics[scale=0.85]{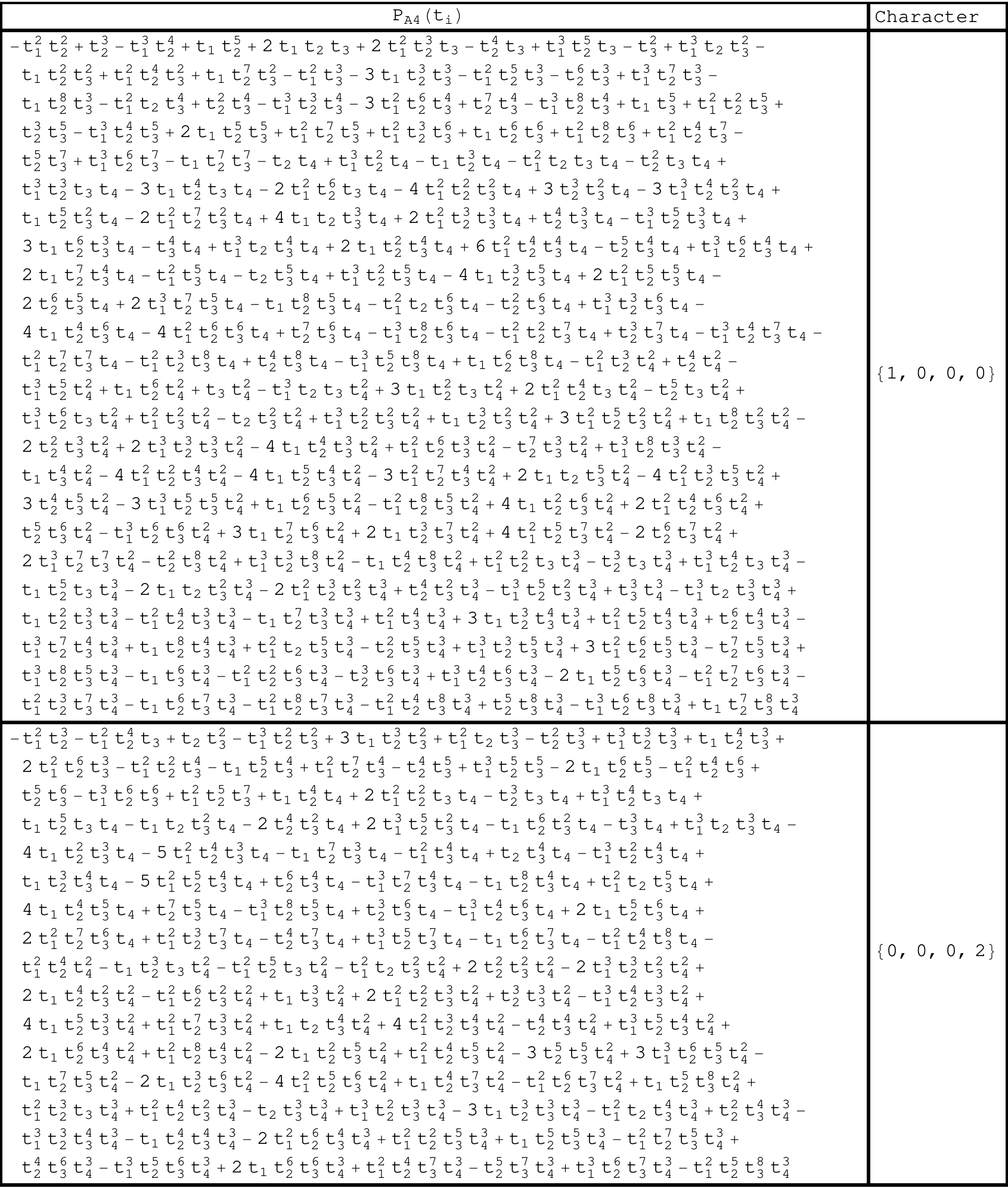}
\clearpage
\pagebreak
\includegraphics[scale=0.85]{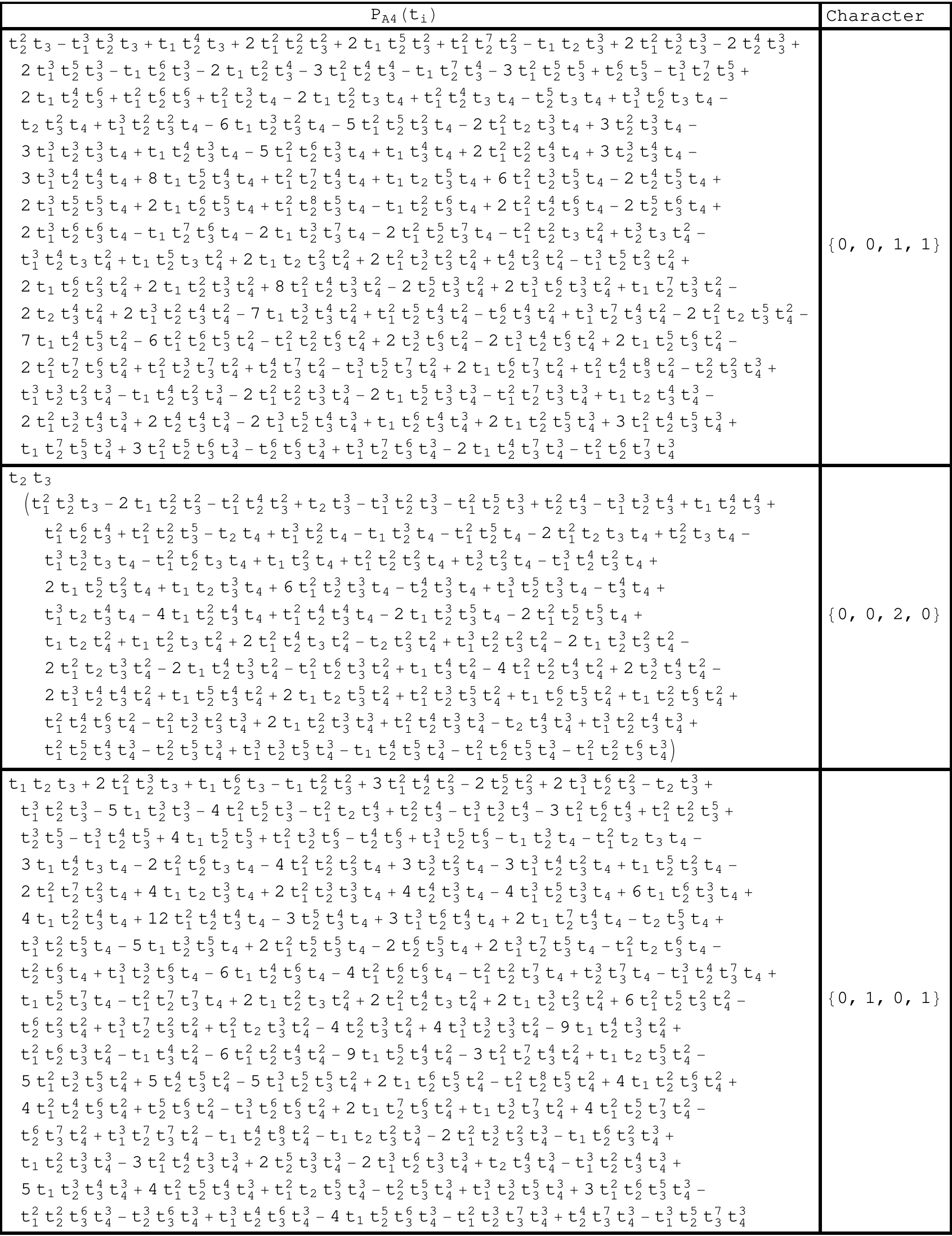}
\clearpage
\pagebreak
\includegraphics[scale=0.85]{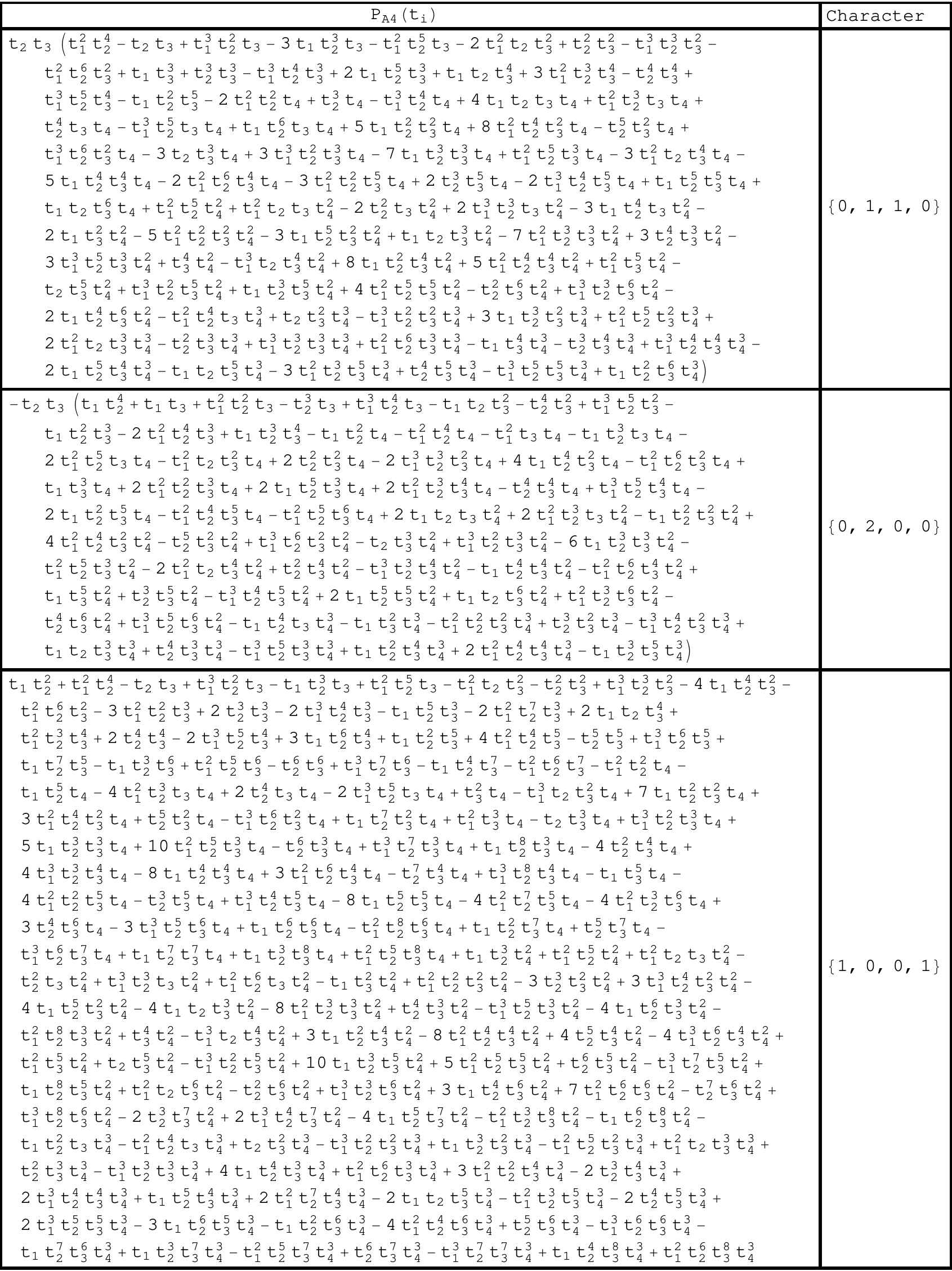}
\clearpage
\pagebreak
\includegraphics[scale=0.85]{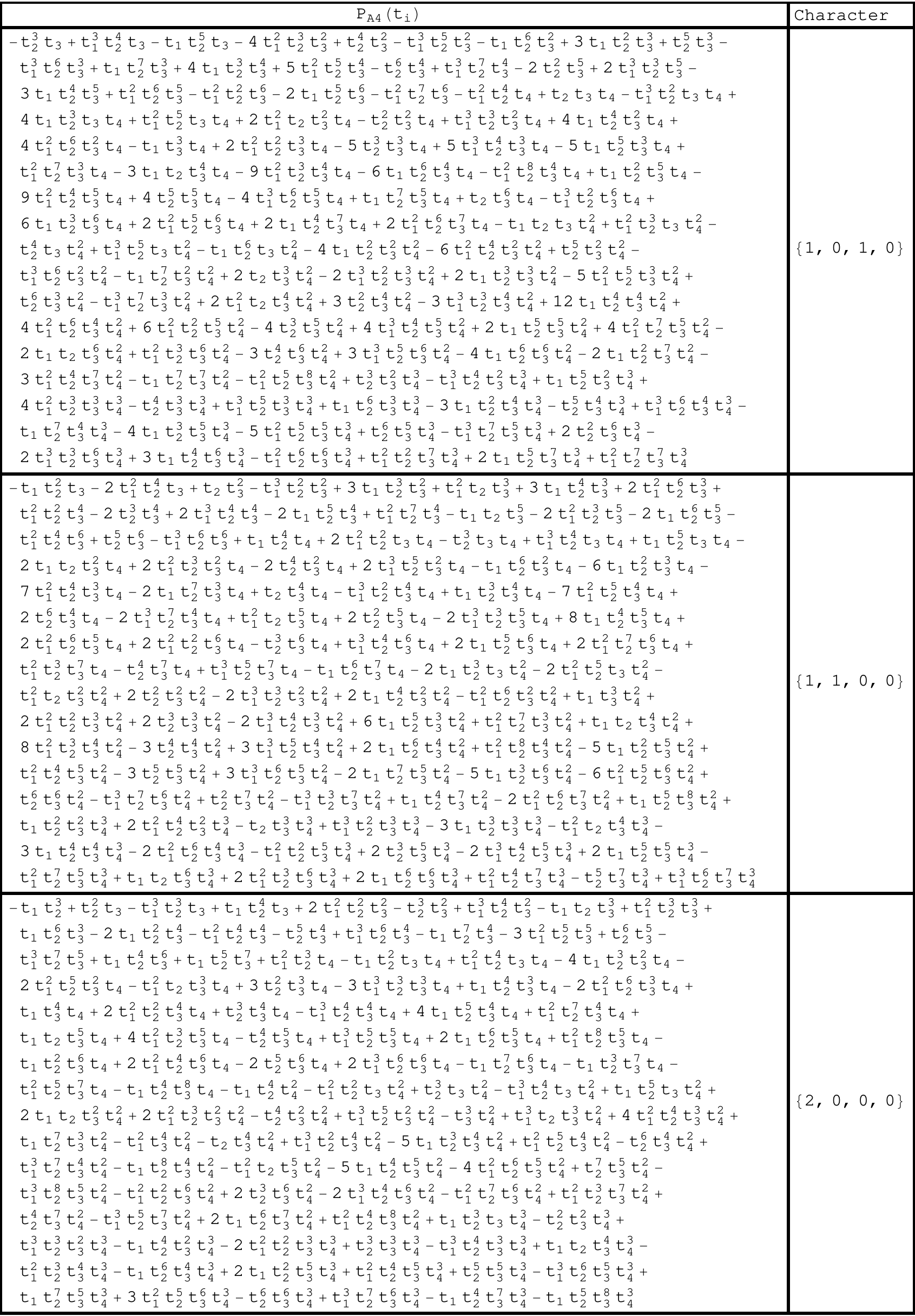}
\clearpage
\pagebreak
\includegraphics[scale=0.85]{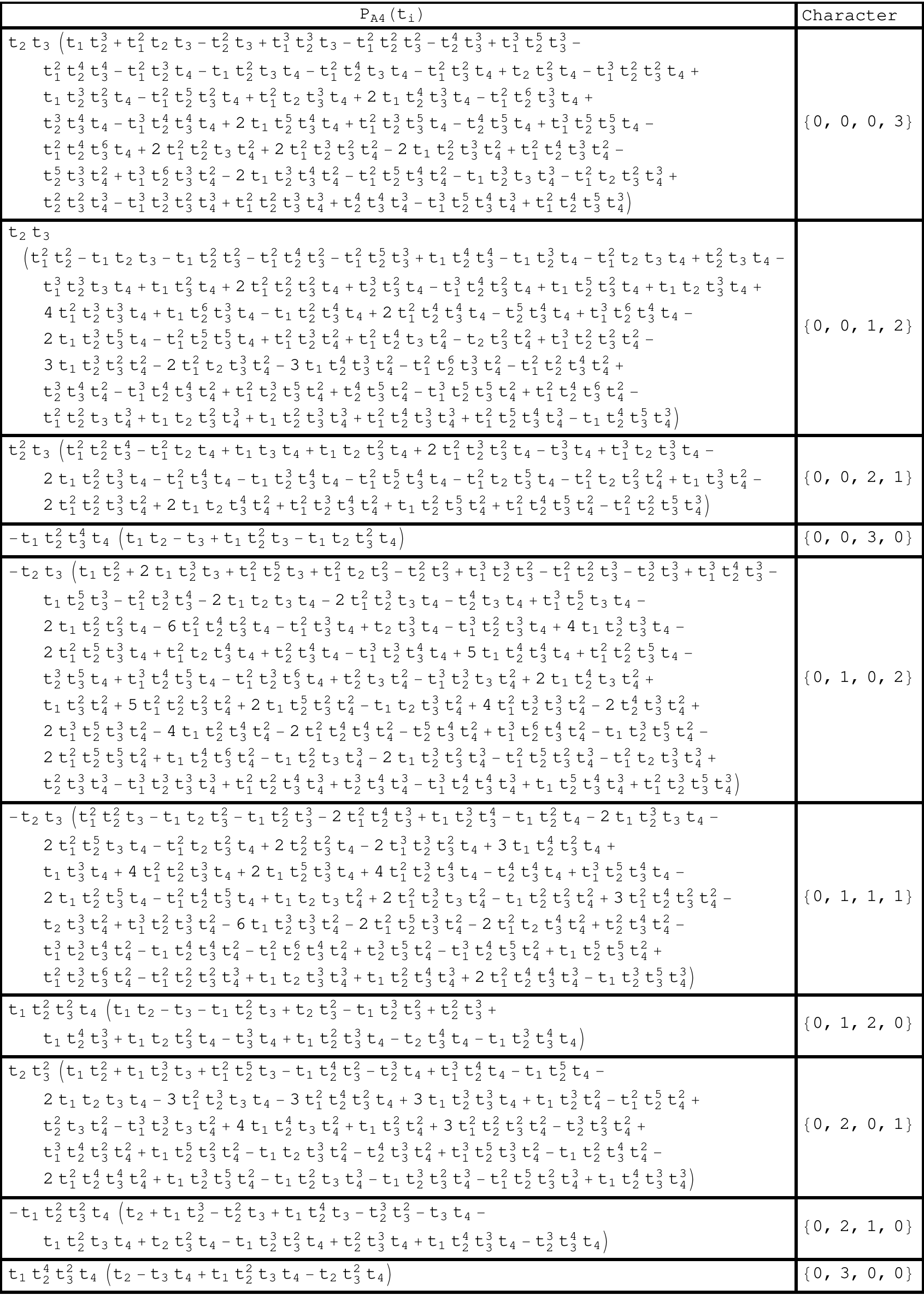}
\clearpage
\pagebreak
\includegraphics[scale=0.85]{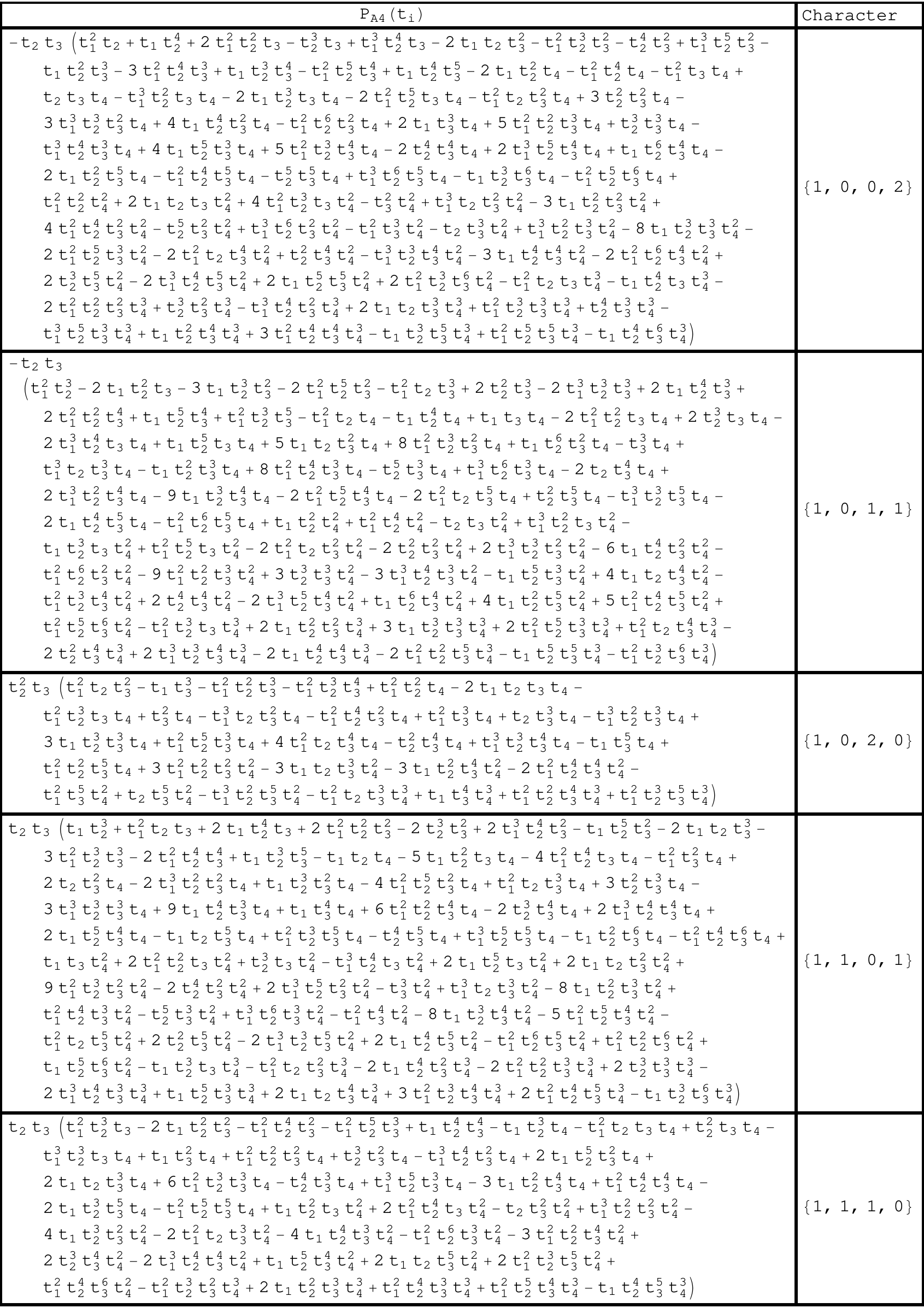}
\clearpage
\pagebreak
\includegraphics[scale=0.85]{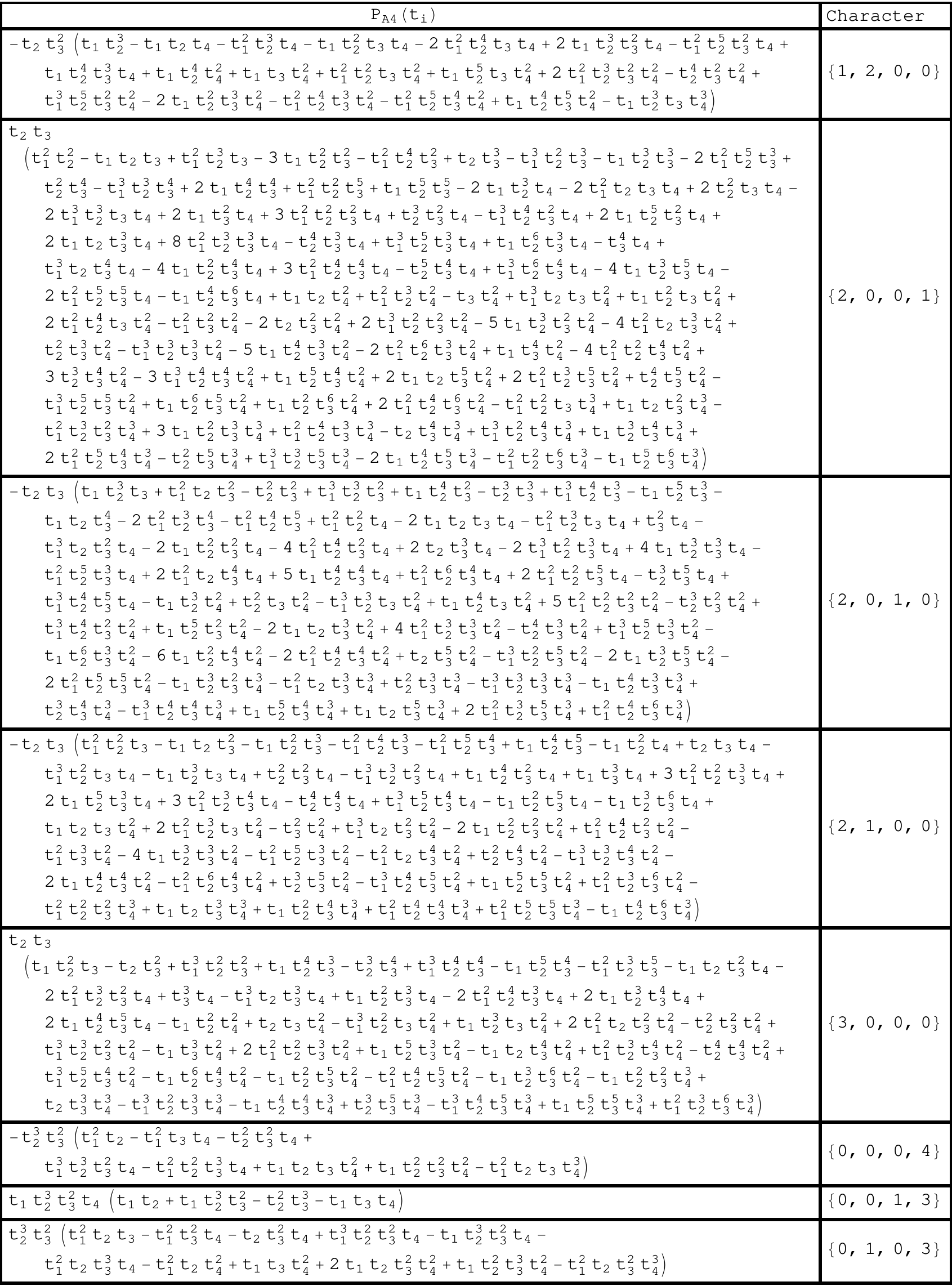}
\clearpage
\pagebreak
\includegraphics[scale=0.85]{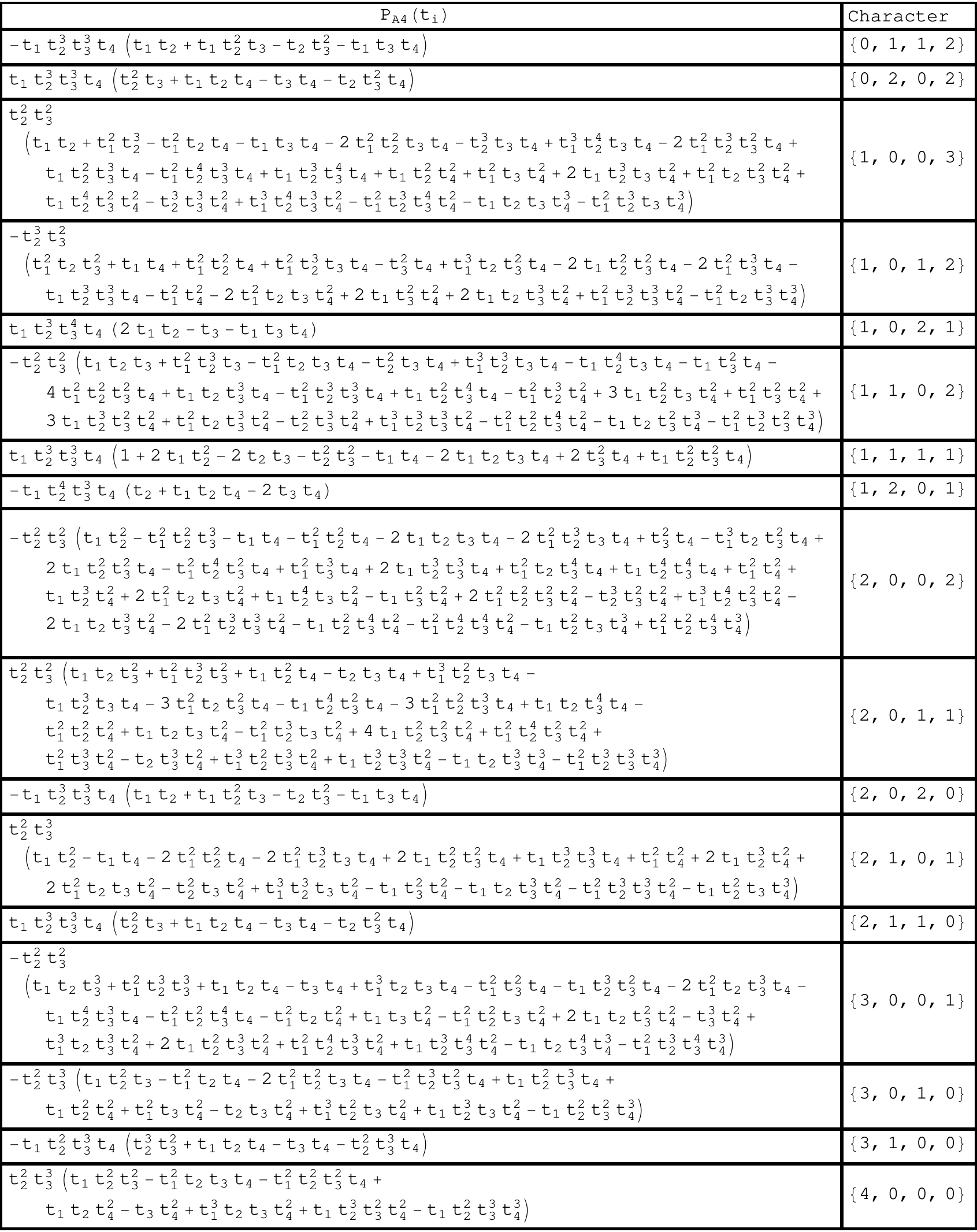}
\clearpage


\subsection{Appendix 5: Generating Function for Dimensions of $A_4$ Irreps}

\includegraphics[scale=0.9]{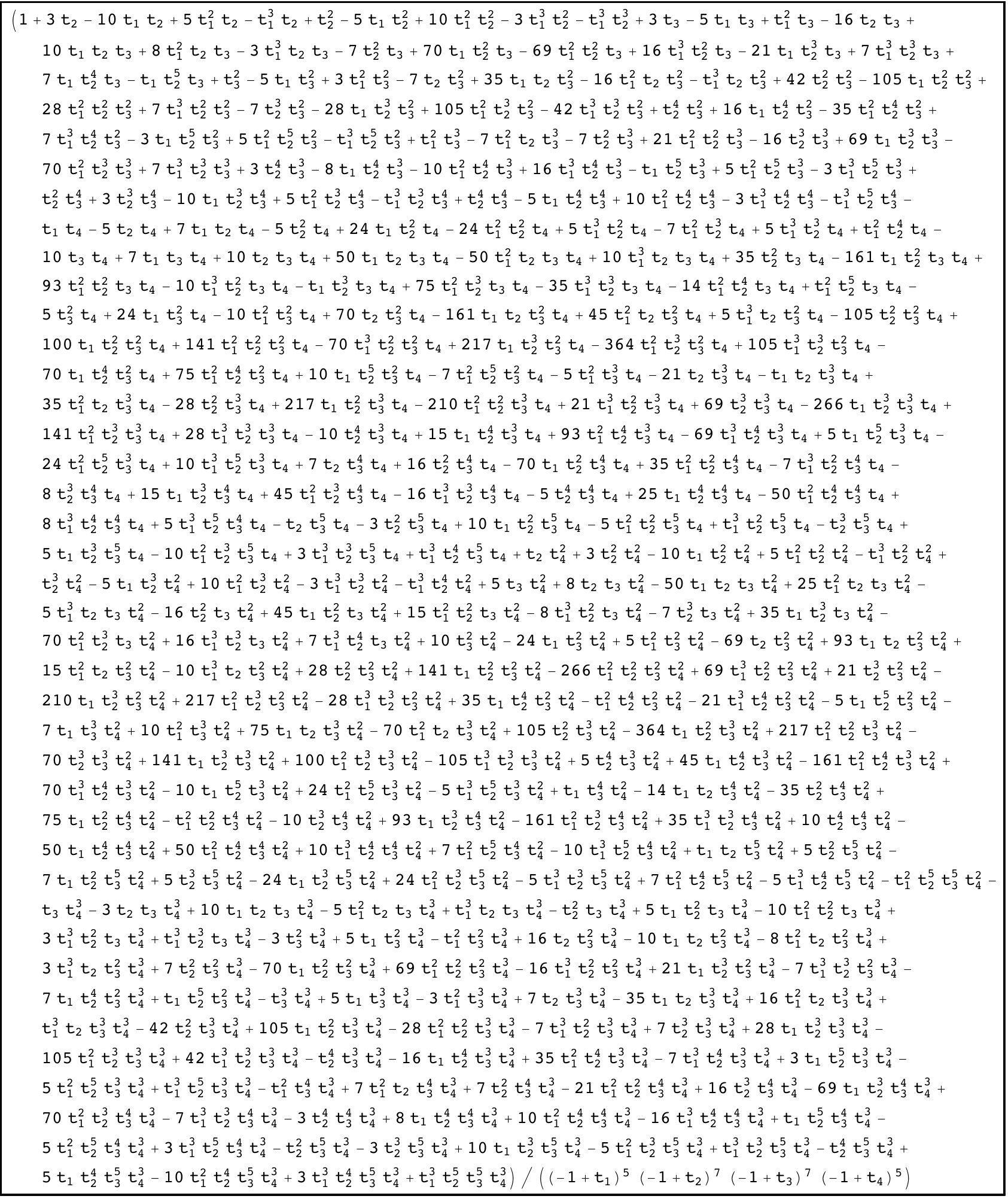}
\clearpage


\bibliographystyle{JHEP}
\bibliography{RJKBibLib}

\providecommand{\href}[2]{#2}\begingroup\raggedright\begin{thebibliography}{10}

\bibitem{Benvenuti:2006qr}
S.~Benvenuti, B.~Feng, A.~Hanany, and Y.-H. He, {\it {Counting BPS Operators in
  Gauge Theories: Quivers, Syzygies and Plethystics}},  {\em JHEP} {\bf 0711}
  (2007) 050, [\href{http://xxx.lanl.gov/abs/hep-th/0608050}{{\tt
  hep-th/0608050}}].

\bibitem{Feng:2007ur}
B.~Feng, A.~Hanany, and Y.-H. He, {\it {Counting gauge invariants: The
  Plethystic program}},  {\em JHEP} {\bf 0703} (2007) 090,
  [\href{http://xxx.lanl.gov/abs/hep-th/0701063}{{\tt hep-th/0701063}}].

\bibitem{Hanany:2008kn}
A.~Hanany and N.~Mekareeya, {\it {Counting Gauge Invariant Operators in SQCD
  with Classical Gauge Groups}},  {\em JHEP} {\bf 0810} (2008) 012,
  [\href{http://xxx.lanl.gov/abs/0805.3728}{{\tt arXiv:0805.3728}}].

\bibitem{Gray:2008yu}
J.~Gray, A.~Hanany, Y.-H. He, V.~Jejjala, and N.~Mekareeya, {\it {SQCD: A
  Geometric Apercu}},  {\em JHEP} {\bf 0805} (2008) 099,
  [\href{http://xxx.lanl.gov/abs/0803.4257}{{\tt arXiv:0803.4257}}].

\bibitem{Hanany:2008sb}
A.~Hanany, N.~Mekareeya, and G.~Torri, {\it {The Hilbert Series of Adjoint
  SQCD}},  {\em Nucl.Phys.} {\bf B825} (2010) 52--97,
  [\href{http://xxx.lanl.gov/abs/0812.2315}{{\tt arXiv:0812.2315}}].

\bibitem{Benvenuti:2010pq}
S.~Benvenuti, A.~Hanany, and N.~Mekareeya, {\it {The Hilbert Series of the One
  Instanton Moduli Space}},  {\em JHEP} {\bf 1006} (2010) 100,
  [\href{http://xxx.lanl.gov/abs/1005.3026}{{\tt arXiv:1005.3026}}].

\bibitem{Hanany:2012dm}
A.~Hanany, N.~Mekareeya, and S.~S. Razamat, {\it {Hilbert Series for Moduli
  Spaces of Two Instantons}},  {\em JHEP} {\bf 1301} (2013) 070,
  [\href{http://xxx.lanl.gov/abs/1205.4741}{{\tt arXiv:1205.4741}}].

\bibitem{Forcella:2008bb}
D.~Forcella, A.~Hanany, Y.-H. He, and A.~Zaffaroni, {\it {The Master Space of
  N=1 Gauge Theories}},  {\em JHEP} {\bf 0808} (2008) 012,
  [\href{http://xxx.lanl.gov/abs/0801.1585}{{\tt arXiv:0801.1585}}].

\bibitem{Forcella:2008eh}
D.~Forcella, A.~Hanany, Y.-H. He, and A.~Zaffaroni, {\it {Mastering the Master
  Space}},  {\em Lett.Math.Phys.} {\bf 85} (2008) 163--171,
  [\href{http://xxx.lanl.gov/abs/0801.3477}{{\tt arXiv:0801.3477}}].

\bibitem{Fuchs:1997bb}
J.~Fuchs and C.~Schweigert, {\em Symmetries, Lie Algebras and Representations}.
\newblock Cambridge University Press, Cambridge, 1997.

\bibitem{Hanany:2008qc}
A.~Hanany, N.~Mekareeya, and A.~Zaffaroni, {\it {Partition Functions for
  Membrane Theories}},  {\em JHEP} {\bf 0809} (2008) 090,
  [\href{http://xxx.lanl.gov/abs/0806.4212}{{\tt arXiv:0806.4212}}].

\bibitem{Feger:2012bs}
R.~Feger and T.~W. Kephart, {\it {LieART - A Mathematica Application for Lie
  Algebras and Representation Theory}},
  \href{http://xxx.lanl.gov/abs/1206.6379}{{\tt arXiv:1206.6379}}.

\bibitem{David-Cox:2007fk}
D.~O. David~Cox, John~Little, {\em Ideals, Varieties and Algorithms}.
\newblock Springer, 2007.

\bibitem{Cvitanovic:2008mw}
P.~Cvitanovic, {\em Group Theory: Birdtracks, Lie Algebras and Exceptional
  Groups}.
\newblock Princeton University Press, Princeton, 2008.

\bibitem{deAzcarraga:1997ya}
J.~de~Azcarraga, A.~Macfarlane, A.~Mountain, and J.~Perez~Bueno, {\it
  {Invariant tensors for simple groups}},  {\em Nucl.Phys.} {\bf B510} (1998)
  657--687, [\href{http://xxx.lanl.gov/abs/physics/9706006}{{\tt
  physics/9706006}}].

\bibitem{Pouliot:2001iw}
P.~Pouliot, {\it {Spectroscopy of gauge theories based on exceptional Lie
  groups}},  {\em J.Phys.} {\bf A34} (2001) 8631--8658,
  [\href{http://xxx.lanl.gov/abs/hep-th/0107151}{{\tt hep-th/0107151}}].

\bibitem{Witten:1982fp}
E.~Witten, {\it {An SU(2) Anomaly}},  {\em Phys.Lett.} {\bf B117} (1982)
  324--328.

\bibitem{Tong:2005un}
D.~Tong, {\it {TASI lectures on solitons: Instantons, monopoles, vortices and
  kinks}},  \href{http://xxx.lanl.gov/abs/hep-th/0509216}{{\tt
  hep-th/0509216}}.

\bibitem{Nekrasov:2004vw}
N.~Nekrasov and S.~Shadchin, {\it {ABCD of instantons}},  {\em
  Commun.Math.Phys.} {\bf 252} (2004) 359--391,
  [\href{http://xxx.lanl.gov/abs/hep-th/0404225}{{\tt hep-th/0404225}}].

\bibitem{Keller:2012da}
C.~A. Keller and J.~Song, {\it {Counting Exceptional Instantons}},  {\em JHEP}
  {\bf 1207} (2012) 085, [\href{http://xxx.lanl.gov/abs/1205.4722}{{\tt
  arXiv:1205.4722}}].

\bibitem{Douglas:1995bn}
M.~R. Douglas, {\it {Branes within branes}},
  \href{http://xxx.lanl.gov/abs/hep-th/9512077}{{\tt hep-th/9512077}}.

\bibitem{Nakajima:2003pg}
H.~Nakajima and K.~Yoshioka, {\it {Instanton counting on blowup. 1.}},  {\em
  Invent.Math.} {\bf 162} (2005) 313--355,
  [\href{http://xxx.lanl.gov/abs/math/0306198}{{\tt math/0306198}}].

\bibitem{Fulton:2004id}
W.~Fulton and J.~Harris, {\em Representation Theory}.
\newblock Springer, New York, 2004.

\bibitem{Hughes:2006ww}
C.~Hughes, ``Haar measure and weyl integration.'' 2006.

\bibitem{Slansky:1981yr}
R.~Slansky, {\it {Group Theory for Unified Model Building}},  {\em Phys.Rept.}
  {\bf 79} (1981) 1--128.

\end{thebibliography}\endgroup


\end{document}